\documentclass[aps,showpacs,twocolumn,superscriptaddress,prb,floatfix, 10pt]{revtex4-1}
\usepackage{graphicx}
\usepackage{subfigure}
\usepackage{amssymb,amsmath,cancel}
\usepackage{comment}
\usepackage{xcolor}
\usepackage{array}

\begin{document}

\title{Topological orders of monopoles and hedgehogs:\\From electronic and magnetic spin-orbit coupling to quarks}

\author{Predrag Nikoli\'c}
\affiliation{Department of Physics and Astronomy,\\George Mason University, Fairfax, VA 22030, USA}
\affiliation{Institute for Quantum Matter at Johns Hopkins University, Baltimore, MD 21218, USA}

\date{\today}

%
%
%

\begin{abstract}

Topological states of matter are, generally, quantum liquids of conserved topological defects. We establish this by constructing and analyzing topological field theories which introduce gauge fields to describe the dynamics of singularities in the original field configurations. Homotopy groups are utilized to identify topologically protected singularities, and the conservation of their protected number is captured by a topological action term that unambiguously obtains from the given set of symmetries. Stable phases of these theories include quantum liquids with emergent massless Abelian and non-Abelian gauge fields, as well as topological orders with long-range quantum entanglement, fractional excitations, boundary modes and unconventional responses to external perturbations. This paper focuses on the derivation of topological field theories and basic phenomenological characterization of topological orders associated with homotopy groups $\pi_n(S^n)$, $n\ge 1$. These homotopies govern monopole and hedgehog topological defects in $d=n+1$ dimensions, and enable the generalization of both weakly-interacting and fractional quantum Hall liquids of vortices to $d>2$. Hedgehogs have not been in the spotlight so far, but they are particularly important defects of magnetic moments because they can be stimulated in realistic systems with spin-orbit coupling, such as chiral magnets and $d=3$ topological materials. We predict novel topological orders in systems with U(1)$\times$Spin($d$) symmetry in which fractional electric charge attaches to hedgehogs. Monopoles, the analogous defects of charge or generic U(1) currents, may bind to hedgehogs via Zeeman effect, or effectively emerge in purely magnetic systems. The latter can lead to spin liquids with different topological orders than that of the RVB spin liquid. Charge fractionalization of quarks in atomic nuclei is also seen as possibly arising from the charge-hedgehog attachment.

\end{abstract}

\maketitle
\tableofcontents

\section{Introduction}

Our understanding of quantum matter rests upon universal behaviors of particles. We can sharply distinguish the states of matter by symmetry, or by their qualitative response to local perturbations. A more subtle distinction is based on non-local state properties, mathematically expressed through topological invariants -- state functions that are not affected by any smooth local transformation. The list of known and envisioned topological systems has been growing steadily since the earliest proposals for spin liquids \cite{anderson73b} and the discovery of quantum Hall states \cite{Klitzing1980, Tsui1982}. These original two-dimensional systems are in most cases shaped by strong interactions between particles, and possess topological order \cite{Wen1990b} -- the ground state spontaneously selects the value of a topological invariant (through a long-range quantum entanglement of many particles \cite{Kitaev2006, Chen2010}) instead of an order parameter that breaks a symmetry. More recent discoveries of topological materials based on spin-orbit coupling \cite{Qi2006, Bernevig2006, Konig2007}, including three-dimensional topological insulators \cite{Fu2007, Moore2007, Chen2009b} and (semi)metals \cite{Ari2010, Burkov2011a, Kuroda2017, Ikhlas2017, Armitage2018}, have inspired explorations of electron interaction effects \cite{Krempa2013, Moon2013} that could potentially produce topological order \cite{Cho2010, Maciejko2010, Hoyos2010, Swingle2011, Levin2011, Maciejko2012, Swingle2012, Juan2014, Maciejko2014, Chan2015, Fradkin2017}. Promising candidates for three-dimensional interacting topological systems include some Kondo insulators \cite{Nikolic2014b, Fuhrman2014}, topological magnetic semimetals and quantum spin-ice materials \cite{Gingras2014}.

\begin{figure}
\subfigure[{}]{\includegraphics[height=1.5in]{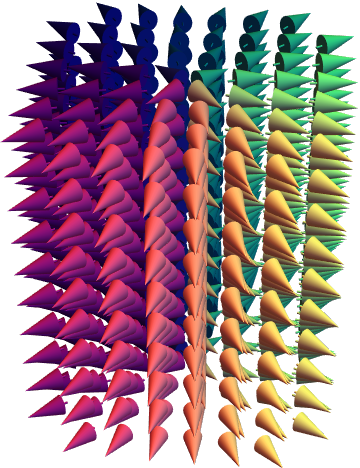}}
\subfigure[{}]{\includegraphics[height=1.5in]{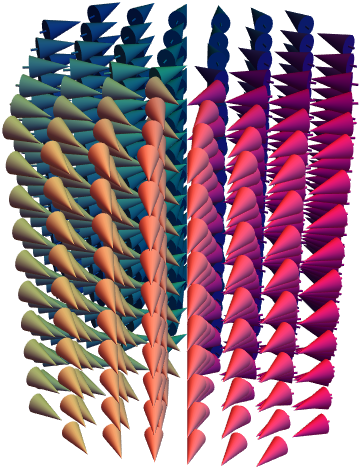}}
\subfigure[{}]{\includegraphics[height=1.5in]{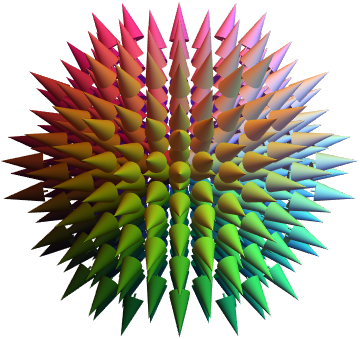}}
\subfigure[{}]{\includegraphics[height=1.5in]{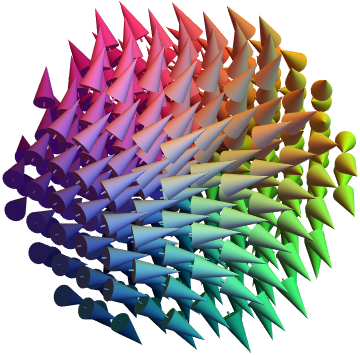}}
\caption{\label{defects}Examples of singular vector field configurations: (a) vortex, (b) antivortex, (c) hedgehog, (d) antihedgehog. Only the point defects like (c) and (d) are topologically protected in $d=3$ dimensions.}
\end{figure}

The purpose of this paper is to derive and analyze topological field theories that describe both conventional and topologically ordered phases of spinor fields. Our ultimate goal is to predict and characterize novel topological orders which may be possible to realize in correlated three dimensional materials. The spinor field $\psi$ represents vector $\hat{\bf n}({\bf r)}$ local degrees of freedom such as spins or staggered moments, and carries a U(1) phase associated with charge currents or an emergent symmetry. The vector field $\hat{\bf n}$ supports hedgehogs as topological defects with a point singularity, shown in Fig.\ref{defects}. The U(1) phase supports vortex singularities, which are topologically protected only in two dimensions. Vortex loops in three dimensions are not topologically protected since they can continuously shrink to a point and vanish. Nevertheless, the diffusion of vortex loops is captured by an emergent U(1) gauge field $A_\mu$, which can support its own quantized point singularities -- topologically protected monopoles. Both monopoles and hedgehogs can be generalized to higher dimensions $d$ and enumerated by integer topological invariants of the homotopy group $\pi_{d-1}(S^{d-1})$. They will be the main protagonists in this paper because topologically ordered phases will be seen as quantum disordered states in which the number of delocalized topological defects is conserved by the mechanism of topological protection.

To make progress, we first formulate a universal approach to topological orders. We apply singular gauge transformations to derive emergent gauge fields from the topological singularities of the physical fields. The flux of such a gauge field is nothing but the invariant of the homotopy group that classifies the singularities \cite{Mermin1979, Nakahara2003, StoneGoldbart2009}. Therefore, a localized singularity becomes the source of a \emph{flux quantum} in a symmetry-broken phase. Quantum fluctuations that restore symmetries can diffuse this flux and give the gauge field its own dynamics. If the flux remains conserved despite the fluctuations, one obtains topological orders whose hierarchy is uniquely determined by the homotopy and symmetry. The emergent gauge field is indistinguishable from a putative fundamental gauge field of the same kind, raising the possibility that a singularity extraction is the fundamental mechanism for the appearance of gauge fields in nature (this echoes the elaborate demonstrations in models \cite{wen02c, wen03b}). Guided by the homotopy classification of topological defects \cite{Mermin1979}, this approach naturally generalizes electron fractionalization to any applicable degrees of freedom in arbitrary dimensions. It transparently identifies a real-space ``magnetic'' field behind any topological state of matter (see Ref.\cite{Frohlich1992, Nikolic2011a, Nikolic2012} for examples with spin-orbit coupling), and stands as an alternative to the popular slave boson method (which introduces by hand the parton fields of a fractionalized electron and a gauge field to suppress the enabled unphysical fluctuations). 

We further extend the previous studies of topological orders by applying the above approach to spinors in arbitrary $d$ dimensions. We predict the existence of new topological orders in systems with U(1)$\times$SU(2) or general U(1)$\times$Spin($d$) symmetry, where a fractional amount of U(1) charge becomes attached to a hedgehog defect of an SU(2) or Spin($d$) order parameter. We reveal various interesting properties of these topological orders related to quantum entanglement, and their notable survival at finite temperatures (in contrast to fractional quantum Hall states). Earlier studies have focused on the attachment of charge to U(1) monopoles, giving rise to dyons in high energy physics \cite{Witten1979, Goldhaber1976, Goldhaber1989, Shnir2005} and magnetoelectric effect in condensed matter physics \cite{Cho2010, Maciejko2010, Hoyos2010, Swingle2011, Levin2011, Maciejko2012, Swingle2012, Juan2014, Maciejko2014, Chan2015, Fradkin2017, Vishwanath2013}; we reproduce some of their results here for completeness. However, we stress that hedgehogs are more physically accessible than monopoles since the spin-orbit coupling in topological materials naturally tends to stimulate their existence. Monopoles can be nucleated and bound to hedgehogs via the same mechanism which binds vortices to skyrmions in some chiral magnets and yields a ``topological'' Hall effect \cite{Nagaosa2010, Nagaosa2012}. Hedgehogs and skyrmions have been found in various chiral magnets \cite{Muhlbauer2009, Fujishiro2019}, perhaps even in a chiral spin liquid state \cite{Machida2010}. Hence, the topological orders based on hedgehogs could exist at least in principle in the systems like chiral magnets and general three-dimensional topological materials. Hedgehogs have been considered in high-energy physics mainly in the context of Higgs fields \cite{Shnir2005}.

A significant portion of our analysis is devoted to the basic characterization of the phases captured by the field theory. Apart from the conventional long-range ordered and gapped disordered phases, we identify a hierarchy of quantum disordered phases with Abelian and non-Abelian massless gauge bosons, as well as topologically ordered incompressible quantum liquids. The former includes the phases familiar from the literature on U(1) spin liquids \cite{Hermele2004a, Savary2016}, and their generalizations to non-Abelian structures and higher dimensions. The topological orders we find form a large hierarchy of fractionalized states in higher dimensions, just like the fractional quantum Hall states in two dimensions. The incompressible quantum liquids of monopoles are more constrained than those of hedgehogs due to the fact that charge attached to a monopole nucleates a quantized angular momentum in the surrounding electromagnetic field \cite{Dirac1931}. Nevertheless, we are not restricted by time-reversal symmetry and hence the monopole liquids we discuss are less constrained than those considered in the recent literature \cite{Fradkin2017}. If the U(1) symmetry emerges at low energies in a purely magnetic system, the obtained fractionalized states are chiral spin liquids with different topological orders than the more familiar resonant-valence-bond (RVB) spin liquids.

Topological quantum entanglement is always evident in the ground state degeneracy, but need not show up in braiding operations. We find that the hedgehog quantum liquids scramble their topological order behind trivial particle-loop braiding (unlike the monopole ones), although more complicated linked-loop braiding \cite{Wang2014a, Jiang2014, Wang2014b, Juan2014, Wang2015} should be explored further. We point out that braiding operations between particles can also be interesting. They are normally cast away because the only topologically protected aspect of particle braiding in higher dimensions is their bosonic or fermionic statistics. However, the fractional quasiparticles with internal degrees of freedom (spin) necessarily live in a long-range entangled state and hence admit non-trivial ``dynamically'' protected braiding operations. Dynamical protection against local noise stems from the finite energy cost of disturbances in an incompressible quantum liquid. While the \emph{topologically} protected particle-loop braiding is Abelian in the theories we consider, a \emph{dynamically} protected braiding statistics specified by additional data about the braiding operation can still be non-Abelian.

On the purely theoretical front, the topological field theory we construct is a variant of the ``background field'' (BF) theory \cite{Cho2010, Vishwanath2013, Chan2015, Fradkin2017} with antisymmetric tensor gauge fields. Here we emphasize a new ingredient of such theories, the ``linking'' Lagrangian terms. These terms arise in the recursive extraction of the gauge fields from topological singularities, and play a crucial role in eliminating unphysical gauge symmetries and shaping the phase diagram. They also enable a certain perspective on some fundamental questions in field theory that we will stumble upon: (i) why all fundamental fields carry the same charge, (ii) can the charge coupled to a non-Abelian gauge field be deconfined, and (iii) why the quarks have fractional charge.

We will analyze the topological orders of hedgehogs and monopoles in an arbitrary number $d\ge 2$ of spatial dimensions for two reasons. First, this generalization will provide a valuable insight and confidence about many unusual results that we obtain (various phenomena occur in all dimensions in qualitatively the same way). Second, we wish to address the important open problem of topological order classification \cite{WenQFT2004, Levin2005, Gu2010, Wen2019a, Wen2019}. Our analysis identifies homotopy as a universal parameter that classifies hierarchies of topological orders, and an obvious first case to study is the well-known infinite sequence of non-trivial homotopy groups $\pi_n(S^n)$ which pertain to the continuous maps from an $n$-sphere to an $n$-sphere.


\subsection{The summary of results and paper layout}

We begin the technical discussion in Section \ref{secQH} by illustrating the main ideas with the familiar example of two-dimensional quantum Hall liquids. Then, we generalize to $d$ dimensions in Section \ref{secGauge} and show how antisymmetric tensor gauge fields capture the singularities of charge and spin currents. Point defects are represented by rank $d-1$ gauge fields in $d$ dimensions, line defects by rank $d-2$, etc, down to rank 1 gauge fields that minimally couple to charge or spin currents. Monopoles and hedgehogs define two separate rank-hierarchies of gauge fields, Abelian and non-Abelian respectively. The conventional part of the effective field theory is formulated in Section \ref{secDyn}. After taking care to not allow unphysical gauge symmetries, we identify the hierarchy of phases where the switch from Coulomb-like to Higgs-like dynamics occurs at some rank $1\le n<d$. The dynamics at the highest rank also admits topologically ordered phases that are protected according to the $\pi_{d-1}(S^{d-1})$ homotopy group. The Coulomb dynamics at rank $k$ features deconfined ``topological'' defects of the rank $k-1$ gauge field. We find that the asymptotically free ``charge'' coupled to a non-Abelian or compact rank $k-1$ gauge field also becomes deconfined in the rank $k$ Coulomb state. This promotion of asymptotic freedom to true freedom ultimately enables the fractionalization of charge and spin, if the homotopy provides an opportunity. In Section \ref{secTopTerm}, we construct the topological Lagrangian density terms consistent with symmetries in order to capture the topological protection in incompressible quantum liquids. Then, Section \ref{secTopOrder} presents the basic analysis of the stable topological orders in the obtained theories.

We show in Section \ref{secFract} that both monopoles and hedgehogs can independently shape topological orders in incompressible quantum liquids. A necessary stability condition is the rational quantization of monopole and hedgehog ``filling factors'', in direct analogy to quantum Hall liquids. These filling factors play a role in the character of fractional quasiparticle excitations (Section \ref{secFract}), the topological ground state degeneracy on non-simply connected manifolds (Section \ref{secTopDeg}), and various aspects of quantum entanglement (Section \ref{secBraid}). The topological ground state degeneracy is the defining property of topological order by the virtue of being the only resilient property against all possible perturbations that leave the energy gap open. This degeneracy is found to have a certain classical character in $d\ge 3$ dimensions (Section \ref{secTopDeg}): it can be infinite in some cases (with a topological sector defined for each value of a classical topological invariant), and it protects the topological order as a thermodynamic phase at finite temperatures in $d\ge 3$.


Further restrictions of topological orders are obtained in Section \ref{secMicroPart} from the requirement that electrons or spin waves be the microscopic degrees of freedom. This reduces the simple hedgehog topological orders to a Laughlin-like sequence of fractional states, while more complicated quantum liquids can arise only hierarchically as in the case of quantum Hall states. Interestingly, the topological orders of hedgehogs scramble their identity in ordinary braiding operations. The fractionalization by monopoles in $d=3$ is more complicated due to the emergent spin of charge-monopole pairs, but represents a more natural generalization of quantum Hall states. We discuss both topologically and dynamically protected manifestations of quantum entanglement in braiding operations. We point out that dynamically protected non-Abelian braiding may be possible owing to the existence of entangled internal degrees of freedom (spin), but leave systematic calculations for future studies.

Topological order of spins without charge degrees of freedom can arise in two forms. First, some mechanism may reduce the full spin symmetry down to U(1). This is a path to both U(1) and gapped spin liquids, here seen to arise from the fluctuations of local spins that remain well-defined at some coarse-grained length-scales instead of being bound into short-range singlets. The ensuing gapped spin liquids, which attach emergent U(1) charge to monopoles, are different from the resonant-valence-bond (RVB) spin liquids (Section \ref{secMicroPart}). The second form obtains in slave-boson theories with spin-orbit coupling \cite{Pesin2010}. A local constraint that controls the fermion density in a Mott insulator introduces an emergent gauge field, so a deconfined charge associated with it can bind to spinon hedgehogs through the spin-orbit coupling. We construct the ensuing topological orders in Section \ref{secMicroPart}, assuming naively that the emergent gauge symmetry is U(1).

We end the analysis with a basic argument supporting the existence of protected soft boundary modes (Section \ref{secEqMot}), and a brief consideration of the interesting topological response -- especially fractional magnetoelectric and Kerr effects that can be expected in the cases of fractionalization by monopoles (Section \ref{secME}). The concluding Section \ref{secConclusions} explores the prospects for realizing monopole and hedgehog topological orders in real systems. We explain why chiral magnets, correlated topological semimetals or insulators, and quantum spin-ice materials are promising candidate materials, which in some cases might be able to stabilize new topological orders. We also speculate that a glimpse of a topological order discussed here might have been already found in nature -- inside atomic nuclei.

Various properties of multi-dimensional theories with tensor gauge fields are presented in appendices, including the forms of non-Abelian Maxwell couplings, duality mappings, canonical formalism and braiding operations.

We use the following conventions in this paper. All discussions employ ``natural'' $\hbar = c = 1$ units, except in the parts of Section \ref{secME} where we switch to Gaussian units. Space-time directions are labeled by Greek indices $\mu,\nu,\lambda\dots$, and spatial directions are labeled by Latin indices $i,j,k\dots$. Repeated indices are summed over, and $\mu=0$ is temporal direction. We mostly work in imaginary time and do not distinguish between upper and lower indices. The use of real time is announced when needed, and further emphasized by separating lower and upper indices. Indices pertaining to internal degrees of freedom are labeled with $a,b,c$.

\clearpage

\section{Example: Quantum Hall liquids}\label{secQH}

Consider a superfluid at zero temperature in $d=2$ dimensions. The order parameter characterizing the ground state is a complex scalar function $\psi = |\psi| e^{i\theta}$ of coordinates ${\bf x}\in \mathbb{R}^2$; its phase $\theta$ is well-defined because the ground state breaks the U(1) symmetry. If a single quantized vortex is placed at the origin, then $\psi$ becomes singular at the origin and $\theta$ winds by $2\pi$ on any loop that encloses the singularity. Let us define a ``singularity gauge field'':
\begin{equation}\label{singA1a}
A_j = -i e^{-i\theta}\partial_j e^{i\theta} \ ,
\end{equation}
or alternatively:
\begin{equation}\label{singA1b}
A_j = \partial_j \theta
\end{equation}
with understanding that the gradient $\partial_j$ is smooth, i.e. blind to $2\pi$ discontinuities of $\theta({\bf x})\in\lbrack 0,2\pi)$. The $2\pi$ phase winding is reflected in the following contour integral on any spatial loop $\mathcal{C}$ that encloses the origin:
\begin{equation}
\oint\limits_{\mathcal{C}} dx_j A_j = 2\pi
\end{equation}
Then, we can use Stokes' theorem to reveal that $A_j$ carries a singular quantized flux:
\begin{equation}
\epsilon_{\mu\nu\lambda} \partial_\mu A_\lambda = 2\pi \delta_{\mu 0} \, \delta({\bf x}) \ .
\end{equation}
Mathematically, $A_j$ can be extracted from $\psi$ by a singular gauge transformation which keeps $\partial_j \theta + A_j$ invariant. If the particles have physical charge $e$, then $A_j$ must be combined with the fundamental electromagnetic gauge field $A_j^{\textrm{em}}$ in the gauge-invariant form $\partial_j^{\phantom{x}} \theta + A_j^{\phantom{x}} - e A_j^{\textrm{em}}$.

We can describe many vortices using a ``singularity gauge field'' $A_j$. This is redundant as long as vortices are not moving, since the superfluid state is accurately described by the order parameter $\psi$ and its phase $\theta$. But, what happens if quantum fluctuations destroy the superfluid long range order by liberating and delocalizing the vortices? The ensuing state with restored U(1) symmetry can no longer be described by a finite complex order parameter $\psi$ because its phase $\theta$ is fluctuating too much to be well-defined. Interestingly, the gauge field $A_j$ can continue to provide a useful description of the state. The gauge flux that was originally singular and associated with quantized vortices can now diffuse and continuously spread in space, producing a smoothly varying ``magnetic'' field $B$:
\begin{equation}\label{singB}
\epsilon_{\mu\nu\lambda} \partial_\nu A_\lambda \to \delta_{\mu 0} \, B({\bf x}) \ .
\end{equation}
Suppose we find $B=0$. The most typical disordered state with $B=0$ is an ordinary Mott insulator. Its proper description requires a lattice, and then a duality mapping \cite{Dasgupta1981, Fisher1989} portrays it as a ``vortex condensate''. The condensation of vortices implies that the number of vortices is not conserved (in any condensate, a well-defined phase will render its canonically conjugate observable, the particle number, undetermined due to Heisenberg uncertainty). This phenomenon can be very easily understood on a lattice even without a detailed duality derivation. The only way to probe the instantaneous presence of a vortex in some region is to analyze the winding of the order parameter's phase $\theta$ on a spatial loop that encloses that region. When fluctuations destroy the original superfluid by generating many vortices and antivortices, one is forced to use only very small probing loops whose size does not exceed the average separation $l$ between vortices and antivortices. In fact, $l$ must be of the order of the lattice constant because no length scale other than the ultra-violet cutoff is available to control the density of vortices and antivortices. Going from one site to another around such a small loop, $\theta$ changes discontinuously and there is no way to distinguish configurations with $0, \pm 2\pi, \pm 4\pi, \dots$, etc. phase winding. Vorticity is quantized in units of $2\pi$, and hence we cannot consider it conserved in a Mott insulator.

Another quantum disordered state is possible in two dimensions: a quantum Hall liquid. It normally takes applying an external magnetic field to stabilize it, so it should be naturally characterized by $B\neq 0$ in (\ref{singB}). Most quantum Hall liquids are fractional and possesses topological order, which means that some defining property of their ground state cannot be disturbed by any smooth and local rearrangement of its degrees of freedom. Going back to vortices in a superfluid, we can easily identify a candidate for one such property, the total vortex charge (vorticity). A single uncompensated vortex in a two-dimensional superfluid costs energy $E\propto\ln(L)$ that scales as the logarithm of the system size $L$. Introducing an antivortex at distance $r$ from the vortex will reduce this energy cost down to a finite value $E\propto\ln(r)$. The infra-red divergent energy barrier to having uncompensated vorticity allows only vortex-antivortex pairs to be created or destroyed, and acts as a powerful agent that conserves the total vortex charge in the system. This conclusion is based on the continuum-limit analysis, which assumes that the order parameter phase $\theta$ is coherent on finite and sufficiently large length-scales $\xi$ in comparison to the lattice constant $a$, and hence avoids the described Mott insulator scenario for flux non-conservation. In addition to pure energy reasons, there is also an entropy component to the conservation of vortex charge: nucleating a single vortex requires adjusting the local degrees of freedom in a macroscopically large portion of the system that extends at least in proportion to the system's linear size $L$. For example, one can smoothly deform a vortex to completely consume its phase winding into a $2\pi$ phase discontinuity across a semi-infinite string that terminates at the singularity. However, the string itself cannot be removed by any smooth transformation. This constitutes a topological protection of the vortex charge. Now, we can imagine a state in which vortices and antivortices move and destroy long-range superfluid coherence, but their total number \emph{remains conserved} and topologically protected. Vortex world-lines are closed loops in 2+1 dimensional space-time. This state is clearly not a plain Mott insulator, and it can be sharply distinguished from a Mott insulator in the thermodynamic limit.

Let us construct an effective action for the quantum liquid with conserved vortex charge in the $L\gg\xi\gg a$ limit. Note that $\xi\sim L$ is a superfluid, and $\xi\sim a$ is a Mott insulator. Formally, the effective action obtains by coarse-graining the microscopic action down to the phase coherence length-scale $\xi$, which involves integrating out all high-energy modes with wavevectors $k>\xi^{-1}$ in the path integral. We argued that the gauge field (\ref{singA1b}) is a useful quantity to describe such a quantum liquid, and so the dynamics may be captured by a certain Maxwell term in the action. However, we should be more concerned about conserving flux. The relevant dynamics for flux conservation is defined at the lattice scale $a\ll\xi$ which is not accessible in the effective theory. Consequently, the effective Lagrangian density must acquire a \emph{topological term} $\mathcal{L}_{\textrm{t}}$ that explicitly implements flux conservation. We cannot microscopically derive $\mathcal{L}_{\textrm{t}}$, but we can construct it by following very stringent fundamental requirements:
\begin{enumerate}
\item $\mathcal{L}_{\textrm{t}}$ may not introduce new degrees of freedom;
\item $\mathcal{L}_{\textrm{t}}$ cannot have any physical effect in conventional states;
\item $\mathcal{L}_{\textrm{t}}$ must not change any symmetries.
\end{enumerate}
Using an independent Lagrange multiplier in the path integral to enforce flux conservation is not the best option because it could violate the first requirement (the Lagrange multiplier would become a dynamical field after an approximate treatment). Instead, we will show that
\begin{equation}\label{LtQHa}
\mathcal{L}_{\textrm{t}} \sim i\, J_\mu \mathcal{J}_\mu
\end{equation}
satisfies all requirements in imaginary time, where $J_\mu$ and $\mathcal{J}_\mu$ are the physical particle and vortex currents respectively:
\begin{equation}
J_\mu \propto \partial_\mu\theta + A_\mu \quad,\quad \mathcal{J}_\mu = \epsilon_{\mu\nu\lambda} \partial_\nu A_\lambda \ .
\end{equation}
If the particles have charge $e$, then one should also include the fundamental electromagnetic gauge field $A_\mu^{\textrm{em}}$ through the replacement $A_\mu^{\phantom{x}} \to A_\mu^{\phantom{x}}-e A_\mu^{\textrm{em}}$ in all formulas (required by gauge invariance). We have implicitly carried out a \emph{singular} gauge transformation (\ref{singA1b}) to transfer vortex singularities from the phase $\theta$ to the gauge field $A_\mu$, while keeping the gauge-invariant current $J_\mu$ unaltered. Vortex configurations are well-defined below the coherence length-scale $\xi$ and a related time-scale, so the phase $\theta$ becomes smooth across distances $\xi$ after the singular gauge transformation. However, rapid vortex motion causes abundant $\theta$ fluctuations at length-scales larger than $\xi$, which are actually featured in the effective field theory. These fluctuations promote $\theta$ into a natural Lagrange multiplier that implements flux conservation after an integration by parts in (\ref{LtQHa}):
\begin{equation}\label{LtQHb}
\mathcal{L}_{\textrm{t}} \sim i\, (\partial_\mu\theta + A_\mu) \mathcal{J}_\mu \to -i \, \theta \, \partial_\mu \mathcal{J}_\mu
  +i \epsilon_{\mu\nu\lambda} A_\mu \partial_\nu A_\lambda \ .
\end{equation}
Integrating out $\theta$ suppresses the gradient $\partial_j \mathcal{J}_j$ of the ``electromagnetic flux''. A non-zero gradient corresponds to ``monopoles'', i.e. events in which the gauge flux $B=\mathcal{J}_0$ is not conserved. The remaining part of $\mathcal{L}_{\textrm{t}}$ is the familiar Chern-Simons coupling known to describe fractional quantum Hall liquids \cite{WenQFT2004} and quantum Hall effect in general through the prescribed inclusion of the physical gauge field $A_\mu^{\textrm{em}}$. The effective Lagrangian density for the dynamics of quantum Hall liquids also contains a Maxwell term:
\begin{equation}
\mathcal{L}_{\textrm{eff}} = \frac{1}{2e^2} (\epsilon_{\mu\nu\lambda}\partial_\nu A_\lambda)^2 + \mathcal{L}_{\textrm{t}} \ .
\end{equation}

The formula (\ref{LtQHa}) shows the essential structure of all topological terms we will construct in this paper. The numerical coefficient to $\mathcal{L}_{\textrm{t}}$ is not yet of concern and needs to be separately determined. The symmetric and simple form of $\mathcal{L}_{\textrm{t}}$ guarantees that no symmetries are changed. Specifically, charge conservation holds just as well as flux conservation, and the explicitly broken time-reversal symmetry is anyway violated by the external magnetic field. Later in this paper we will elaborate the topological term construction and derive it in a more robust form which also manifestly satisfies the second requirement.

\section{The hierarchy of singularity gauge fields}\label{secGauge}

Here we generalize the vortex formalism of quantum Hall liquids to topological defects in $d$ spatial dimensions (${\bf x}\in\mathbb{R}^d$). We are interested in degrees of freedom given by vector fields $\hat{\bf n}({\bf x})$ of \emph{fixed magnitude}. A $d$-dimensional vector $\hat{\bf n}$ can label spin coherent states of a spinor field $\psi$ in the Spin($d$) representation and naturally describe magnetic moments. A 2-dimensional vector $\hat{\bf n} = \hat{\bf x}\cos\theta + \hat{\bf y}\sin\theta$ is equivalent to the overall U(1) phase $\theta$ of the same complex spinor $\psi$ and associated with charge currents.

Homotopy groups enumerate the topologically inequivalent classes (or sectors) of field configurations, and thus classify topological defects. A well-known sequence of homotopy groups $\pi_n(S^n)=\mathbb{Z}$, $n=1,2,3,\dots$ comes with integer-valued topological invariants, while the homotopy groups $\pi_k(S^n)=\lbrace 0\rbrace$ for $k<n$ are trivial \cite{Nakahara2003}. A $d$-dimensional vector field of fixed norm $\hat{\bf n}\in S^{d-1}$ can have only point-like topologically protected singularities in $d$-dimensional space, because $\pi_k(S^{d-1})=\lbrace 0\rbrace$ when $k<d-1$. The protected singularity is a ``hedgehog'' topological defect characterized by an integer winding number $N\in \pi_{d-1}(S^{d-1})$. In $d=2$ dimensions, a hedgehog is equivalent to a vortex -- the topologically protected singularity of a complex scalar field that carries charge currents. Interestingly, there is a generic mechanism to extend the singularities of $n$ dimensional vector fields to higher dimensions $d>n$. We will analyze here only one instance of this dimensional extension, which starts from U(1) vortices and leads to point-like monopoles in $d$ dimensions characterized by the $\pi_{d-1}(S^{d-1})$ homotopy group.

If a field $f({\bf x})$ belongs to a topological space $X$ and lives in $d$ dimensions, then the total ``topological charge'' $N\in \pi_{d-1}(X)$ of its point defects inside a sphere $S^{d-1}$ is a topological invariant of the map $f:X\to S^{d-1}$. We can generally express this invariant as a gauge field flux through $S^{d-1}$. In the case of $\pi_{d-1}(S^{d-1})$,
\begin{eqnarray}\label{top-charge-1}
N &=& \frac{1}{q} \oint\limits_{S^{d-1}} A = 
  \frac{1}{q} \oint\limits_{S^{d-1}} \left( \bigwedge_{i=1}^{d-1} dx_{j_i} \right) A_{j_1\cdots j_{d-1}} \nonumber \\
  &\equiv& \frac{1}{q} \oint\limits_{S^{d-1}} d^{d-1}x \, \epsilon_{j_1\cdots j_{d-1}} A_{j_1\cdots j_{d-1}} \ ,
\end{eqnarray}
where $q$ is the topological charge quantum. The first notation is the integral of a $d-1$ form $A$ on the oriented manifold $S^{d-1}$ surrounding the singularity. The quantity $A_{j_1\cdots j_{d-1}}$ is a rank $d-1$ antisymmetric tensor which represents the ``singularity gauge field''. Throughout this paper, we will adopt the second notation based on a conventional integral in flat space-time, where the antisymmetrization and local projection of indices to the oriented integration manifold is carried out by the antisymmetric tensor $\epsilon_{j_1\cdots j_{d-1}}$. Using Stokes-Cartan theorem, we can convert (\ref{top-charge-1}) into an integral over the volume bounded by $S^{d-1}$:
\begin{equation}\label{top-charge-2}
N = \frac{1}{q} \int dA \equiv \frac{1}{q} \int d^{d}x \, \epsilon_{ij_1\cdots j_{d-1}} \partial_i A_{j_1\cdots j_{d-1}} \ .
\end{equation}
The gauge flux
\begin{equation}\label{top-flux-1}
\mathcal{J}_0({\bf x}) = \epsilon_{ij_1\cdots j_{d-1}} \partial_i A_{j_1\cdots j_{d-1}} = q \sum_i N_i \delta({\bf x}-{\bf x}_i)
\end{equation}
has only quantized point singularities in any classical field configuration. The goal in the remainder of this section is to precisely define the singularity gauge field in the context of charge and spin dynamics.

\subsection{Monopoles}\label{secGaugeMon}

For our purposes, monopoles are point topological defects arising from charge currents. The antisymmetric tensor gauge field of rank $n=d-1$ is amenable to smooth gauge transformations
\begin{equation}\label{gauge-transf-mon}
A_{j_{1}\cdots j_{n}}\to A_{j_{1}\cdots j_{n}}+\frac{1}{n}\sum_{i=1}^{n}(-1)^{i-1}\partial_{j_{i}}A_{j_{1}\cdots j_{i-1}j_{i+1}\cdots j_{n}}
\end{equation}
that preserve the flux (\ref{top-flux-1}). The quantity that specifies the gauge transformation in the sum on the right-hand side is an antisymmetric tensor of rank $n-1$. Using this relationship, we can recursively introduce antisymmetric tensors at all ranks, from $1$ to $d-1$, and regard them as gauge fields. At rank $n=1$, we find the familiar U(1) gauge field that transforms as:
\begin{equation}
A_j \to A_j + \partial_j \theta \ .
\end{equation}
A singular gauge transformation (\ref{singA1a}) can transfer quantized vorticity from $\theta$ to $A_j$. In a coherent superfluid state, the U(1) phase $\theta$ of the order parameter is well-defined and we do not need any gauge field to specify the state. But, if fluctuations destroy the long-range order, the phase $\theta$ becomes ill-defined and we may be able to describe a non-trivial state of diffused vortices only using a well-defined gauge field $A_j$. Such a state has its own degree of coherence if the gauge flux of $A_j$ is smooth and static. However, $A_j({\bf x})$ can develop its own singularities. These would be ``magnetic'' monopoles in $d=3$ dimensions, but appear multi-dimensional if $d>3$. It is natural to describe them by a rank 2 gauge field, which transforms as:
\begin{equation}\label{gauge-transf-mon-2}
A_{ij} \to A_{ij} + \frac{1}{2} (\partial_i A_j - \partial_j A_i)
\end{equation}
according to the rule (\ref{gauge-transf-mon}). Here we recognize the usual electromagnetic field tensor $F_{ij} = \partial_i A_j - \partial_j A_i$ that can easily describe an isolated monopole in $d=3$ with magnetic field $B_k$ given by:
\begin{equation}\label{d3-mon}
F_{ij} = \epsilon_{ijk} B_k \quad,\quad \partial_k B_k({\bf x}) = 2\pi \delta ({\bf x} - {\bf x}_0) \ .
\end{equation}
We can transfer the singularities of $A_j$ into $A_{ij}$ through a singular version of the rank 2 gauge transformation (\ref{gauge-transf-mon-2}). This formally requires the appearance of $A_j$ inside exponential functions, analogous to the placement of $\theta$ in (\ref{singA1a}). A \emph{compact lattice gauge theory} discussed in Section \ref{secDynPhaseDiag} satisfies this requirement, but the continuum limit used here will suffice for most of our purposes. Next, in $d>3$ dimensions we can imagine a state in which these rank 2 singularities proliferate and move, rendering $A_j$ ill-defined. A new degree of coherence can be established in a state where the flux of $A_{ij}$ remains static and continuously distributed in space. Clearly, we can repeat this exercise by considering the singularities of $A_{ij}$ and defining a rank 3 gauge field with transformations:
\begin{equation}
A_{ijk} \to A_{ijk} + \frac{1}{3}(\partial_{i}A_{jk}-\partial_{j}A_{ik}+\partial_{k}A_{ij}) \ .
\end{equation}
Proceeding recursively, we eventually reach the highest rank $d-1$ where the gauge field $A_{j_{1}\cdots j_{d-1}}$ describes the actual point-like monopole singularities in $d$ dimensions. Conversely, the rank $n$ gauge field describes $d-n-1$ dimensional singularities.

It naively seems that the entire hierarchy of gauge fields can be ultimately derived from a single scalar function $\theta({\bf x})$ by singular gauge transformations:
\begin{eqnarray}\label{singA1c}
A_j &=& \partial_j \theta \\
A_{ij} &=& \frac{1}{2} (\partial_i A_j - \partial_j A_i) \nonumber \\
&\vdots& \nonumber \\
A_{j_{1}\cdots j_{n}} &=& \frac{1}{n}\sum_{i=1}^{n}(-1)^{i-1}\partial_{j_{i}}A_{j_{1}\cdots j_{i-1}j_{i+1}\cdots j_{n}} \ . \nonumber
\end{eqnarray}
However, this leads to a familiar problem. Even though $A_{j_1\cdots j_n}$ is perfectly capable of carrying finite rank $n$ flux
\begin{equation}\label{top-flux-2}
\mathcal{J}_{0 k_1\cdots k_{d-n-1}}({\bf x}) = \epsilon_{ik_1\cdots k_{d-n-1} j_1\cdots j_n} \partial_i A_{j_1\cdots j_n} \ ,
\end{equation}
as the example (\ref{d3-mon}) shows, it ends up carrying zero flux when we derive it from an analytic lower-rank gauge field according to (\ref{singA1c}). We can deal with this problem by generalizing Dirac strings attached to monopoles.

Consider a point-like monopole at the origin. The intrinsic rank $d-1$ gauge field $A_{j_1\cdots j_{d-1}}$ near the monopole should carry flux $\mathcal{J}_0({\bf x}) = q\, \delta({\bf x})$, but then it can't have the form produced by (\ref{singA1c}). In order to convert $A_{j_1\cdots j_{d-1}}$ to the form mandated by (\ref{singA1c}), we must add to it the gauge field of a semi-infinite Dirac string that terminates at the monopole and feeds it the flux $q$. After this string attachment, there are no more sources and drains of flux, so we formally get $\mathcal{J}_0({\bf x})=0$. And, if the Dirac string is physically unobservable, then we still have a proper isolated monopole for all practical purposes. The monopole-string combination allows us to represent $A_{j_1\cdots j_{d-1}}$ solely in terms of $A_{j_1\cdots j_{d-2}}$. Similarly, we must recursively define Dirac attachment at every other rank $n$ in order to relate $A_{j_1\cdots j_n}$ to $A_{j_1\cdots j_{n-1}}$.

Start with a Dirac string terminated at a monopole in an $n+1$ dimensional manifold $\mathcal{M}_{n+1}$. Let us separate the full gauge field into the intrinsic monopole $A'_{j_1\cdots j_n}$ and Dirac string $A''_{j_1\cdots j_n}$ parts. The monopole is a topological defect of the $\pi_{n}(S^{n})$ homotopy group. We can compute its topological charge $I'_n=q$ from $A'_{j_1\cdots j_n}$ by integrating (\ref{top-charge-1}) on an $n$ sphere $S^n\subset\mathcal{M}_{n+1}$ that encloses the monopole:
\begin{equation}\label{top-charge-3}
I'_n = \oint\limits_{S^{n}} d^{n}x \, \epsilon^{\phantom{,}}_{j_1\cdots j_{n}} A'_{j_1\cdots j_{n}} \ .
\end{equation}
Recall that $\epsilon$ always projects its spatial indices onto the integration manifold. By Stokes-Cartan theorem,
\begin{equation}\label{top-charge-4}
I'_n = \int\limits_{\mathcal{M}_{n+1}} d^{n+1}x \, \partial^{\phantom{,}}_i \Phi'_i \quad,\quad 
  \Phi'_i \equiv \epsilon^{\phantom{,}}_{ij_1\cdots j_{n}} A'_{j_1\cdots j_{n}} \ .
\end{equation}
If we used the full gauge field $A=A'+A''$ to compute (\ref{top-charge-3}), we would obtain zero because the monopole and string together present no sources and drains of flux. Consequently, we can alternatively compute the monopole's topological charge $I''_n=q$ by integrating the string part $A''_{j_1\cdots j_n}$ on any ``flat'' $n$ dimensional manifold $\mathcal{M}_n$ that intersects the string at a single point:
\begin{eqnarray}
I''_n &=& \int\limits_{\mathcal{M}_{n}} d^{n}x \, \epsilon^{\phantom{,}}_{j_1\cdots j_{n}} A''_{j_1\cdots j_{n}} \\
  &=& \int\limits_{\mathcal{M}_{n}} d^{n}x \, \epsilon^{\phantom{,}}_{j_1\cdots j_{n}} 
        \frac{1}{n} \sum_{k=1}^{n} (-1)^{k-1} \partial^{\phantom{,}}_{j_k} A''_{j_1\cdots j_{k-1}j_{k+1}\cdots j_{n}} \nonumber \\
  &=& \int\limits_{\mathcal{M}_{n}} d^{n}x \, \epsilon^{\phantom{,}}_{ij_1\cdots j_{n-1}} 
        \partial^{\phantom{,}}_{i} A''_{j_1\cdots j_{n-1}} \nonumber \ .
\end{eqnarray}
We used (\ref{singA1c}) to express the rank $n$ gauge field in terms of the rank $n-1$ gauge field, and obtained the expression:
\begin{equation}
I''_n = \int\limits_{\mathcal{M}_{n}} d^{n}x \, \partial^{\phantom{,}}_i \Phi''_i \quad,\quad 
  \Phi''_i \equiv \epsilon^{\phantom{,}}_{ij_1\cdots j_{n-1}} A''_{j_1\cdots j_{n-1}}
\end{equation}
analogous to (\ref{top-charge-4}) but defined in one lower dimension, i.e. $I''_n = I'_{n-1}$. This indicates that the projection of the Dirac string onto $\mathcal{M}_{n}$ is a lower-dimensional $\pi_{n-1}(S^{n-1})$ monopole living in $\mathcal{M}_{n}$. We can now recursively restart this analysis from $\mathcal{M}_{n}$, by attaching a Dirac string to the projected monopole strictly within $\mathcal{M}_{n}$. In fact, in order to establish relationships (\ref{singA1c}) at lower ranks, we must continuously stack many manifolds $\mathcal{M}_{n}$ that intersect the original string at all possible places, and attach a reduced rank string inside each $\mathcal{M}_{n}$. When we reach the lowest rank 1, we obtain the final integrals:
\begin{equation}
I'_1 = \oint\limits_{S^1} dx \, A'_j = \oint\limits_{S^1} dx \, \partial^{\phantom{,}}_j \theta = 2\pi N
\end{equation}
that establish $q=2\pi$. In conclusion, the monopole charge quantum is $q=2\pi$ in all spatial dimensions $d$.

Physically observable Dirac attachments have tension and lead to the confinement of monopoles into small neutral clusters. Monopoles can exist as free topological defects only in compact gauge theories where the quantized Dirac attachments become unobservable.

\subsection{Hedgehogs}\label{secGaugeHed}

Let $\hat{\bf n}({\bf x})$ be a field of $d$-dimensional unit-vectors with components $\hat{n}^a$ ($a\in\lbrace 1,\dots,d\rbrace$). The topological defects of spins are characterized by the gauge field
\begin{equation}\label{singA2a}
A_{j_1\cdots j_{d-1}} = \frac{1}{(d-1)!} \epsilon_{a_{0}a_{1}\cdots a_{d-1}}\;\hat{n}^{a_{0}}\prod_{i=1}^{d-1} \partial_{j_{i}}\hat{n}^{a_{i}} \ .
\end{equation}
The integral (\ref{top-charge-1}) with this gauge field is quantized as an integer if we choose $q=S_{d-1}$ to be the area of a unit $d-1$ sphere,
\begin{equation}
S_{n}=\frac{2\pi^{(n+1)/2}}{\Gamma\left(\frac{n+1}{2}\right)} \ .
\end{equation}
The corresponding hedgehog flux (\ref{top-flux-1}) is singular and quantized in units of $q=S_{d-1}$ when the ground state possesses long-range magnetic order.

We can parametrize the vector field $\hat{\bf n}({\bf x})$ using a set of angles $\theta_j({\bf x})$, $j\in\lbrace 1,\dots,d-1\rbrace$:
\begin{eqnarray}\label{n-vs-theta}
\hat{n}^0 &=& \cos\theta_1 \\
\hat{n}^1 &=& \sin\theta_1 \cos\theta_2 \nonumber \\
\hat{n}^2 &=& \sin\theta_1 \sin\theta_2 \cos\theta_3 \nonumber \\
 &\vdots& \nonumber \\
\hat{n}^{d-2} &=& \sin\theta_1 \cdots \sin\theta_{d-2} \cos\theta_{d-1} \nonumber \\
\hat{n}^{d-1} &=& \sin\theta_1 \cdots \sin\theta_{d-2} \sin\theta_{d-1} \nonumber
\end{eqnarray}
on the domain $\theta_j\in \lbrack 0,\pi \rbrack$ for $j<d-1$ and $\theta_{d-1}\in\lbrack 0,2\pi)$. Then (see Appendix \ref{appHedgehogs}):
\begin{equation}\label{singA2b}
A_{j_1\cdots j_{d-1}} = \frac{\epsilon_{k_{1}\cdots k_{d-1}}}{(d-1)!} \prod_{i=1}^{d-1} (\sin\theta_{i})^{d-1-i}\, \partial_{j_{i}} \theta_{k_{i}} \ .
\end{equation}
Specifically, in naturally accessible dimensions:
\begin{eqnarray}
d=2 \quad &\cdots& \quad A_i = \partial_i\theta_1 \\
d=3 \quad &\cdots& \quad A_{ij} = \frac{1}{2}\sin\theta_1\Bigl\lbrack (\partial_i\theta_1)(\partial_j\theta_2)
  - (\partial_i\theta_2)(\partial_j\theta_1) \Bigr\rbrack \nonumber
\end{eqnarray}
We can also define:
\begin{eqnarray}\label{Phi1}
\Phi_i &=& \epsilon_{ij_{1}\cdots j_{d-1}} A_{j_{1}\cdots j_{d-1}} \\
  &=& \epsilon_{ij_{1}\cdots j_{d-1}} \prod_{k=1}^{d-1}(\sin\theta_{k})^{d-1-k}\,\partial_{j_{k}}\theta_{k} \ , \nonumber
\end{eqnarray}
and observe that identifying $\theta_j ({\bf x})$ with the spherical coordinate system angles $\theta'_j$ at ${\bf x} = (|{\bf x}|, \theta'_1, \dots, \theta'_{d-1})$ yields:
\begin{equation}
\theta^{\phantom{,}}_j({\bf x}) = \theta'_j \quad\Rightarrow\quad \Phi^{\phantom{,}}_i({\bf x}) = \frac{x_i}{|{\bf x}|^d} \ .
\end{equation}
The topological charge (\ref{top-charge-2}) is extracted via a Gauss' law in terms of $\Phi_i$:
\begin{equation}
N = \frac{1}{S_{d-1}} \int d^{d}x \, \partial_i \Phi_i = 1 \ ,
\end{equation}
showing that the gauge flux $\partial_i \Phi_i = S_{d-1} \delta({\bf x})$ is singular. In order to obtain any other quantized topological charge $N\neq 1$, we only need to tweak the relationship between $\theta_{d-1}$ and the corresponding spherical coordinate system angle:
\begin{eqnarray}\label{top-hedgehog-angle}
&& (\forall j<d-1)\quad \theta^{\phantom{,}}_j({\bf x}) = \theta'_j \quad,\quad \theta^{\phantom{,}}_{d-1}({\bf x}) = N \theta'_{d-1} \nonumber \\
&& \quad\Rightarrow\quad \Phi^{\phantom{,}}_i({\bf x}) = N\frac{x_i}{|{\bf x}|^d} \ .
\end{eqnarray}
Note that only $\theta_{d-1}$ can be modified this way because all components of $\hat{\bf n}$ are periodic functions of it on the full $2\pi$ interval. $N$ is required to be an integer in order for $\hat{\bf n}$ to be single-valued and smooth everywhere in space.

The structure and properties of the hedgehog gauge field $A_{j_1\cdots j_{d-1}}$ are completely analogous to those of the monopole gauge field; only the flux quantum $q$ is different. Likewise, it is possible to define an entire hierarchy of spin-related gauge fields at different ranks, which is analogous to the hierarchy of charge-related Abelian gauge fields $A_{j_1\cdots j_n}$. This will become useful when we construct and analyze the effective field theory. The hierarchy ends with $A_{j_1\cdots j_{d-1}}$ and starts at rank 1 where the gauge field is minimally coupled to currents. The expression for spin current can be obtained from the prototype Lagrangian density of magnetic degrees of freedom
\begin{equation}
\mathcal{L}=\frac{K}{2}(\partial_{\mu} \hat{n}^{a})(\partial_{\mu}\hat{n}^{a}) \ ,
\end{equation}
which has rotational symmetry. Infinitesimal rotations $\hat{n}^{a}\to \hat{n}^{a}+\delta \hat{n}^{a}$, 
\begin{equation}
\delta \hat{n}^{a}=\epsilon_{abc_{1}\cdots c_{d-2}}\hat{n}^{b}\delta\omega^{c_{1}\cdots c_{d-2}} + \mathcal{O}(\delta\omega^2)
\end{equation}
are generated by an antisymmetric tensor $\delta\omega^{c_{1}\cdots c_{d-2}}$, and so by Noether's theorem we find a conserved spin current:
\begin{equation}
j_{\mu}^{\phantom{,}}\propto\pi_{\mu}^{a}\delta \hat{n}^{a}=K\,\epsilon_{abc_{1}\cdots c_{d-2}}(\partial_{\mu}^{\phantom{,}} \hat{n}^{a}) \hat{n}^{b}
  \delta\omega^{c_{1}\cdots c_{d-2}} \ ,
\end{equation}
where $\pi_{\mu}^{a}=\delta\mathcal{L}/\delta\partial_{\mu}\hat{n}^{a} = K\partial_{\mu}^{\phantom{,}}\hat{n}^{a}$ is the canonical momentum. The tensor $\delta\omega$ has $d(d-1)/2$ degrees of freedom corresponding to choices of independent two-dimensional rotation planes in $d$ dimensional space (the two omitted indices in $\delta\omega$ specify the plane). Therefore, we identify $d(d-1)/2$ different spin currents which take the following form after normalization and symmetrization:
\begin{eqnarray}\label{spin-current}
j_{\mu}^{c_{1}\cdots c_{d-2}} &=& \frac{1}{2}\epsilon_{abc_{1}\cdots c_{d-2}}^{\phantom{,}}
  \Bigl\lbrack \hat{n}^{a}(\partial_{\mu}^{\phantom{,}}\hat{n}^{b})-\hat{n}^{b}(\partial_{\mu}^{\phantom{,}}\hat{n}^{a})\Bigr\rbrack \nonumber \\
&=& \epsilon_{abc_{1}\cdots c_{d-2}}^{\phantom{,}}\,\hat{n}^{a}(\partial_{\mu}^{\phantom{,}}\hat{n}^{b}) \ .
\end{eqnarray}
The rank 1 gauge field must be minimally coupled to this, so it must carry the same internal spin indices. The effective Lagrangian density must contain a gauge-invariant combination $j_{\mu}^{c_{1}\cdots c_{d-2}} + A_{\mu}^{c_{1}\cdots c_{d-2}}$, so we can envision a singular gauge transformation that preserves the Lagrangian density:
\begin{equation}\label{singA2c}
A_{\mu}^{c_{1}\cdots c_{d-2}} \to \epsilon_{abc_{1}\cdots c_{d-2}}^{\phantom{,}}\,\hat{n}^{a}(\partial_{\mu}^{\phantom{,}}\hat{n}^{b}) \ .
\end{equation}
The purpose of this transformation is again to transfer the singularities of the matter field onto gauge fields, so that we could keep track of their dynamics even when quantum fluctuations diffuse them. As an example, consider the configuration $\hat{\bf n}=\hat{{\bf x}}\cos\phi+\hat{{\bf y}}\sin\phi$ in $d=3$ dimensions expressed in terms of the azimuthal angle $\phi$. It represents a ``vortex'' line stretching along the $z$-direction with singularity at $(x,y)=0$, shown in Fig.\ref{defects}(a). Specifying the plane for $\hat{\bf n}$ near the singularity requires one internal spin index $c_1$. Note that this singularity is not topologically protected because the “vortex” can be smoothly deformed into a uniform $\hat{\bf n}$ configuration, by tilting $\hat{\bf n}$ toward $\hat{\bf z}$ without ever reshaping the singular line.

In order to build the hierarchy of gauge fields, we must start from (\ref{singA2c}), carry out a rank-promotion procedure at every rank, and arrive at (\ref{singA2a}) at the highest rank $d-1$. Clearly, each rank promotion needs to consume one spin index and introduce one spatial index. This leaves only one option for generating gauge fields by singular gauge transformations:
\begin{eqnarray}\label{singN1}
A_{\lambda_1}^{c_{2}\cdots c_{d-1}} &\to& \epsilon_{c_0\cdots c_{d-1}}^{\phantom{,}}\,\hat{n}^{c_0}(\partial_{\lambda_1}^{\phantom{,}}\hat{n}^{c_1}) \\
A_{\lambda_1\lambda_2}^{c_{3}\cdots c_{d-1}} &\to& \frac{1}{2} \epsilon_{c_0\cdots c_{d-1}}^{\phantom{,}}\,\hat{n}^{c_0}
    (\partial_{\lambda_1}^{\phantom{,}}\hat{n}^{c_1})(\partial_{\lambda_2}^{\phantom{,}}\hat{n}^{c_2}) \nonumber \\
  &\vdots& \nonumber \\
A_{\lambda_1\cdots \lambda_{d-1}} &\to& \frac{1}{(d-1)!} \epsilon_{c_0\cdots c_{d-1}}\;\hat{n}^{c_0}\prod_{i=1}^{d-1} \partial_{\lambda_{i}}\hat{n}^{c_{i}} \nonumber
\end{eqnarray}
All gauge fields are antisymmetric both with respect to their upper and lower indices, and the presence of upper indices makes them non-Abelian. Apart from being relevant to spin dynamics in the presence of spin-orbit coupling, the rank 1 and 2 gauge fields have been of interest in the context of non-Abelian monopoles in high-energy physics \cite{Shnir2005}. The best we can do to relate a rank $n$ gauge field to the lower rank one is:
\begin{equation}
A_{\lambda_{1}\cdots\lambda_{n}}^{c_{n+1}\cdots c_{d-1}} = \frac{1}{n}(\partial_{\lambda_{n}}^{\phantom{,}}\hat{n}_{\phantom{|}}^{c_{n}})
  A_{\lambda_{1}\cdots\lambda_{n-1}}^{c_{n}c_{n+1}\cdots c_{d-1}} \ .
\end{equation}
This is a much more relaxed relationship than the one for monopoles (\ref{singA1c}) due to the $\partial_{\lambda_{n}}\hat{n}^{c_n}$ factor. Quantum fluctuations that destroy long-range order will effectively uncorrelate the gauge fields at different ranks through rapid changes of $\hat{n}^{c_n}$. For this reason, hedgehogs do not come with Dirac strings attached.

\section{Effective field theory and dynamics}\label{secDyn}

Our goal is to describe topologically non-trivial dynamics of strongly interacting particles represented by a spinor field $\psi$. The appropriate field theory will have the imaginary time Lagrangian density
\begin{equation}\label{L1}
\mathcal{L} = \mathcal{L}_{\textrm{d}} + \mathcal{L}_{\textrm{t}}
\end{equation}
constrained by symmetries, where $\mathcal{L}_{\textrm{d}}$ governs conventional dynamics and $\mathcal{L}_{\textrm{t}}$ is a topological term responsible for conserving topological charge in incompressible quantum liquids. In order to simplify discussion, we will assume relativistic dynamics and work with a conventional part of the Lagrangian density such as
\begin{equation}\label{L2}
\mathcal{L}_{\textrm{d}} = \frac{1}{2} \bigl\vert (\partial_\mu + iA_\mu)\psi \bigr\vert^2 - t|\psi|^2 + u|\psi|^4 + \cdots \ .
\end{equation}
Since our main focus are insulating states, most of the analysis will be applicable to non-relativistic dynamics as well.

Charge and spin currents carried by the field $\psi$ can have singular configurations, which we now know how to extract into gauge fields. The lowest-dimensional point singularities are described by the highest rank gauge field with $d-1$ space-time indices. The gauge fields that couple minimally to currents have a single space-time index and describe $d-2$ dimensional singular domains. Lastly, $d-1$ dimensional domain walls that separate space into disconnected regions are singularities of the matter field itself (the corresponding rank 0 gauge field would not carry any space-time indices). In order to capture possible quantum diffusion of these singularities, we need to construct an effective theory in terms of the gauge fields, which obtains from (\ref{L1}) upon coarse-graining to a certain coherence length-scale $\xi$. We will postpone the discussion of the topological term $\mathcal{L}_{\textrm{t}}$ to Section \ref{secTopTerm}, and focus here on the effective theory derived from (\ref{L2}) and expressed in terms of the gauge fields. We will initially rely on symmetries to separately construct the effective Lagrangian densities for charge dynamics in Section \ref{secDynCharge} and spin dynamics in Section \ref{secDynSpin}. Following each symmetry construction, we will argue that quantum fluctuations indeed dynamically generate the constructed Lagrangian terms at higher ranks. The final segment of this discussion in Section \ref{secDynPhaseDiag} is about the phase diagram of the effective theory. There we address the very important issues of defect and charge deconfinement, which are required for the existence of topological order and critically dependent on the field theory regularization.

\subsection{Abelian charge dynamics}\label{secDynCharge}

Lagrangian density can contain only gauge invariant scalar combinations of fields. Generally, the Abelian gauge fields introduced in Section \ref{secGaugeMon} can be involved in two kinds of couplings at every rank $n$:
\begin{eqnarray}\label{CM1}
\mathcal{L}_{\textrm{C}n} &=& \frac{\kappa_n}{2} \bigl( j_{\lambda_1\cdots\lambda_n} + A_{\lambda_1\cdots\lambda_n} \bigr)^2 \\
\mathcal{L}_{\textrm{M}n} &=& \frac{1}{2(d-n)!\,e_n^2}
  \bigl( \epsilon_{\mu_1\cdots\mu_{d-n}\nu\lambda_1\cdots\lambda_n} \partial_\nu A_{\lambda_1\cdots\lambda_n} \bigr)^2 \ .\nonumber
\end{eqnarray}
The first term $\mathcal{L}_{\textrm{C}n}$ minimally couples the gauge field to a current, and the second Maxwell term $\mathcal{L}_{\textrm{M}n}$ contains only the gauge field and captures the energy density of flux. The ``conserved'' current at rank $n$ must have the form of a pure gauge:
\begin{equation}\label{cur1}
j_{\lambda_1\cdots\lambda_n} = \sum_{i=1}^{n}(-1)^{i-1}\partial_{\lambda_i} \theta_{\lambda_1\cdots\lambda_{i-1}\lambda_{i+1}\cdots\lambda_n}
\end{equation}
dictated by the rank $n$ gauge transformations derived from (\ref{gauge-transf-mon}):
\begin{eqnarray}\label{gt1}
\theta_{\lambda_1\cdots\lambda_{n-1}} &\to& \theta_{\lambda_1\cdots\lambda_{n-1}} - \delta\theta_{\lambda_1\cdots\lambda_{n-1}} \\
A_{\lambda_1\cdots\lambda_n} &\to& A_{\lambda_1\cdots\lambda_n} +
  \sum_{i=1}^{n}(-1)^{i-1}\partial_{\lambda_i} \delta\theta_{\lambda_1\cdots\lambda_{i-1}\lambda_{i+1}\cdots\lambda_n} \ . \nonumber
\end{eqnarray}
If all currents were independent degrees of freedom, the theory would have an independent gauge symmetry at every rank. However, the gauge symmetries at $n>1$ ranks are unphysical. We must introduce additional rank linking terms to remedy this problem:
\begin{equation}\label{link1}
\mathcal{L}_{\textrm{L}n} = \Lambda_n\left(\theta_{\lambda_1\cdots\lambda_n}+\frac{1}{n}A_{\lambda_1\cdots\lambda_n}\right)^2 \ .
\end{equation}
The links $\mathcal{L}_{\textrm{L}n}$ break the gauge transformations (\ref{gt1}) and remove the current independence at ranks $n>1$. The physical U(1) gauge symmetry residing at rank 1 is spared, and the physical charge current $j_\mu = \partial_\mu\theta$ remains an independent degree of freedom because the matter field $\theta$ never appears in (\ref{link1}). In that manner, we obtain the full Lagrangian density
\begin{equation}\label{L3}
\mathcal{L}_{\textrm{d}} = \sum_{n=1}^{d-1} \Bigl( \mathcal{L}_{\textrm{C}n} + \mathcal{L}_{\textrm{M}n} + \mathcal{L}_{\textrm{L}n} \Bigr)
\end{equation}
with correct symmetries and degrees of freedom, featuring the gauge fields that describe all possible kinds of singularities. We may also integrate out all $\theta_{\lambda_1\cdots\lambda_n}$ fields with $n\ge 1$ and write:
\begin{equation}\label{L3b}
\mathcal{L}^{\phantom{,}}_{\textrm{d}} = \sum_{n=1}^{d-1} \Bigl( \mathcal{L}'_{\textrm{C}n} + \mathcal{L}^{\phantom{,}}_{\textrm{M}n} \Bigr)
\end{equation}
\vspace{-0.2in}
\begin{equation}
\mathcal{L}'_{\textrm{C}n} \!=\! \frac{\kappa'_n}{2} \!\left\lbrack \frac{1}{n} \sum_{i=1}^{n}(-1)^{i-1}\partial_{\lambda_i}^{\phantom{,}}
         A_{\lambda_1\cdots\lambda_{i-1}\lambda_{i+1}\cdots\lambda_n}^{\phantom{,}} \!-\! A_{\lambda_1\cdots\lambda_n}^{\phantom{,}}\right\rbrack^2 \nonumber
\end{equation}

The effective field theory (\ref{L3b}) has the necessary ingredients to describe the phases with either confined or deconfined monopoles -- even if we regard it as being strictly \emph{non-compact}. When $\kappa'_n$ is large, the gauge fields at ranks $n-1$ and $n$ become dynamically related according to (\ref{singA1c}) and every rank $n$ singularity must have a Dirac attachment. This confines the singularities because Dirac attachments have a finite tension expressed through the Maxwell terms in a non-compact theory. In the opposite limit of sufficiently small $\kappa'_n$, the system gains more free energy density from the entropy of fluctuations than from the energy of linking the gauge fields across ranks. Dirac attachments become unnecessary and the singularities become deconfined. Specifically, consider substituting a vanishing rank $n-1$ gauge field $A_{\lambda_1\cdots\lambda_{n-1}}=0$ in $\mathcal{L}'_{\textrm{C}n}$. Now we find by dimensional analysis that a singular configuration of $A_{\lambda_1\cdots\lambda_n}$ at rank $n$, without a Dirac attachment, costs at most
\begin{equation}
E_n^{\phantom{,}} = \int d^d x\, \mathcal{L}'_{\textrm{C}n} \sim E^{\phantom{,}}_{\textrm{UV}} + \kappa'_n R^{d-2n}
\end{equation}
energy, where $R$ is an infra-red cutoff length scale and $E_{\textrm{UV}}$ is an ultra-violet contribution. The singularity of rank $n$ occupies a $d-n-1$ dimensional manifold, so its energy per unit manifold area scales as $R^{1-n}$, plus a constant that comes from $E_{\textrm{UV}}$ (we assume that the theory is regularized in the ultra-violet limit). Therefore, the price for having a singularity is paid only locally when $n>1$, and deconfined singularities without Dirac attachments can be entropically stimulated with small $\kappa'_n$.

The higher rank Lagrangian terms in (\ref{L3}) or (\ref{L3b}) arise dynamically from the lower rank terms in the process of coarse-graining. Starting from the basic coupling of a current to a gauge field
\begin{equation}\label{L4a}
\mathcal{L}=\frac{\kappa}{2}\left(j_{\mu}+A_{\mu}\right)^{2}+\cdots \ ,
\end{equation}
we are free to separate the smooth matter field fluctuations $\theta$ from singular vortex ones $j'_{\mu}$
\begin{equation}\label{singJ1}
j_{\mu}^{\phantom{,}} = \partial_{\mu}^{\phantom{,}}\theta + j'_{\mu}
\end{equation}
using some arbitrary convention for fixing the gauge of $j'_{\mu}$ (i.e. we use the same particular algorithm to calculate a definite $j'_{\mu}$ from any given configuration of vortices). Integrating out the smooth $\theta$ in the path-integral would result in an effective Lagrangian for $j'_{\mu} + A^{\phantom{,}}_\mu$ which must have a Maxwell term due to gauge invariance. If we integrate only certain \emph{short wavelength modes} of $\theta$ in (\ref{L4a}), we also preserve the coarse-grained coupling between the current and the gauge field:
\begin{eqnarray}\label{L4c}
\mathcal{L}' &=& \frac{\kappa'}{2}\left(\partial_\mu^{\phantom{,}}\theta+j'_\mu+A_{\mu}^{\phantom{,}}\right)^{2} \\
  && +K'\left\lbrack\frac{1}{2}\left(\partial_{\mu}^{\phantom{,}} j'_{\nu}
     -\partial_{\nu}^{\phantom{,}}j'_{\mu}\right) + \frac{1}{2}F_{\mu\nu}^{\phantom{x}}\right\rbrack^{2}+\cdots \ . \nonumber
\end{eqnarray}
We may complete a singular gauge transformation to absorb $j'_{\mu}$ into $A^{\phantom{,}}_\mu$, and finish the coarse-graining step by integrating out the short-wavelength fluctuations of the gauge field. The next rank in $d\ge 3$ dimensions is generated by another round of a singular gauge transformation and coarse-graining. The rank 1 gauge field makes a ``matter'' field at rank 2 ($\theta_\mu \sim A_\mu$), and the rank 1 Maxwell term has the form of a rank 2 current-gauge field coupling. Separate the smooth $\theta^{\phantom{'}}_\mu$ and singular monopole $j'_{\mu\nu}$ fluctuations of rank 2 ``matter''
\begin{equation}\label{singJ2}
j_{\mu\nu}^{\phantom{,}} = \partial_\mu\theta_\nu - \partial_\nu\theta_\mu + j'_{\mu\nu} \ ,
\end{equation}
mirroring (\ref{singJ1}), then integrate out the short-wavelength fluctuations of $\theta_\mu$. This produces a rank 2 Maxwell term in the effective Lagrangian density, with an emergent gauge field $A^{\phantom{,}}_{\mu\nu} \sim j'_{\mu\nu}$. Repeating these steps recursively generates analogous dynamics at all higher ranks. However, the emergent ``charge'' quantization at all ranks derives from the topological quantization of vorticity at rank 1. 

The derivation of gauge fields from the singularities of matter fields can explain why all particles that couple to the same gauge field have the same unit of charge quantization (as is the case in the standard model of particle physics). Consider several complex scalar fields $\psi_{1},\dots,\psi_{n}$. Carry out singular gauge transformations for every $1\le j\le n$ in order to extract singularities from the matter field phases into gauge fields $A_{j}$ according to $A_{j\mu}\psi_{j}=-i\partial_{\mu}\psi_{j}$. The resulting current terms in the Lagrangian density read $\kappa_{j}|(\partial_{\mu}+iA_{\mu j})\psi_{j}|^{2}$. Now assume that the dynamics has only one global U(1) symmetry. This locks all singularity gauge fields $A_{j\mu}=A_{\mu}+\delta A_{j\mu}$ to a single free gauge field $A_{\mu}$, allowing only small gapped fluctuations $\delta A_{j\mu}$. If we integrate out $\delta A_{j\mu}$ and also the short-length fluctuations of $\psi_{j}$, we obtain a coarse-grained theory with current terms $\kappa_{j}|(\partial_{\mu}+iA_{\mu})\psi_{j}|^{2}$ involving only $A_\mu$. Coarse-graining also produces a Maxwell term $(1/2e^2)(\epsilon_{\mu\nu\lambda} \partial_{\nu}A_{\lambda})^{2}$. By renormalizing $A_{\mu}$, we can bring $e$ inside the current terms $\kappa_{j}|(\partial_{\mu}+ieA_{\mu})\psi_{j}|^{2}$ where it clearly plays the role of a single quantized charge coupling for all matter fields. Particles with charge $2e$, etc, are bound states of the elementary ones. Fractional quantization of charge is also possible, but requires a special dynamical state of topological defects that we discuss later.

\subsection{Non-Abelian spin dynamics}\label{secDynSpin}

Here we construct the effective Lagrangian density
\begin{equation}\label{L5}
\mathcal{L}_{\textrm{d}} = \sum_{n=1}^{d-1} \Bigl( \mathcal{L}_{\textrm{C}n} + \mathcal{L}_{\textrm{M}n} + \mathcal{L}_{\textrm{L}n} \Bigr)
\end{equation}
for the dynamics of spin currents and their singularities using the same symmetry principles as in the previous section. We expect:
\begin{eqnarray}\label{CM2}
\mathcal{L}^{\phantom{,}}_{\textrm{C}n} &=& \frac{k_n}{2}
  \bigl( j_{\lambda_1\cdots\lambda_n}^{a_{n+1}\cdots a_{d-1}} + A_{\lambda_1\cdots\lambda_n}^{a_{n+1}\cdots a_{d-1}} \bigr)^2 \\
\mathcal{L}_{\textrm{M}n} &=& \frac{1}{2g_n^2} \mathcal{J}_{\mu_{1}\cdots\mu_{d-n}}^{a_{n+1}\cdots a_{d-1}}
  \mathcal{J}_{\mu_{1}\cdots\mu_{d-n}}^{a_{n+1}\cdots a_{d-1}} \ . \nonumber
\end{eqnarray}
All non-Abelian gauge fields $A_{\lambda_1\cdots\lambda_n}^{a_{n+1}\cdots a_{d-1}}$ are initially generated by singular gauge transformations (\ref{singN1}) from the same physical matter field $\hat{\bf n}$. However, when the singularities of $\hat{\bf n}$ diffuse by fluctuations, the gauge fields at all ranks acquire independent dynamics that goes beyond the limitations of (\ref{singN1}). The residual smooth fluctuations of $\hat{\bf n}$ are captured by currents $j_{\lambda_1\cdots\lambda_n}^{a_{n+1}\cdots a_{d-1}}$ that minimally couple to the gauge fields. We can regard the currents as independent degrees of freedom, and include the linking terms in the Lagrangian density
\begin{equation}\label{link2}
\mathcal{L}_{\textrm{L}n} = \Lambda'_{n}\left(j_{\lambda_{1}\cdots\lambda_{n}}^{a_{n+1}\cdots a_{d-1}}
  +\frac{\mathcal{A}}{n}\;(\partial_{\lambda_{n}}^{\phantom{x}}\hat{n}_{\phantom{\lambda}}^{a_{n}})
    A_{\lambda_{1}\cdots\lambda_{n-1}}^{a_{n}a_{n+1}\cdots a_{d-1}}\right)^{2}
\end{equation}
in order to have a single gauge symmetry at rank 1. We formally define
\begin{equation}
A^{a_1\cdots a_{d-1}} = -\epsilon_{a_0a_1\cdots a_{d-1}} \hat{n}^{a_0}
\end{equation}
in consideration of the formula (\ref{spin-current}) for spin current, and the operator $\mathcal{A}$ that antisymmetrizes the space-time indices
\begin{equation}
\mathcal{A} f_{\lambda_1\cdots\lambda_n} = \frac{1}{n!} \sum_{\mathcal{P}}^{1\cdots n} (-1)^{\mathcal{P}} f_{\lambda_{\mathcal{P}(1)}\cdots\lambda_{\mathcal{P}(n)}} \ ,
\end{equation}
where $\mathcal{P}$ is a permutation and $(-1)^{\mathcal{P}}$ its parity. Note that large values of $k^{\phantom{,}}_m$ and $\Lambda'_m$ at ranks $m\le n$ suppress the diffusion of singularities and pin the currents to:
\begin{equation}\label{spin-current-rank-n}
j_{\lambda_1\cdots\lambda_n}^{a_{n+1}\cdots a_{d-1}} \to \frac{1}{n!} \epsilon_{a_0\cdots a_{d-1}}
  \hat{n}^{a_0}\prod_{i=1}^{n} (\partial_{\lambda_{i}}\hat{n}^{a_{i}}) \ .
\end{equation}

If we integrate out all currents in (\ref{L5}), we obtain a more economic version of the effective theory:
\begin{equation}\label{L5b}
\mathcal{L}^{\phantom{,}}_{\textrm{d}} = \sum_{n=1}^{d-1} \Bigl( \mathcal{L}'_{\textrm{C}n} + \mathcal{L}^{\phantom{,}}_{\textrm{M}n} \Bigr)
\end{equation}
\vspace{-0.2in}
\begin{equation}
\mathcal{L}'_{\textrm{C}n} \!=\! \frac{k'_n}{2} \!\left\lbrack \frac{\mathcal{A}}{n}\;
  (\partial_{\lambda_{n}}^{\phantom{x}}\hat{n}_{\phantom{\lambda}}^{a_{n}})A_{\lambda_{1}\cdots\lambda_{n-1}}^{a_{n}a_{n+1}\cdots a_{d-1}}
  - A_{\lambda_1\cdots\lambda_n}^{a_{n+1}\cdots a_{d-1}} \right\rbrack^2 \nonumber \ .
\end{equation}
The Maxwell terms $\mathcal{L}_{\textrm{M}n}$ depend only on the gauge fields through non-Abelian fluxes $\mathcal{J}_{\mu_{1}\cdots\mu_{d-n}}^{a_{n+1}\cdots a_{d-1}}$ whose space-time indices are compatible with (\ref{top-flux-2}) and internal indices correspond to those of the gauge field. The gauge field curl is still an essential component of flux. However, the non-Abelian gauge invariance of Maxwell terms requires additional non-linear flux components, except at the highest rank $n=d-1$ where the flux is Abelian in any number of dimensions $d$:
\begin{equation}\label{top-current}
\mathcal{J}_\mu = \epsilon_{\mu\nu\lambda_1\cdots\lambda_{d-1}} \partial_{\nu} A_{\lambda_1\cdots\lambda_{d-1}} \ .
\end{equation}
We can determine the expressions for fluxes by working exclusively with singular gauge fields (\ref{singN1}) and considering their transformations under smooth deformations of the vector field $\hat{\bf n}$. Such deformations amount to smooth gauge transformations that cannot move or reshape the singularities, and hence do not affect the Maxwell Lagrangian density. Detailed derivation of the fluxes is shown in Appendix \ref{appNonAbelianMaxwell}. Here we only state the most useful non-trivial result for rank $n=1$ in $d=3$:
\begin{equation}\label{Maxwell-NA-d3-n1}
\mathcal{J}_{\mu\nu}^{a} = \epsilon_{\mu\nu\alpha\beta}^{\phantom{,}} \left(\partial_{\alpha}^{\phantom{,}} A_{\beta}^{a}
  -\epsilon_{abc}^{\phantom{,}}A_{\alpha}^{b}A_{\beta}^{c}\right) \ .
\end{equation}
This form is familiar from the non-Abelian SU(2) gauge theory:
\begin{equation}
\mathcal{J}_{\mu\nu}^{a}=\frac{1}{2}\epsilon_{\mu\nu\alpha\beta}^{\phantom{x}}F_{\alpha\beta}^{a}\quad,\quad
  F_{\alpha\beta}^{a}=\partial_{\alpha}^{\phantom{x}}A_{\beta}^{a}-\partial_{\beta}^{\phantom{x}}A_{\alpha}^{a}
    -g\epsilon_{abc}^{\phantom{x}}A_{\alpha}^{b}A_{\beta}^{c} \nonumber
\end{equation}
with gauge charge $g=2$ corresponding to the choice $|\hat{\bf n}|=1$. The value of $g$ is determined by the spin representation generators, which also determine $|\hat{\bf n}|$: if we had chosen to work with the minimal SU(2) representation $|\hat{\bf n}| = \frac{1}{2}$, we would have obtained $g=1$.

The fundamental microscopic Lagrangian describes only the rank 1. All higher ranks of the effective theory arise dynamically in a coarse-graining procedure. The technical demonstration of this claim is postponed to Appendix \ref{appNonAbelianranks} due to its complexity. There we also discuss in more detail the singular gauge transformations of non-Abelian gauge theories.

Classical vector field $\hat{\bf n}$ configurations can be topologically non-trivial even without singularities. Such configurations are generalized skyrmions. If a $d+1$ dimensional vector field lives in a $d$ dimensional space, then we can formally define a rank $d$ gauge field
\begin{equation}
A_{j_1\cdots j_d} = \frac{1}{d!} \epsilon_{c_0\cdots c_d} \hat{n}^{c_0} \prod_{i=1}^d \partial_{j_i} \hat{n}^{c_i}
\end{equation}
and compute its skyrmion number with the following volume integral over entire space:
\begin{equation}
N = \frac{1}{S_d} \int\limits_{V} d^dx \, \epsilon_{j_1\cdots j_{d}} A_{j_1\cdots j_d} \ .
\end{equation}
This is quantized if the space can be effectively compactified, for example by the virtue of $\hat{\bf n}$ having the same constant value at all points far away from the origin. However, skyrmions enjoy topological protection only in the \emph{classical} continuum limit. A skyrmion can be smoothly deformed into a mostly uniform field configuration whose spatial variations are confined to a finite volume. Then, a \emph{quantum} tunneling process, or instanton, can flip it into a topologically trivial state. Formally, one does not have enough space-time indices to construct a topological current (\ref{top-current}) from $A_{\lambda_{1}\cdots\lambda_{d}}$ and a Lagrangian term that conserves it. Instantons are governed by a remnant of the Maxwell term in Lagrangian density:
\begin{eqnarray}
\mathcal{L}_{\textrm{i}} = \frac{\Gamma}{2} (\partial_{0}\Phi_{0})^{2} 
\end{eqnarray}
where $\Phi_{0}$ is the rank $d$ ``dual'' gauge field (\ref{Phi1}):
\begin{equation}
\Phi_0 = \epsilon_{0 j_1\cdots j_d} A_{j_1\cdots j_d} \ .
\end{equation}
Instantons look like quantized ``hedgehogs'' $\partial_{\mu}\Phi_{\mu}\neq0$ in space-time. They unavoidably proliferate, and then their coarse-grained dynamics involves arbitrary local fluctuations of the real scalar field $\Phi_{0}$, which spoils the quantization of skyrmion number $N$ in the ground state.

\subsection{Essential phase diagram}\label{secDynPhaseDiag}

The effective field theories given by the Lagrangian densities (\ref{L3}) and (\ref{L5}) have rich phase diagrams. We will argue that a proper regularization enables a hierarchy of phases featuring Higgs-like and Coulomb-like gauge field dynamics at different ranks $n$, up to $n=d-1$ in $d$ spatial dimensions.

The plain continuum limit Lagrangians written in the previous sections always penalize gauge flux through Maxwell terms. This is a problem when we want to describe topologically ordered phases with deconfined monopoles. The solution to this problem is a compact gauge theory. If we put a dimensionless gauge field $\mathcal{A}_{\mu_1\cdots\mu_n} = a^n A_{\mu_1\cdots\mu_n}$ on a lattice, where $a$ is the lattice constant, then a compact Abelian Maxwell term in the action can be symbolically written as:
\begin{equation}\label{compactMaxwell}
S_{M\textrm{n}} = -\beta_n \sum_{\lbrace\mu\rbrace}
  \cos \Bigl(\epsilon_{\mu_1\cdots\mu_{d-n}\nu\lambda_1\cdots\lambda_n} \Delta_\nu \mathcal{A}_{\lambda_1\cdots\lambda_n} \Bigr) \ .
\end{equation}
The summation runs over all oriented $n+1$ dimensional ``plaquettes'' of the space-time lattice (with discretized time). It takes $d-n$ ordered indices $\mu$ to specify a ``plaquette'' orientation. The symbol inside the cosine is a placeholder for the sum over the oriented $n$ dimensional ``edges'' of the given ``plaquette'', where $\Delta_\mu f_i = f_{i+\mu} - f_i$ is the discrete lattice derivative of $f$ in the direction $\mu$ computed at the lattice site $i$. The lattice gauge field $\mathcal{A}$ is an angle variable that lives on the oriented ``edge'' specified by its indices. For example, the cubic 2+1D space-time lattice has square plaquettes with four corners $1,2,3,4$ whose orientation is specified by a single index $\mu$ (perpendicular to the plaquette); the lattice curl inside the cosine is $\mathcal{A}_{12}+\mathcal{A}_{23}+\mathcal{A}_{34}+\mathcal{A}_{41}$ if we relabel the gauge fields living on the oriented plaquette's edges by the initial and final site of the edge. The continuum limit $a \to 0$ of (\ref{compactMaxwell}) with a proper choice of the dimensionless coupling $\beta_n$ is the non-compact Abelian Maxwell term. Taking the continuum limit, i.e. expanding the cosine to quadratic order, is permissible only if $\beta_n$ is large so that the fluctuating values of $\mathcal{A}$ are small.

\begin{figure}
\includegraphics[width=2in]{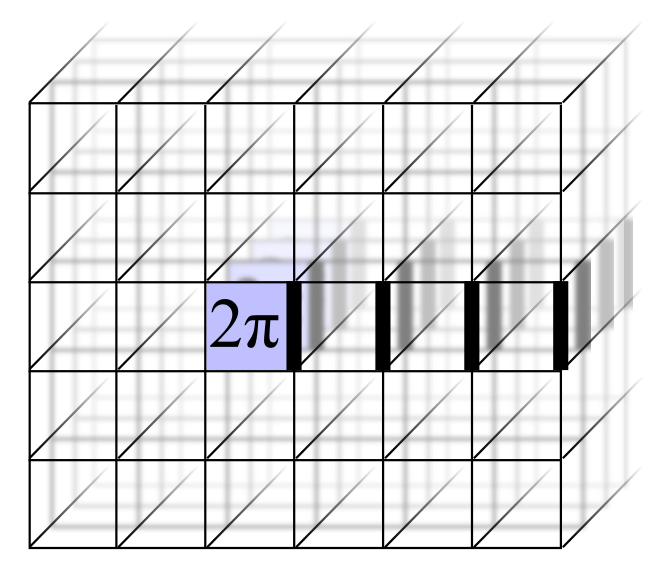}
\caption{\label{CompactDiracString}An unobservable Dirac string on the cubic lattice. The gauge field $A_{ij}=-A_{ji}$ lives on lattice bonds between neighboring sites $i,j$ and equals zero everywhere except on the thick bonds, where it is $2\pi$. The lattice curl of the gauge field is defined on lattice plaquettes as $\epsilon_{\mu\nu\lambda}\partial_\nu A_\lambda \equiv A_{12} + A_{23} + A_{34} + A_{41}$ if the sites of a square plaquette are labeled $1,2,3,4$ in the counter-clockwise sense. Magnetic flux is zero on all plaquettes except the shaded one and the ones parallel to it on the same vertical column. The flux value of $2\pi$ lives in the compact Maxwell term (\ref{compactMaxwell}) and has no physical impact.}
\end{figure}

The benefit of the compact Maxwell term is that a $2\pi$ flux quantum on a ``plaquette'' is physically unobservable and constitutes a pure-gauge configuration (see Fig.\ref{CompactDiracString}). This gives freedom to monopoles. Consider a $d=3$ dimensional system. We can insert a monopole by generating an appropriate rank 2 field configuration $\mathcal{A}_{\mu\nu}\neq 0$. This monopole can interact with charge currents only if its presence affects the rank 1 gauge field $\mathcal{A}_\mu$ through rank linking. However, the induced rank 1 gauge field of a monopole necessarily comes with a Dirac string. If the gauge dynamics is non-compact, then the string costs a finite energy per unit length and confines the monopoles to small topologically neutral clusters. In contrast, a compact theory makes the Dirac string invisible by collecting all of its quantized flux through a single column of plaquettes -- monopoles can be free and charged particles can experience them.

A byproduct of monopole proliferation in pure rank 1 compact gauge theories is charge confinement. Monopoles are abundant when $\beta_1$ is small and the plain continuum limit of (\ref{compactMaxwell}) cannot be justified. Then, the lattice dynamics features an angle-valued gauge field $\mathcal{A}$ whose canonically conjugate electric field $\mathcal{E}$ must be integer-valued. This field lives on the lattice links, so electric flux comes in the form of quantized strings that terminate at the locations of charged particles according to Gauss' law. One could say that monopole fluctuations gap out the electric field -- electric flux lines cost energy in proportion to their length, so charged particles are confined \cite{Polyakov1987}. This phenomenon does not occur in the disordered phase of a system as simple as our reference model of neutral bosons hopping on a lattice. So, how can we avoid it despite introducing gauge fields by singular gauge transformations? The key new feature of the present theory is the presence of multiple gauge field ranks and links between them. The confinement of rank 1 charge is avoided because the rank-linking term in the action modifies Gauss' law. Charged particles can interact directly with the deconfined rank 2 gauge field in the disordered phase, and not act as sources of the costly rank 1 electric flux. We elaborate this mechanism later in this section, in the context of a non-Abelian gauge theory. Another possible deconfinement mechanism is tied to the \emph{frustrated} compact gauge theories that naturally describe certain frustrated magnets \cite{sachdev00, Hermele2004a, nikolic:024401, nikolic:064423}. Here, entropy effects keep charges free even in the strong coupling regime with small $\beta_1$, as seen in solvable theoretical models \cite{motrunich02, Senthil2002, wen03b, Hermele2004a} and numerical calculations \cite{Shannon2012, Banerjee2008}.

The goal of this paper is to explore topologically ordered phases, and the practical feasibility of this task currently relies on the continuum limit. Therefore, we will not emphasize any further the compact formulation of the theory. Instead, it will be understood that the continuum limit theory requires an ultra-violet regularization that renders quantized Dirac attachments unobservable, and such a lattice regularization is indeed available. Note, however, that a regularization lattice is not necessarily the microscopic lattice of the system.

Constructing non-compact Maxwell terms with non-Abelian gauge fields is a more challenging task. One could define dimensionless non-Abelian gauge fields that operate on particle spinors, and construct the Maxwell terms from the traces of the products of Peierls factors $W=\exp(i \epsilon_{\mu_1\cdots\mu_{d-n} \nu \lambda_1\cdots\lambda_n} \Delta_\nu \mathcal{A}_{\lambda_1\cdots\lambda_n})$. This works fine on two-dimensional plaquettes because their oriented boundary is one-dimensional and uniquely represented by the order of $W$ factors under the trace. However, it is unclear how to unambiguously generalize this to higher dimensions and accommodate rank $n>1$ fields. Fortunately, a compact regularization is not needed for non-Abelian gauge fields: hedgehogs do not come with Dirac strings attached.

Now that we have defined a regularization where it is needed, we can proceed with the phase diagram analysis. Let us characterize the dynamics of the rank $n$ gauge field as Higgs-like if its fluctuations are suppressed, and Coulomb-like if its fluctuations are abundant. We will shortly make this characterization precise, with a provision which is not emphasized in the plain continuum formulations of the effective theory. When we introduce a gauge field at rank $n$ by a singular gauge transformation, this gauge field must have a strictly quantized and localized flux in the Higgs $n$ state at every position. The formal agent of flux quantization is either an explicit constraint in the path integral measure, or $\beta_n \to \infty$ in the compact gauge theory. Without this modification of the effective theory, the artificially introduced gauge field would gap out the gapless modes of the ``matter'' field as a part of the Anderson-Higgs mechanism. The explicit constraints on the gauge fields are not needed only in the topologically ordered phases which we ultimately pursue. Also, we will not tackle the important and difficult question of what stabilizes the phases with Higgs dynamics at intermediate ranks. Such phases feature emergent gauge boson excitations and definitely require significant and perhaps intricate interactions \cite{wen02c, wen03b} between simple microscopic degrees of freedom (the phase transitions involving scalars and emergent gauge fields can be first order \cite{Halperin1974} and hence beyond reach of the basic renormalization group treatment in scalar theories). Our goal will be merely to identify and characterize these phases from the perspective of singularity dynamics.

A Higgs state at rank $m$ implies a Higgs state at all higher ranks $n>m$. In a generalization of the usual Higgs mechanism, the rank $n$ gauge field $A(n)$ is suppressed into a Higgs state by the condensation of the current $j(n)$ it minimally couples to. Moreover, $A(n)$ is suppressed if any current $j(m)$ at a lower rank $m\le n$ condenses. This is a consequence of the origin of gauge fields in the matter field singularities. A condensation of $j(m)$ either expels or localizes all of its singularities, making them costly and preventing their diffusion which could give rise to soft gauge modes at higher ranks. Formally, the simplified effective theories (\ref{L3b}) and (\ref{L5b}) replace currents $j(n)$ with constructs involving linked gauge fields $A(n-1)$, so suppressed fluctuations of $A(n-1)$ in a Higgs $n-1$ state amount to matter condensation at rank $n$. The Higgs mechanism then propagates recursively to all higher ranks where it gaps out the gauge fields.

Similarly, a Coulomb state at rank $n$ implies a Coulomb state at all lower ranks $m<n$. When the rank $n$ gauge field $A(n)$ fluctuates abundantly in its Coulomb $n$ state, then the singularities of the lower rank current $j(n-1)$ have necessarily proliferated and diffused. The gauge field $A(n-1)$ is gapped out by Coulomb mechanism (deconfinement of defects), and its Coulomb dynamics recursively propagates down the ranks in the Lagrangian densities (\ref{L3b}) and (\ref{L5b}). Note that the absence of a gauge symmetry at rank $n$ does not automatically induce a Higgs $n$ state because the lower rank Coulomb dynamics provides no bias for an ``order parameter'' at rank $n$.

As a consequence of these relationships between ranks, each conventional phase of the effective theory corresponds to a sequence $\mathcal{G}_n = C_{1}C_{2}\cdots C_{n}H_{n+1}\cdots H_{d-1}$ of Coulomb $C_{n}$ and Higgs $H_{n}$ types of dynamics at consecutive ranks, with a switch from Coulomb to Higgs dynamics at one particular rank $n$. These phases are sharply defined in the thermodynamic limit. Only the gauge field at the last Coulomb-like rank $n$ is spared from both Higgs and Coulomb mechanisms, and remains massless with an infinite penetration depth. There is one exception to this rule in the compact gauge theory. We show in Appendix \ref{appDuality} that the rank $d-1$ gauge field is gapped in the all-Coulomb phase $\mathcal{G}_{d-1} = C_{1}\cdots C_{d-1}$. In the non-Abelian case, we naively expect that the matter coupled to the massless gauge field at rank $n$ is confined and free only asymptotically. However, matter at lower ranks $m<n$ is truly free, as we discuss at the end.

This distinction between phases can also be characterized by the confinement of singularity defects. A rank $n$ defect in $d$ dimensions is a $d-n-1$ dimensional excitation characterized by the $\pi_n(S^n)$ homotopy group (with understanding that only point-defects at rank $d-1$ are topologically protected). Confined defects are closed neutral manifolds of finite size, typically small due to their high energy cost per unit manifold area. A Higgs $n$ state features gapped fluctuations of confined defects, and in that sense conserves the defect charge. A deconfined state at rank $n$ is characterized by abundant, arbitrarily large and possibly open manifolds of $d-n-1$ dimensional defects, and in that sense can be a defect condensate.

As a physically relevant example, consider neutral spinless bosons in $d=3$ dimensions. $\mathcal{G}_0 = H_1H_2$ is a superfluid phase with Goldstone modes and confined vortices. The phase $\mathcal{G}_2 = C_1C_2$ is a fully gapped conventional Mott insulator with uncorrelated fluctuations. The phase $\mathcal{G}_1 = C_1H_2$ is unconventional: the rank 1 matter field is gapped and coupled to an emergent U(1) electrodynamics with deconfined vortices and confined monopoles. This is identified with the U(1) spin liquid in magnetic systems \cite{Hermele2004a}. In the analogous case of \emph{spin} dynamics, $\mathcal{G}_0$ is a magnet, $\mathcal{G}_2$ a gapped paramagnet, and $\mathcal{G}_1$ a paramagnet with an emergent non-Abelian gauge field and asymptotic freedom for particles. The phases with prominent gauge field dynamics are obviously realized in our world, as described by the standard model of particle physics.

Special phases $C_1^{\phantom{,}}\cdots T_{d-1}$ with topological order can be stabilized by topological protection: any change of the total topological charge of point defects requires crossing an infinite free energy barrier in infinite systems. Such phases are incompressible quantum liquids of abundant but non-condensed monopoles and hedgehogs. The rank $d-1$ gauge field remains gapped as if it lived in a Higgs state, and keeps the lower rank gauge field gapped via the Coulomb mechanism, thus propagating the gapped dynamics recursively down to the rank 1. We will discuss these kinds of phases in Section \ref{secTopOrder} and show that they have deconfined fractional quasiparticles in which a rationally quantized amount of charge or spin is bound to a topological defect. 

We have already established the possibility of topological defect deconfinement. We now show that this also leads to particles' ``charge'' (spin) deconfinement at lower ranks even in the non-Abelian gauge theory. An ordinary non-Abelian theory in $d=3$
\begin{equation}
\mathcal{L}=\frac{1}{2}\left\vert (\partial_{\mu}^{\phantom{x}}+ig\gamma^{a}A_{\mu}^{a})\psi\right\vert ^{2}-\frac{1}{4}F_{\mu\nu}^{a}F^{\mu\nu,a}
\end{equation}
featuring a field tensor
\begin{equation}
F_{\mu\nu}^{a}=\partial_{\mu}^{\phantom{x}}A_{\nu}^{a}-\partial_{\nu}^{\phantom{x}}A_{\mu}^{a}-g\,f^{abc}A_{\mu}^{b}A_{\nu}^{c}
\end{equation}
has the stationary-action field equation
\begin{equation}
J^{\mu,a}=\partial_{\nu}^{\phantom{a}}F^{\mu\nu,a}-gf^{abc}A_{\nu}^{b}F^{\mu\nu,c}
\end{equation}
that identifies a particle with charge $g$ (spin) as a source of the gauge flux. Charge is confined at least in the strong-coupling limit. In contrast, the non-Abelian effective theory (\ref{L5b}) in $d=3$ yields the following stationary condition by variations of the rank 1 gauge field:
\begin{equation}
J_{\mu}^{a} = \partial_{\nu}^{\phantom{,}}F_{\mu\nu}^{a} - g\epsilon^{abc}A_{\nu}^{b}F_{\mu\nu}^{c}
  +k'_{2}(\partial_{\nu}^{\phantom{,}}\hat{n}^{a})j_{\mu\nu}^{\phantom{,}}
\end{equation}
where $g=2$ and
\begin{equation}
j_{\mu\nu}^{\phantom{,}} = \frac{1}{2}(A_{\mu}^{a}\partial_{\nu}^{\phantom{,}}\hat{n}^{a}-A_{\nu}^{a}\partial_{\mu}^{\phantom{,}}\hat{n}^{a})
  -A_{\mu\nu}^{\phantom{,}} \ .
\end{equation}
Now, we can avoid attaching the rank 1 gauge flux to a particle and instead attach a rank 2 flux:
\begin{equation}
J_{\mu}^{a} \to k'_2 (\partial_{\nu}^{\phantom{x}}\hat{n}^{a})j_{\mu\nu}^{\phantom{x}}\quad,\quad F_{\mu\nu}^{a}\to0 \ .
\end{equation}
This is an option only if the gauge field $A_{\mu\nu}$ is not dynamically suppressed by the confinement of its flux. Very roughly, we get a Gauss law type of relationship $\langle\hat{n}^{a}\rangle \partial_i^{\phantom{,}} A_{0i}^{\phantom{,}} \sim J_0^{a} \sim \delta({\bf x})$ for a static point source $J_{0}^{a}$, and an infra-red convergent energy cost through the Abelian $A_{\mu\nu}$ Maxwell term. Note that inserting a definite spin $J_{0}^{a}$ necessarily creates a region with a non-zero average $\langle\hat{n}^{a}\rangle$ despite large fluctuations of $\hat{n}^{a}$ in an incompressible quantum liquid. Effectively, the rank 2 flux can screen charge from the rank 1 flux and preempt charge confinement.

The above argument can be readily generalized to compact Abelian gauge theories and higher dimensions. However, a compact gauge theory in $d=2$ dimensions does not have a rank 2 gauge field that could deconfine charges. Instanton events \cite{Polyakov1987}, identified as space-time ``monopoles'' in the literature on spin liquids \cite{sachdev02e, herbut03, Hermele2004}, confine the particles at rank 1, including any fractional partons of an electron. A weaker logarithmic charge confinement ``by vortices'' occurs even in the continuum-limit situations, through the unbounded Coulomb potential $V(r)\sim \ln(r)$ between static charges a distance $r$ apart. It seems naively that charge deconfinement in $d=2$ is possible only if topological defects are suppressed by a Higgs mechanism. Of course, the truth is more complicated and interesting. Two-dimensional deconfinement without a Higgs mechanism is experimentally evident in fractional quantum Hall states, and it has been theoretically established in certain U(1) spin liquids of Dirac spinons \cite{Hermele2004, Grover2014}.

\section{Topological Lagrangian term}\label{secTopTerm}

Here we construct the topological Lagrangian density term $\mathcal{L}_\textrm{t}$ of the effective field theory. Its role is to implement the topological $\pi_{d-1}(S^{d-1})$ charge conservation in the continuum limit description of incompressible quantum liquids. This is necessary only at the highest gauge theory rank $n=d-1$ in $d$ dimensions, because a Maxwell term, which normally controls defect confinement, is absent from the Lagrangian density at rank $d$.

The total topological charge $N$ of point defects contained in a certain volume is given by (\ref{top-charge-2}). $N$ is conserved if
\begin{equation}
\partial_0 N = \frac{1}{q}\int d^{d}x\;\epsilon_{ij_1\cdots j_{d-1}}\partial_0\partial_i A_{j_1\cdots j_{d-1}} = 0 \ ,
\end{equation}
or equivalently $\partial_0 \mathcal{J}_0 = 0$ expressed using the topological current
\begin{equation}\label{top-current-1}
\mathcal{J}_\mu = \epsilon_{\mu\nu\lambda_1\cdots\lambda_{d-1}} \partial_\nu A_{\lambda_1\cdots\lambda_{d-1}}
\end{equation}
However, this still allows instantaneous creation and annihilation of arbitrarily separated defect-antidefect pairs. In order to be consistent with local dynamics, we must promote the condition for topological charge conservation into:
\begin{equation}\label{top-charge-conserv}
\partial_\mu \mathcal{J}_\mu = 0 \ .
\end{equation}

One way to implement the topological charge conservation involves introducing an auxiliary Lagrange multiplier field $\Lambda$ into the path integral and writing the topological Lagrangian term as:
\begin{equation}
\mathcal{L}_{\textrm{t}} \sim i\, \Lambda \, \partial_\mu \mathcal{J}_\mu \ .
\end{equation}
Any world-lines that violate (\ref{top-charge-conserv}) will destructively interfere and cancel their contributions to the path integral. However, this is not adequate because the topological charge is forcefully conserved regardless of the underlying dynamics, even if the particles are localized. The only remedy for this problem is to use an existing degree of freedom as a Lagrange multiplier. The next section will describe the main construction principles for a topological term that:
\begin{enumerate}
\item does not introduce new degrees of freedom;
\item has no physical effect in conventional states;
\item respects all symmetries.
\end{enumerate}
Section \ref{secTopTermSpinor} then derives the topological term directly from a spinor field that represents a vector field $\hat{\bf n}$ in $d$ dimensions using the Spin($d$) group. Finally, we consider symmetry properties and restrictions for topological terms in Section \ref{secTopTermSym}.

\subsection{Topological term preliminaries}\label{secTopTermPrelim}

Section \ref{secQH} has already hinted the following topological term in the Lagrangian density:
\begin{equation}\label{top-term-2}
\mathcal{L}_{\textrm{t}} = i\, K_d \, J_\mu \mathcal{J}_\mu \ ,
\end{equation}
where $K_d$ is a coupling constant that we will determine later. The particles' gauge-invariant charge current $J_\mu$ is an existing degree of freedom, so $\mathcal{L}_{\textrm{t}}$ satisfies the above criterion 1. The conventional states for the criterion 2 are typically superfluids and Mott insulators. A topological defect in a superfluid phase always has a well defined core from which the particles are expelled. Therefore, the presence of a static defect with density $\mathcal{J}_0 \neq 0$ at some location implies the absence of particles $J_0 = 0$ at that location, leading to $\mathcal{L}_{\textrm{t}} = 0$ in a superfluid. Similarly, if we reverse the roles played by the canonical particle number operator $n$ and its conjugate phase $\theta$, we find that the presence of a particle with density $J_0 \neq 0$ at some location in a Mott insulator implies a local expulsion of topological defects $\mathcal{J}_0 = 0$, again leading to $\mathcal{L}_{\textrm{t}} = 0$. In this sense, the topological term (\ref{top-term-2}) satisfies the criterion 2. For now, we will assume that the dynamical part $\mathcal{L}_{\textrm{d}}$ of the Lagrangian density has the same symmetries as (\ref{top-term-2}). If that is not the case, we will have to modify the topological term in order to fix its symmetries and satisfy the criterion 3. We will discuss how this can be done in Section \ref{secTopTermSym}.

The Lagrange multiplier that implements topological charge conservation $\partial_\mu \mathcal{J}_\mu = 0$ is hidden within the charge current, as revealed in Section \ref{secQH}. It works only in unconventional incompressible quantum liquids where abundant quantum fluctuations allow point-defects and particles to occupy the same location (with resolution determined by the coarse-grained length scale $\xi$). Note that incompressibility of both particle and defect densities is crucial -- if either can adjust, it will adjust to avoid a costly overlap between particles and defects. The symmetry (duality) between particle and defect currents in (\ref{top-term-2}) simultaneously reaffirms the particle charge conservation $\partial_\mu J_\mu = 0$. We can extract the currents from appropriate spinor fields $\psi$ for particles and $\Psi$ for point-defects
\begin{eqnarray}\label{currents-def}
J_{\mu} &=& -\frac{i}{2}\Bigl\lbrack\psi^{\dagger}(\partial_{\mu}\psi)-(\partial_{\mu}\psi^{\dagger})\psi\Bigr\rbrack + A_\mu |\psi|^2 \\
\mathcal{J}_{\mu} &=& -\frac{i}{2}\Bigl\lbrack\Psi^{\dagger}(\partial_{\mu}\Psi)-(\partial_{\mu}\Psi^{\dagger})\Psi\Bigr\rbrack + \mathcal{A}_\mu |\Psi|^2 \ . \nonumber
\end{eqnarray}
to show the charge conservation mechanism. Incompressibility implies frozen amplitudes of $\psi$ and $\Psi$, so that only the phases $\phi,\Phi$ in $\psi=|\psi|e^{i\phi}$ and $\Psi=|\Psi|e^{i\Phi}$ are free to fluctuate, producing effectively:
\begin{equation}
J_\mu = |\psi|^2 (\partial_\mu \phi + A_\mu) \quad,\quad \mathcal{J}_\mu = |\Psi|^2 (\partial_\mu \Phi + \mathcal{A}_\mu) \ .
\end{equation}
Substituting in (\ref{top-term-2}) yields
\begin{eqnarray}\label{top-term-2b}
\mathcal{L}_{\textrm{t}} &=& i\, K_d |\psi|^2 \, (\partial_\mu\phi + A_\mu) \mathcal{J}_\mu \\
  &\to& -i\, K_d |\psi|^2 \, \phi \; \partial_\mu\mathcal{J}_\mu + i\, K_d |\psi|^2 \, A_\mu \mathcal{J}_\mu \nonumber \\[0.1in]
\mathcal{L}_{\textrm{t}} &=& i\, K_d |\Psi|^2 \, (\partial_\mu\Phi + \mathcal{A}_\mu) J_\mu \nonumber \\
  &\to& -i\, K_d |\Psi|^2 \, \Phi \; \partial_\mu J_\mu + i\, K_d |\Psi|^2 \, \mathcal{A}_\mu J_\mu \nonumber
\end{eqnarray}
after an integration by parts, so $\phi$ and $\Phi$ can act as Lagrange multipliers that implement the conservation of topological and particle charge respectively. 

Let us scrutinize the conservation mechanism more carefully. Both $\phi$ and $\Phi$ are angles. Integrating out $\phi\in\lbrack 0,2\pi)$ in (\ref{top-term-2b}) gives us:
\begin{eqnarray}
&& \int\limits _{0}^{2\pi}\mathcal{D}\phi\,\exp\left\lbrace \int d^{d+1}x\,iK_d|\psi|^2\phi\,\partial_{\mu}\mathcal{J}_{\mu}\right\rbrace \\
&& \quad \propto \prod_{{\bf x}}\frac{\sin(\pi K_d|\psi|^2 d^{d+1}x\,\partial_{\mu}\mathcal{J}_{\mu})}{K_d|\psi|^2 d^{d+1}x\,\partial_{\mu}\mathcal{J}_{\mu}}
  \to \prod_{{\bf x}}\delta(\partial_{\mu}\mathcal{J}_{\mu}) \nonumber
\end{eqnarray}
in the following qualitative sense. The final Dirac delta function of $\partial_\mu \mathcal{J}_\mu$ is formally obtained from the integral over $\phi$ only when the dimensionless number $K_d|\psi|^2 d^{d+1}x\,\partial_{\mu}\mathcal{J}_{\mu}$ is an integer. This condition is indeed satisfied by the microscopic quantization of topological charge, as we will now show by discretizing the integral. Let $d^{d}x$ be the volume that contains a single particle, and $l^{d}$ be the volume that contains a single topological defect. Consider a state with $n$ topological defects per particle, i.e. with the ``filling factor'' $\nu=1/n$. Since $d^{d}x=nl^{d}$, we can interpret
\begin{eqnarray}
K_d|\psi|^2 d^{d+1}x\,\partial_{\mu}\mathcal{J}_{\mu} &\sim& K_{d}|\psi|^2\,nl^{d}\,\Delta_{\mu}\mathcal{J}_{\mu} \\
  &=& K_{d}|\psi|^2\,n q\,\Delta_{\mu}\mathcal{J}'_{\mu} \nonumber
\end{eqnarray}
with $\Delta_{\mu}=dx\partial_{\mu}$ being a discrete derivative on the scale $dx$, and $q$ the unit of topological charge (flux quantum). We defined an integer-valued defect current $\mathcal{J}'_{\mu}=l^{d} \mathcal{J}_{\mu}/q$ based on the fact that the flux density $\mathcal{J}_{0}$ makes $\mathcal{J}_{0}/q$ the number density of topological defects. The quantized topological current has no divergence if $K_{d}|\psi|^2nq\in\mathbb{Z}$, i.e. $K_{d}|\psi|^2=(\nu/q)\times\textrm{integer}$. Later, when we consider topological orders, we will reproduce this relationship in a proper field-theoretical manner.

\subsection{Topological term from spinor fields}\label{secTopTermSpinor}

The goal of this section is to construct the topological Lagrangian term (\ref{top-term-2}) directly from a spinor field $\psi$ of particles. Such a construction is possible because the particle field contains all information about the currents and topological defects. We will develop the basic idea here, and analyze symmetry restrictions and extensions in the next Section \ref{secTopTermSym}. To begin with, the spinor $\psi$ has to represent a U(1) phase $\theta$ for charge dynamics and a vector field $\hat{\bf n}$ for spin dynamics. The vector $\hat{\bf n}$ must be $d$ dimensional with fixed magnitude in order to have topologically protected hedgehog defects in $d$ spatial dimensions. Therefore, we will use a coherent state complex spinor representation of the Spin($d$) group, which generalizes spin to $d$ dimensions.

The generators $\gamma^a$ of the Spin($d$) group are $d$ Dirac matrices that obey the Clifford anticommutator algebra:
\begin{equation}
\lbrace\gamma^{a},\gamma^{b}\rbrace=2\delta_{ab} \ .
\end{equation}
The angular momentum operators that generate rotations in $ab$ planes:
\begin{equation}\label{ang-momentum}
J_{ab}=-\frac{i}{4}\lbrack\gamma^{a},\gamma^{b}\rbrack
\end{equation}
can be used to rotate a fixed reference spinor $\psi_0$ into a coherent state whose spin points along $\hat{\bf n} = (\theta_1,\dots,\theta_{d-1})$:
\begin{equation}\label{spinor}
\psi(\hat{\bf n})=e^{-iJ_{d-1,d}\theta_{d-1}}\cdots e^{-iJ_{2,3}\theta_{2}}e^{-iJ_{1,2}\theta_{1}}e^{i\phi}\,\psi_0 \ .
\end{equation}
The spherical coordinate system angles $\theta_i$ and $\hat{\bf n}$ are related according to (\ref{n-vs-theta}). The last angle $\phi$ is not associated with any generator and defines a U(1) phase for charge currents.

The main ingredient of the topological Lagrangian term $\mathcal{L}_{\textrm{t}}$ is the topological current (\ref{top-current-1}) that involves the rank $d-1$ gauge field. How can we extract this gauge field from the spinor $\psi$? For example, if we use the Abelian singular gauge transformations (\ref{singA1c}) recursively from the rank $d-1$ down to rank 1, we naively obtain the following relationship between the Abelian gauge field and the spinor's U(1) phase $\phi$:
\begin{equation}
A_{\mu_1\cdots\mu_{d-1}} = \frac{1}{(d-1)!}\left(\epsilon_{a_{1}\cdots a_{d-1}}\prod_{i=1}^{n}\partial_{\mu_{a_{i}}}\right)\!\phi \ .
\end{equation}
This expression applies an antisymmetrized product of derivatives on $\phi$. Any analytic function $\phi({\bf x})$ automatically yields $A_{\mu_1\cdots\mu_{d-1}}=0$, so this expression can have meaning only if we define a rigorous rule for applying the derivatives on singular functions. We will define such a rule by generalizing the familiar two-dimensional case. When we extract a vortex $\psi(r,\phi)=e^{i\phi}$ expressed using the polar angle $\phi$ into a gauge field $A_i = \partial_i \phi$, then the magnetic flux $B = \epsilon_{ij} \partial_i A_j \sim \epsilon_{ij} \partial_i \partial_j \phi = 2\pi \delta({\bf x})$ integrates as:
\begin{equation}
2\pi = \int\limits_{B^2} \! d^2 x \, B = \int\limits_{B^2} \! d^2 x \, \epsilon_{ij} \partial_i \partial_j \phi
  \equiv \oint\limits_{S^1} \! dx_i \, A_i = \oint\limits_{S^1} \! dx_i \, \partial_i \phi \ . \nonumber
\end{equation}
The first integral is defined on a disk, or a 2-ball $B^2$ that contains the vortex singularity, and we formally rewrite it using the double derivative notation. In order to calculate this integral, we apply Stokes theorem on the loop (1-sphere $S^1$) that bounds $B^2$. The ensuing loop integral with one less derivative is well-defined.

Now consider general expressions
\begin{equation}\label{Fsing}
\mathcal{F}_{n} = \psi^{\dagger}\epsilon_{\mu_{1}\cdots\mu_{n}}\partial_{\mu_{1}}\cdots\partial_{\mu_{n}}\psi
\end{equation}
for $1\le n \le d$ involving the spinor (\ref{spinor}), and integrals
\begin{equation}
I_{k,n} = \!\! \int\limits_{B^{n}(k)} \!\!\! d^{n}x\,\mathcal{F}_{n}
\end{equation}
defined on $n$ dimensional ball domains $B^n(k)$ indexed by $k$. The integrals $I_{k,n}$ can be sensitive only to the $\pi_{n-1}(S^{n-1})$ singularities of $\mathcal{F}_{n} (\theta_1, \dots, \theta_{d-1}, \phi)$, which are point-like in an $n$ dimensional domain. Let us start from the highest rank in $d$ dimensions. Consider one $\pi_{d-1}(S^{d-1})$ point-singularity embedded inside a small ball $B^d(1) \subset \mathbb{R}^d$ with a sphere boundary $S^{d-1}(1)$. We anticipate that $\mathcal{F}_{d}$ is proportional to the Dirac function $\delta^{d}({\bf x})$ at the singularity, and hence properly characterized by $I_{1,d}$. All singularities that we integrate are formally characterized by appropriate distributions like $\delta^{d}({\bf x})$. Let us define
\begin{equation}\label{Iprime}
I'_{1,d} = \!\! \int\limits_{B^{d}(1)} \!\!\!d^{d}x\,(\partial_{\mu_{1}}\psi^{\dagger})\epsilon_{\mu_{1}\cdots\mu_{d}}\partial_{\mu_{2}}\cdots\partial_{\mu_{d}}\psi
\end{equation}
and apply Stokes-Cartan theorem:
\begin{eqnarray}\label{I-Stokes-Cartan}
I_{1,d} &=& \!\! \int\limits_{B^{d}(1)} \!\!\! d^{d}x\,\partial_{\mu_{1}}(\psi^{\dagger}\epsilon_{\mu_{1}\cdots\mu_{d}}\partial_{\mu_{2}}\cdots\partial_{\mu_{d}}\psi)
      - I'_{1,d} \\
  &=& \!\! \oint\limits_{S^{d-1}(1)} \!\!\! d^{d-1}x\,\psi^{\dagger}\epsilon_{\mu_{1}\cdots\mu_{d-1}}\partial_{\mu_{1}}\cdots\partial_{\mu_{d-1}}\psi - I'_{1,d} \ . 
      \nonumber
\end{eqnarray}
The displayed integral over $S^{d-1}(1)$ contains the function $\mathcal{F}_{d-1}$ which is singular by construction and zero away from the singularities. Thus, we can focus on the finite patches $B^{d-1}(k)\subset S^{d-1}(1)$ that contain one singularity of $\mathcal{F}_{d-1}$ each:
\begin{equation}\label{I-recursive-1}
I_{1,d} + I'_{1,d} = \sum_k \int\limits_{B^{d-1}(k)} \!\!\! d^{d-1}x\,\mathcal{F}_{d-1} = \sum_{k} I_{k,d-1} \ .
\end{equation}
The original $\pi_{d-1}(S^{d-1})$ point-singularity of $\mathcal{F}_d$ does not reside on $S^{d-1}(1)$, yet it is detected in lower-dimensional singular integrals over $B^{d-1}(k)$. This is possible only if singular strings of $\mathcal{F}_{d-1}$ emanate from the point singularity and intersect $S^{d-1}(1)$. Note that $S^{d-1}(1)$ is of arbitrary size, and multiple strings lead to multiple intersection points embedded inside the balls $B^{d-1}(k)$.

Now we can show that the residual $d$ dimensional integral $I'_{1,d}$ does not contribute to the topological Lagrangian term. The integral (\ref{Iprime}) contains the same antisymmetrized derivatives that are applied on $\psi$ in $\mathcal{F}_{d-1}$, so its value can build up only from the points on the strings where $\mathcal{F}_{d-1}$ is singular. We will work in the continuum limit for simplicity, assuming that some regularization procedure is available to rescue the usual rules of calculus when needed. Let us change the integration variables ${\bf x}\in B^d(1)$ into a ``radius'' $s$ that scans a singular string $\mathcal{S}$ and $y_1,\dots,y_{d-1}$ that span a shell $S^{d-1}$ locally perpendicular to the string at $s$. Since $\partial_s$ commutes with all $\partial_{y_i}$, integrating out $y_1,\dots,y_{d-1}$ has a chance to produce a finite spinor $\Psi(s)$ from the antisymmetrized $\partial_{\mu_2} \cdots \partial_{\mu_d} \psi$ in (\ref{Iprime}) only if all directions $\mu_2,\dots,\mu_d$ are tangential to $S^{d-1}$. Hence, we need
\begin{equation}
\partial_{\mu_1}\psi^{\dagger} \equiv \partial_s \psi^{\dagger} = i \psi^{\dagger} \left\lbrack \sum_{i=1}^{d-1} (\partial_s \theta_i) \Gamma_i - (\partial_s \phi)
  \right\rbrack
\end{equation}
in the immediate vicinity of the strings, where each operator
\begin{eqnarray}
\Gamma_i &=& \Bigl(e^{iJ_{i+1,i+2}\theta_{i+1}}\cdots e^{iJ_{d-1,d}\theta_{d-1}}\Bigr)^{\dagger} J_{i,i+1} \nonumber \\
  && \quad \times \Bigl(e^{iJ_{i+1,i+2}\theta_{i+1}}\cdots e^{iJ_{d-1,d}\theta_{d-1}}\Bigr)
\end{eqnarray}
is independent of $\theta_i$. In the presence of multiple strings $\mathcal{S}_k$ we get:
\begin{equation}
I'_{1,d} = i \sum_k \int\limits_{\mathcal{S}_k} ds \left\lbrack \sum_{i=1}^{d-1} (\partial_s^{\phantom{,}} \theta_i^{\phantom{,}}) \bar{I}_{k,d-1}^{(i)}
  - (\partial_s^{\phantom{,}} \phi) \bar{I}_{k,d-1}^{\phantom{,}} \right\rbrack \ .
\end{equation}
The scalar factors
\begin{eqnarray}
\bar{I}_{k,d-1}^{\phantom{,}} &=& \!\! \int\limits_{B^{d-1}(k)} \!\!\! d^{d-1}y\, \psi^{\dagger}
  \epsilon_{s\mu_{2}\cdots\mu_{d}}^{\phantom{,}}\partial_{\mu_{2}}^{\phantom{,}}\cdots\partial_{\mu_{d}}^{\phantom{,}}\psi \\
\bar{I}_{k,d-1}^{(i)} &=& \!\! \int\limits_{B^{d-1}(k)} \!\!\! d^{d-1}y\, \psi^{\dagger}\Gamma_i^{\phantom{,}}
  \epsilon_{s\mu_{2}\cdots\mu_{d}}^{\phantom{,}}\partial_{\mu_{2}}^{\phantom{,}}\cdots\partial_{\mu_{d}}^{\phantom{,}}\psi \nonumber
\end{eqnarray}
involve singular integrands and hence cannot possibly depend on the values of $\phi({\bf x})$ and $\theta_i({\bf x})$ respectively away from the strings, ${\bf x} \notin \mathcal{S}_k$. These scalars are not even arbitrary complex numbers, so their invariance under global U(1) and Spin($d$) rotations also prohibits a dependence on $\phi({\bf x})$ and $\theta_i({\bf x})$ on the local string, ${\bf x} \in \mathcal{S}_k$. Therefore, we can treat them as constants:
\begin{eqnarray}
I'_{1,d} &=& i \sum_k \int\limits_{\mathcal{S}_k} \left\lbrack \sum_{i=1}^{d-1} (d\theta_i^{\phantom{,}}) \bar{I}_{k,d-1}^{(i)}
  - (d\phi) \bar{I}_{k,d-1}^{\phantom{,}} \right\rbrack \\
  &=& i \sum_k \left\lbrack \sum_{i=1}^{d-1} (\Delta\theta_i^{\phantom{,}}) \bar{I}_{k,d-1}^{(i)}
  - (\Delta\phi) \bar{I}_{k,d-1}^{\phantom{,}} \right\rbrack \ .\nonumber
\end{eqnarray}
Here, $\Delta\theta_i$ and $\Delta\phi$ are the respective angle differences between the opposite ends of the string segments inside the integration domain. If $\theta_i$ or $\phi$ varied slowly, we would need to deal with the consequences of $I'_{1,d}$ being possibly finite. However, we need a topological Lagrangian term only in the effective theory that describes the coarse-grained dynamics of an incompressible quantum liquid. The spatial and temporal variations of $\theta_i$, $\phi$ average out to zero on the coarse-graining scale $\xi$ (which is larger than the average spatial separation between mobile singularities), causing $I'_{1,d}$ to effectively vanish.

Given that $I'_{1,d}\to 0$ is irrelevant upon coarse-graining in incompressible quantum liquids, the relationship (\ref{I-recursive-1}) directly connects a $d$ dimensional integral $I_{1,d}$ on a ball $B^d \subset \mathbb{R}^d$ to integrals $I_{k,d-1}$ defined on $d-1$ dimensional balls (disks) $B^{d-1} \subset S^{d-1}$ that live on the boundary $S^{d-1}$ of the original $B^d$. Such a connection separately holds for every point-like topological defect in $d$ dimensional space. We can carry out identical analysis starting from each $I_{k,d-1}$ on its own $B^{d-1}$, and relate it to similar $I_{k,d-2}$. Clearly, we can proceed recursively down to the lowest rank, by constructing a tree-graph in which a node at any rank $2<n\le d$ represents an integral $I_{i,n}$ equal to the sum of $I_{j,n-1}$ at lower ranks:
\begin{equation}\label{I-recursive-2}
I_{i,n} = \sum_{j} I_{j,n-1} \ .
\end{equation}
If we explicitly calculate the well-defined integrals
\begin{equation}\label{I-recursive-3}
I_{j,2} = \!\! \int\limits_{B^{2}(j)} \!\!\! d^{2}x\,\mathcal{F}_{2} = \!\! \oint\limits_{S^{1}(j)} \!\!\! dx\, \epsilon_\mu \psi^\dagger \partial_\mu \psi
\end{equation}
on loops that bound $B^{2}(j)$, we can recover the highest rank integrals $I_d$ which are related to (\ref{top-charge-1}) and extract the topological charge of monopoles and hedgehogs. 

Charge currents contribute (\ref{I-recursive-3}) through the U(1) phase $\phi$:
\begin{equation}
I_{j,2} \xrightarrow{\textrm{charge}} i|\psi|^2 \!\!\oint\limits_{S^{1}(j)}\!\!\! dx\, \epsilon_\mu \partial_\mu \phi = i|\psi|^2 \!\!\oint\limits_{S^{1}(j)}\!\!\! d\phi = 2\pi i|\psi|^2 N_j \ , \nonumber
\end{equation}
where $N_j\in \mathbb{Z}$ is the winding number of $\phi$ on the loop. We assumed that $|\psi|^2$ is finite, incompressible and constant on the loop length scales. The formula (\ref{I-recursive-2}) recursively collects all such winding numbers into $I_d$, which is designed to detect the topological charge of a point-like monopole in $d$ dimensions. The collection pattern is identical to that of Dirac attachments discussed in Section \ref{secGaugeMon}. If the spin structure of $\psi$ is smooth, we find
\begin{equation}\label{Icharge}
I_d \xrightarrow{\textrm{charge}} \int\limits_{B^d} d^dx\, \epsilon_{\mu_1\cdots\mu_d} \psi^\dagger \partial_{\mu_1}\cdots\partial_{\mu_d}\psi = 2\pi i |\psi|^2 N
\end{equation}
for a uniform $|\psi|^2\neq 0$, where $N$ is the total monopole charge contained within $B^d$.

Spin degrees of freedom $\hat{{\bf n}}(\theta_{1}\cdots\theta_{d-1})$ can contribute to (\ref{I-recursive-3}) only through the angle $\theta_{d-1}\in\lbrack 0,2\pi)$, since the other angles $\theta_j\in\lbrack 0,\pi\rbrack$ are topologically inert on loops. A vortex singularity of $\theta_{d-1}$ can exist only at positions where $\theta_{d-2}=0$ or $\theta_{d-2}=\pi$, because $\hat{{\bf n}}$ given by (\ref{n-vs-theta}) must be single-valued. At such positions, the value of $\theta_{d-1}$ does not distinguish different vectors $\hat{{\bf n}} = \hat{\bf x}^a (\psi^\dagger \gamma^a \psi)$. Hence, the spin coherent state $\psi$ can only acquire a U(1) phase factor $e^{i\theta}$ under rotation $\exp(-i J_{d-1,d} \theta_{d-1})$. In other words, $\psi$ is an eigenspinor of $J_{d-1,d}$ at the vortex singularities of $\theta_{d-1}$, with an eigenvalue $S \cos(\theta_{d-2}) = \pm S$ that depends on the Spin($d$) representation. The integral (\ref{I-recursive-3}) is most easily calculated on an infinitesimal loop around the vortex singularity, starting from (\ref{spinor}) and $\partial_\mu \psi = -i (\partial_\mu \theta_{d-1}) J_{d-1,d} \psi + \cdots$ where we ignore the topologically trivial angle variations:
\begin{eqnarray}
&& I_{j,2} \xrightarrow{\textrm{spin}} -i(\psi^\dagger J_{d-1,d} \psi) \!\!\oint\limits_{S^{1}(j)}\!\!\! dx\, \epsilon_\mu \partial_\mu \theta_{d-1} \nonumber \\
&& \qquad\quad = -2\pi i|\psi|^2 S \cos(\theta_{d-2}) N_j \ .
\end{eqnarray}
The winding number $N_j$ of $\theta_{d-1}$ is related to the topological charge of a hedgehog defect in the $\hat{\bf n}$ configuration, as naively indicated in the formula (\ref{top-hedgehog-angle}) and the surrounding discussion. When we go one rank up, the integral $I_{i,3}$ is defined on the $S^2$ sphere which contains multiple loops indexed by $j$. The total vorticity $\sum_j N_j = 0$ of $\theta_{d-1}$ must vanish on that closed manifold $S^2$. However, vortices and antivortices are necessarily attached to the opposite ``poles'' $\theta_{d-2}=0$ and $\theta_{d-2}=\pi$ respectively. One could visualize this situation by imagining vortex lines of $\theta_{d-1}$ that stretch in a three-dimensional space and go through the $S^2$ sphere. In this manner, the factor $\cos(\theta_{d-2})$ in the above formula constructively adds the opposite vortex charges on $S^2$, and effectively translates into a factor of $2$ if we collect the numbers $N_j$ only from the ``north'' poles. Analogous factors of $2$ appear each time we move one rank up in the hierarchy because every $n$ dimensional hedgehog configuration on an $n$ sphere always has at least one ``north'' $\theta_n = 0$ and ``south'' $\theta_n = \pi$  pole where the lower-rank angles $\theta_i$, $i<n$ form an $n-1$ dimensional hedgehog of their own (living on a lower-dimensional sphere $S^{n-1} \subset S^n$ centered at the pole). The formula (\ref{n-vs-theta}) illustrates this mathematically. Most generally, every point singularity of $\mathcal{F}_n$ in an $n$ dimensional domain terminates a number of south-pole and north-pole singular strings of $\mathcal{F}_{n-1}$ in the same domain, which together carry a vanishing $\pi_{n-1}(S^{n-1})$ and a non-vanishing $\pi_n(S^n)$ topological charge. The latter is equal to twice the $\pi_{n-1}(S^{n-1})$ charge collected from north poles only, and both are immune to smooth field transformations. In the end, we find:
\begin{equation}\label{Ispin}
I_d \xrightarrow{\textrm{spin}} \int\limits_{B^d} d^dx\, \epsilon_{\mu_1\cdots\mu_d} \psi^\dagger \partial_{\mu_1}\cdots\partial_{\mu_d}\psi
  = -2^{d-1}\pi i |\psi|^2 S N
\end{equation}
for a uniform $|\psi|^2\neq 0$, where $S$ is a representation-dependent eigenvalue of the rotation generators $J_{ab}$ and $N$ is the total hedgehog charge contained within $B^d$.

Finally we are well positioned to explore the following form of the topological Lagrangian term:
\begin{equation}\label{top-term-3}
\mathcal{L}_{\textrm{t}} \propto i\,\epsilon_{\mu\nu\lambda_1\cdots\lambda_{d-1}} \psi^\dagger \partial_\mu \partial_\nu \partial_{\lambda_1} \cdots
   \partial_{\lambda_{d-1}} \psi \ .
\end{equation}
The structure of derivatives is compatible with (\ref{top-term-2}). The rank $d-1$ gauge field should clearly emerge by coarse-graining the antisymmetrized ($\mathcal{A}$) expression
\begin{equation}\label{sing-gauge-field}
A_{\lambda_1\cdots\lambda_{d-1}} \propto \mathcal{A}\, \psi^\dagger \partial_{\lambda_1} \cdots \partial_{\lambda_{d-1}} \psi 
\end{equation}
because its quantized integral (\ref{top-charge-1}) extracts the topological charge enclosed within a sphere $S^{d-1}$ in the same fashion as the integral $I_{d}$ treated to Stokes-Cartan theorem in (\ref{I-Stokes-Cartan}). The topological current (\ref{top-current-1}) similarly emerges by coarse-graining
\begin{eqnarray}\label{top-current-2}
\mathcal{J}_{\mu} &=& \epsilon_{\mu\nu\lambda_1\cdots\lambda_{d-1}} \partial_\nu A_{\lambda_1\cdots\lambda_{d-1}} \\
  &\propto& \epsilon_{\mu\nu\lambda_1\cdots\lambda_{d-1}}\, \psi^\dagger \partial_\nu \partial_{\lambda_1} \cdots \partial_{\lambda_{d-1}} \psi \ . \nonumber
\end{eqnarray}
Note that the derivative $\partial_\nu$ is initially applied on (\ref{sing-gauge-field}) \emph{externally}, but it can be pulled inside, between $\psi^\dagger$ and $\psi$, since $\mathcal{J}_{\mu}$ always lives in space-time integrals and we found that the integrals $I'_n$ such as (\ref{Iprime}) can be ignored. Lastly, the topological Lagrangian term (\ref{top-term-2}) is:
\begin{eqnarray}
\mathcal{L}_{\textrm{t}} &=& i K_d\,J_{\mu}\mathcal{J}_{\mu}
    \propto \frac{1}{2}\Bigl\lbrack\psi^{\dagger}(\partial_{\mu}\psi)-(\partial_{\mu}\psi^{\dagger})\psi\Bigr\rbrack \mathcal{J}_{\mu} \nonumber \\
  &=& \frac{1}{2}\Bigl\lbrack\psi^{\dagger}(\partial_{\mu}\mathcal{J}_{\mu}\psi)-(\partial_{\mu}\psi^{\dagger}\mathcal{J}_{\mu})\psi\Bigr\rbrack \nonumber \\
  &=& \psi^{\dagger}(\partial_{\mu}\mathcal{J}_{\mu}\psi)-\frac{1}{2}\partial_{\mu}(\psi^{\dagger}\mathcal{J}_{\mu}\psi)
    -\frac{1}{2}\psi^{\dagger}(\partial_{\mu}\mathcal{J}_{\mu})\psi \nonumber \\
  &\to& \psi^{\dagger}(\partial_{\mu}\mathcal{J}_{\mu}\psi) \ .
\end{eqnarray}
The absence of a U(1) gauge field $A_\mu$ in the charge current $J_\mu$ is deliberate and in the spirit of extracting defect fluctuations from matter fields; $A_\mu$ can be generated by a singular gauge transformation. Behind the arrow, we removed a total derivative and the vanishing contribution of $\partial_\mu \mathcal{J}_\mu \to 0$ in incompressible quantum liquids. The last remaining piece amounts to (\ref{top-term-3}) after observing that the omitted  proportionality constant in (\ref{top-current-2}) must include a factor of $|\psi|^{-2}$. $\mathcal{J}_\mu$ is intrinsically composed from the angles $\theta_i$ and $\phi$ that $\psi$ depends on, and does not scale in proportion to $|\psi|^2$.

The precise proportionality constant in (\ref{top-term-3}) can be determined using (\ref{Icharge}) and (\ref{Ispin}). We know that the topological charges featured in these formulas obtain from the same gauge fields that live in the topological Lagrangian term. However, we have clearly identified two independent topological charges, one for monopoles and one for hedgehogs. We must relate them to two different gauge fields at rank $d-1$, according to (\ref{top-charge-1}):
\begin{eqnarray}
&& \textrm{monopoles:} \quad N_{\textrm{m}}^{\phantom{,}} = \frac{1}{2\pi} \oint\limits_{S^{d-1}} d^{d-1}x \, \epsilon_{j_1\cdots j_{d-1}}^{\phantom{,}}
    A^{\textrm{m}}_{j_1\cdots j_{d-1}} \nonumber \\
&& \textrm{hedgehogs:} \quad N_{\textrm{h}}^{\phantom{,}} = \frac{1}{S_{d-1}} \oint\limits_{S^{d-1}} d^{d-1}x \, \epsilon_{j_1\cdots j_{d-1}}^{\phantom{,}}
    A^{\textrm{h}}_{j_1\cdots j_{d-1}} \nonumber
\end{eqnarray}
Here, $A^{\textrm{m}}$ and $A^{\textrm{h}}$ are the final members of the Abelian and non-Abelian gauge field hierarchies respectively. Since (\ref{top-term-3}) renders $\mathcal{L}_{\textrm{t}} \propto I_d$, but extracts both topological charges from the same spinor field, we can write:
\begin{eqnarray}\label{top-term-4}
\mathcal{L}_{\textrm{t}} &=& \frac{i\,K_{d}S_{d-1}}{2^{d-1}\pi S}\,\psi^{\dagger}\epsilon_{\mu_{1}\cdots\mu_{d+1}}\partial_{\mu_{1}}\cdots\partial_{\mu_{d+1}}\psi
        \nonumber \\
  &\to& i K_{d}^{\phantom{,}}\, J_\mu^{\phantom{,}} \left(\mathcal{J}_{\mu}^{\textrm{h}}-\frac{S_{d-1}}{2^{d-1}\pi S}\,\mathcal{J}_{\mu}^{\textrm{m}}\right)
\end{eqnarray}
with hedgehog and monopole currents:
\begin{eqnarray}
\mathcal{J}_{\mu}^{\textrm{h}} &=& \epsilon_{\mu\nu\lambda_1\cdots\lambda_{d-1}}^{\phantom{,}} \partial_\nu^{\phantom{,}} A_{\lambda_1\cdots\lambda_{d-1}}^{\textrm{h}}
    \\
\mathcal{J}_{\mu}^{\textrm{m}} &=& \epsilon_{\mu\nu\lambda_1\cdots\lambda_{d-1}}^{\phantom{,}} \partial_\nu^{\phantom{,}} A_{\lambda_1\cdots\lambda_{d-1}}^{\textrm{m}}
    \nonumber \ .
\end{eqnarray}

It is interesting to note that the topological charges of any additional vector fields $\hat{\bf m}$ embedded into the spinor $\psi$ would also be automatically governed by the topological term (\ref{top-term-4}). If such a vector field spans a vector space with fewer dimensions than $d$, its topological defects would need to be enriched by Dirac attachments similar to those of the U(1) monopoles.

\subsection{Designing topological terms to meet symmetry requirements}\label{secTopTermSym}

The topological Lagrangian term (\ref{top-term-2}) is manifestly invariant under translations and rotations in space and time. Its other important symmetry properties are transformations under time reversal $x_0\to-x_0$ and mirror reflection $x_i\to-x_i$. We will analyze them in real rather than imaginary time. The fields transform as:
\begin{eqnarray}
&& \psi(x_0,{\bf x}) \xrightarrow{x_0\to-x_0} \mathcal{C}\mathcal{I}_0\psi(-x_0,{\bf x}) \\
&& \psi(x_0,{\bf x}) \xrightarrow{x_i\to-x_i} \mathcal{I}_i\psi(x_0,x_1,\dots,-x_i,\dots,x_d) \nonumber \ ,
\end{eqnarray}
where $\mathcal{C}$ performs complex conjugation of numerical factors ($\mathcal{C}w\mathcal{C}^\dagger = w^*$), and $\mathcal{I}_\mu$ carries out $\mu$-inversion on the spinor degrees of freedom. This leads to the following transformations
\begin{eqnarray}\label{transf-charge-current}
J_0 \xrightarrow{x_{0}\to-x_{0}} J_0 \quad &,& \quad J_0 \xrightarrow{x_{i}\to-x_{i}} J_0 \\
J_j \xrightarrow{x_{0}\to-x_{0}} -J_j \quad &,& \quad J_j \xrightarrow{x_{i}\to-x_{i}} (-1)^{\delta_{ij}} J_j \nonumber
\end{eqnarray}
of the charge current (\ref{currents-def}), and applies to both relativistic and non-relativistic cases. The Abelian rank 1 gauge field must transform the same way by gauge invariance. For all higher ranks, we can use relationships (\ref{singA1c}) to deduce:
\begin{eqnarray}
&& A_{\mu_{1}\cdots\mu_{n}}\xrightarrow{x_{0}\to-x_{0}}-\left\lbrack \prod_{k=1}^{n}(-1)^{\delta_{\mu_{k},0}}\right\rbrack A_{\mu_{1}\cdots\mu_{n}} \\
&& A_{\mu_{1}\cdots\mu_{n}}\xrightarrow{x_{i}\to-x_{i}}+\left\lbrack \prod_{k=1}^{n}(-1)^{\delta_{\mu_{k},i}}\right\rbrack A_{\mu_{1}\cdots\mu_{n}} \ . \nonumber 
\end{eqnarray}
Hence, the monopole current extracted from (\ref{top-current-1}) transforms as:
\begin{eqnarray}\label{transf-top-current}
\mathcal{J}_{0}\xrightarrow{x_{0}\to-x_{0}}-\mathcal{J}_{0} \quad&,&\quad\mathcal{J}_{0}\xrightarrow{x_{i}\to-x_{i}}-\mathcal{J}_{0} \\
\mathcal{J}_{j}\xrightarrow{x_{0}\to-x_{0}}+\mathcal{J}_{j} \quad&,&\quad \mathcal{J}_{j}\xrightarrow{x_{i}\to-x_{i}}-(-1)^{\delta_{ij}}\mathcal{J}_{j} \ . \nonumber
\end{eqnarray}
Combining (\ref{transf-charge-current}) and (\ref{transf-top-current}) yields non-trivial transformations of the monopole topological term (\ref{top-term-2}), which takes form $\mathcal{L}_{\textrm{t}} = -K_d J_\mu \mathcal{J}^\mu$ (without a factor of $i$) in the real time path integral:
\begin{equation}\label{transf-top-term-monopole}
\mathcal{L}_{\textrm{t}}\xrightarrow{x_{0}\to-x_{0}}-\mathcal{L}_{\textrm{t}} \quad,\quad
  \mathcal{L}_{\textrm{t}}\xrightarrow{x_{i}\to-x_{i}}-\mathcal{L}_{\textrm{t}} \ . 
\end{equation}
The topological term breaks time reversal and mirror symmetries. This is the behavior of a Chern-Simons coupling in $d=2$, and it generalizes to higher dimensions. We will show later that the coupling constant $K_d$ depends on the scalar gauge flux at rank $d-1$, which generalizes the magnetic field of $d=2$.

The topological Lagrangian $\mathcal{L}_{\textrm{t}}$ governs the dynamics of hedgehogs as well, so we should also analyze the time reversal and mirror reflections of the spin currents (\ref{spin-current}) and non-Abelian gauge fields. The latter must transform the same as generalized currents (\ref{spin-current-rank-n}), which in turn depends on how the vector field $\hat{\bf n}$ transforms. There are two characteristic transformation rules for $\hat{\bf n}$, pseudovector (P) and vector (V), which are consistent with the expected rank 2 tensor transformations of the angular momentum (\ref{ang-momentum}):
\begin{eqnarray}
\textrm{P:}\quad && \hat{n}^{a}\xrightarrow{x_{0}\to-x_{0}}-\hat{n}^{a}
  \quad,\quad\hat{n}^{a}\xrightarrow{x_{i}\to-x_{i}}-(-1)^{\delta_{ia}}\hat{n}^{a} \nonumber \\
\textrm{V:}\quad && \hat{n}^{a}\xrightarrow{x_{0}\to-x_{0}}\hat{n}^{a}
  \quad,\quad\hat{n}^{a}\xrightarrow{x_{i}\to-x_{i}}(-1)^{\delta_{ia}}\hat{n}^{a} \ . \nonumber 
\end{eqnarray}
The rank $n$ non-Abelian gauge field transforms as:
\begin{eqnarray}\label{transf-gauge-NA}
\textrm{P:}\quad && A_{\mu_{1}\cdots\mu_{n}}^{a_{n+1}\cdots a_{d-1}}\xrightarrow{x_{0}\to-x_{0}} \\
  && ~ (-1)^{n-1}\left\lbrack \prod_{k=1}^{n}(-1)^{\delta_{\mu_{k},0}}\right\rbrack A_{\mu_{1}\cdots\mu_{n}}^{a_{n+1}\cdots a_{d-1}} \nonumber \\[0.1in]
                 && A_{\mu_{1}\cdots\mu_{n}}^{a_{n+1}\cdots a_{d-1}}\xrightarrow{x_{i}\to-x_{i}} \nonumber \\
  && ~ (-1)^{n}\left\lbrack \prod_{k=n+1}^{d-1}(-1)^{\delta_{a_{k},i}}\right\rbrack \left\lbrack \prod_{k=1}^{n}(-1)^{\delta_{\mu_{k},i}}\right\rbrack
    A_{\mu_{1}\cdots\mu_{n}}^{a_{n+1}\cdots a_{d-1}} \nonumber
\end{eqnarray}
\begin{eqnarray}
\textrm{V:}\quad && A_{\mu_{1}\cdots\mu_{n}}^{a_{n+1}\cdots a_{d-1}}\xrightarrow{x_{0}\to-x_{0}} \nonumber \\
  && ~ \left\lbrack \prod_{k=1}^{n}(-1)^{\delta_{\mu_{k},0}}\right\rbrack A_{\mu_{1}\cdots\mu_{n}}^{a_{n+1}\cdots a_{d-1}} \nonumber \\[0.1in]
                 && A_{\mu_{1}\cdots\mu_{n}}^{a_{n+1}\cdots a_{d-1}}\xrightarrow{x_{i}\to-x_{i}} \nonumber \\
  && ~ -\left\lbrack \prod_{k=n+1}^{d-1}(-1)^{\delta_{a_{k},i}}\right\rbrack \left\lbrack \prod_{k=1}^{n}(-1)^{\delta_{\mu_{k},i}}\right\rbrack
    A_{\mu_{1}\cdots\mu_{n}}^{a_{n+1}\cdots a_{d-1}} \nonumber \ .
\end{eqnarray}
Then, the hedgehog current transforms as:
\begin{eqnarray}
\textrm{P:}\quad && \mathcal{J}_{0}\xrightarrow{x_{0}\to-x_{0}}(-1)^{d}\mathcal{J}_{0}
  \quad,\quad \mathcal{J}_{0}\xrightarrow{x_{i}\to-x_{i}}(-1)^{d}\mathcal{J}_{0} \nonumber \\
&& \mathcal{J}_{j}\xrightarrow{x_{0}\to-x_{0}}-(-1)^{d}\mathcal{J}_{j}
  \quad,\quad \mathcal{J}_{j}\xrightarrow{x_{i}\to-x_{i}}(-1)^{d+\delta_{ij}}\mathcal{J}_{j} \nonumber \\
\textrm{V:}\quad && \mathcal{J}_{0}\xrightarrow{x_{0}\to-x_{0}}\mathcal{J}_{0}
  \quad,\quad\mathcal{J}_{0}\xrightarrow{x_{i}\to-x_{i}}\mathcal{J}_{0} \nonumber \\
&& \mathcal{J}_{j}\xrightarrow{x_{0}\to-x_{0}}-\mathcal{J}_{j}
  \quad,\quad \mathcal{J}_{j}\xrightarrow{x_{i}\to-x_{i}}(-1)^{\delta_{ij}}\mathcal{J}_{j} \ ,
\end{eqnarray}
and the topological Lagrangian density $\mathcal{L}_{\textrm{t}} = -K_d J_\mu \mathcal{J}^\mu$ in real time behaves according to:
\begin{eqnarray}\label{transf-top-term-hedgehog}
\textrm{P:}\quad && \mathcal{L}_{\textrm{t}}\xrightarrow{x_{0}\to-x_{0}}(-1)^{d}\mathcal{L}_{\textrm{t}}
  \quad,\quad \mathcal{L}_{\textrm{t}}\xrightarrow{x_{i}\to-x_{i}}(-1)^{d}\mathcal{L}_{\textrm{t}} \nonumber \\
\textrm{V:}\quad && \mathcal{L}_{\textrm{t}}\xrightarrow{x_{0}\to-x_{0}}\mathcal{L}_{\textrm{t}}
  \quad,\quad \mathcal{L}_{\textrm{t}}\xrightarrow{x_{i}\to-x_{i}}\mathcal{L}_{\textrm{t}} \ . 
\end{eqnarray}
We immediately observe that this is consistent with the behavior of monopoles (\ref{transf-top-term-monopole}) only in odd-dimensional $d$ spaces when the spin is a pseudovector.

One might be concerned whether the symmetry properties of the topological Lagrangian expressed using spinors are the same as the properties deduced above. In the real time path integral, this topological Lagrangian transforms as:
\begin{eqnarray}
&& \mathcal{L}_{\textrm{t}} \propto \, \epsilon^{\mu\nu\lambda_1\cdots\lambda_{d-1}} \psi^\dagger \partial_\mu \partial_\nu \partial_{\lambda_1} \cdots
   \partial_{\lambda_{d-1}} \psi \\
&& ~~ \xrightarrow{x_0\to-x_0} \epsilon^{\mu\nu\lambda_1\cdots\lambda_{d-1}} (\psi^\dagger\mathcal{T}^\dagger) \partial_\mu \partial_\nu \partial_{\lambda_1}
   \cdots \partial_{\lambda_{d-1}} (\mathcal{T}\psi) \nonumber \\
 && \qquad\qquad = \epsilon^{\mu\nu\lambda_1\cdots\lambda_{d-1}} \psi^\dagger (\mathcal{T}^\dagger \partial_\mu \partial_\nu \partial_{\lambda_1}
   \cdots \partial_{\lambda_{d-1}} \mathcal{T}) \psi \nonumber
\end{eqnarray}
under time reversal $\mathcal{T}$. Here we interpret the transformation either in the Schrodinger or Heisenberg picture. In the Schrodinger picture, a simple scalar field transforms as $\mathcal{T}\psi(x_0,{\bf x}) = \psi(-x_0,{\bf x})$, so that $\mathcal{L}_{\textrm{t}} \to -\mathcal{L}_{\textrm{t}}$ as previously found for monopoles. Interpreted in the Heisenberg picture, this implies the transformation $\mathcal{T}^\dagger \tau \mathcal{T} = - \tau$ of the singular operator
\begin{equation}\label{sing-tau}
\tau = \epsilon_{\mu\nu\lambda_1\cdots\lambda_{d-1}} \partial_\mu \partial_\nu \partial_{\lambda_1} \cdots \partial_{\lambda_{d-1}} \ . 
\end{equation}
For spinors that represent a vector field $\hat{\bf n}$, we apply the time reversal $\mathcal{T}$ on (\ref{spinor}). In the Schrodinger picture,
\begin{equation}
\mathcal{T}\psi = \mathcal{T} e^{-iJ_{d-1,d}\theta_{d-1}} \mathcal{T}^{\dagger} \cdots \mathcal{T} e^{-iJ_{2,3}\theta_{2}} \mathcal{T}^{\dagger} \ .
\end{equation}
Since both the angular momentum operators $J_{ab}$ and the factors of $i$ change sign under time reversal, all rotation operators $e^{-iJ_{i,i+1}\theta_{i}}$ stay the same apart from $\theta_{i}(x_{0})\to\theta_{i}(-x_{0})$. However, the reference spinor $\psi_0 = \psi(0,\theta_2,\theta_3,\dots,\theta_{d-1})$ transforms as $\mathcal{T}\psi(0,\cdots)=\psi(\pi,\cdots)$ in the case of pseudovectors (P), and $\mathcal{T}\psi_0 = \psi_0$ in the case of vectors (V). Pseudovector transformations require $\mathcal{T}\gamma^{a}\mathcal{T}^{\dagger}=-\gamma^{a}$, while vector transformations are $\mathcal{T}\gamma^{a}\mathcal{T}^{\dagger}=+\gamma^{a}$; neither one of them affects the rank 2 tensor transformations of the angular momentum (\ref{ang-momentum}). After all manipulations, one finds for pseudovectors:
\begin{eqnarray}
\textrm{P:}\quad && \mathcal{T}\psi(\theta_{1}\cdots,\theta_{d-2},\theta_{d-1}) \\
&& \quad = \psi(\pi-\theta_{1}\cdots,\pi-\theta_{d-2},\pi+\theta_{d-1}) \nonumber \ .
\end{eqnarray}
Applying the derivatives from $\mathcal{L}_{\textrm{t}}$ on $\mathcal{T}\psi$ has the same effect under time reversal as the transformations we deduced for the hedgehog $\mathcal{L}_{t}$. So, generally, the transformations of $\mathcal{L}_{t}$ expressed in terms of currents and spinors are always the same. However, this is unusual because the Heisenberg picture now implies that there is no unique symmetry transformation for the singular operator (\ref{sing-tau}). $\mathcal{T} \tau \mathcal{T}^{\dagger}$ depends on the dimensionality $d$ and the type of spinor singularities that this operator is applied to. We conclude that the singular operator $\tau$ does not necessarily have a definite parity under time reversal and mirror reflections, i.e. it can have different parities in distinct Fock subspaces.

The intrinsic dynamics of the system need not feature the same reduced or broken symmetries as the above topological term. Then, the topological order, if stable, must be described by a different topological term $\mathcal{L}'_{\textrm{t}}$ with compatible symmetries. We will consider one example of an alternative topological term $\mathcal{L}'_{\textrm{t}}$ that can be constructed from the spinor fields. The degrees of freedom and their topological defects are the same as before, so we may only couple different currents to the topological defect current $\mathcal{J}_\mu^{\phantom{,}}$ in $\mathcal{L}'_{\textrm{t}}$. Consider spin currents (\ref{spin-current}) and a pure spin-related topological Lagrangian:
\begin{equation}\label{top-term-alt}
\mathcal{L}'_{\textrm{t}} = -K_d^{a_2\cdots a_{d-1}} j_\mu^{a_2\cdots a_{d-1}} \mathcal{J}^\mu
\end{equation}
in real time, whose coupling $K_d^{a_2\cdots a_{d-1}}$ necessarily breaks spin rotation symmetry. Obviously, this can be useful only if the spin dynamics actually has reduced symmetry, with two unbiased spin directions equivalent to U(1). The time reversal and mirror reflections of $\mathcal{L}'_{\textrm{t}}$ for pseudovectors (P) and vectors (V) are found to be:
\begin{eqnarray}
\textrm{P:}\quad && \mathcal{L}'_{t}\xrightarrow{x_{0}\to-x_{0}}-(-1)^{d}\mathcal{L}'_{t} \nonumber \\
  && \mathcal{L}'_{t}\xrightarrow{x_{i}\to-x_{i}}-(-1)^{d}\left\lbrack \prod_{k=2}^{d-1}(-1)^{\delta_{a_{k},i}}\right\rbrack \mathcal{L}'_{t} \nonumber \\
\textrm{V:}\quad && \mathcal{L}'_{t}\xrightarrow{x_{0}\to-x_{0}}-\mathcal{L}'_{t} \nonumber \\
  && \mathcal{L}'_{t}\xrightarrow{x_{i}\to-x_{i}}-\left\lbrack \prod_{k=2}^{d-1}(-1)^{\delta_{a_{k},i}}\right\rbrack \mathcal{L}'_{t} \nonumber \ .
\end{eqnarray}
We see, for example, that $\mathcal{L}'_{\textrm{t}}$ could be appropriate for topological orders in $d=3$ with time-reversal symmetry, which is absent in the original construction. A similar idea was used in a two-dimensional setting \cite{Nikolic2012} to describe spin-orbit-coupled fractional topological insulators. In terms of the spinor fields, we would write in real time:
\begin{eqnarray}
\mathcal{L}'_{\textrm{t}} &\propto& -K_d^{a_2\cdots a_{d-1}} \epsilon_{a_0 a_1 a_2 \cdots a_{d-1}} \epsilon^{\mu\nu\lambda_1\cdots\lambda_{d-1}} \\
  && \times \psi^\dagger \gamma^{a_0} \gamma^{a_1} \partial_\mu \partial_\nu \partial_{\lambda_1} \cdots \partial_{\lambda_{d-1}} \psi \ . \nonumber
\end{eqnarray}

Obviously, the symmetries of the possible topological terms have certain restrictions determined by the nature of fields and their topological defects. These are reflected on the possible symmetries of topologically ordered ground states.

\section{Topological order}\label{secTopOrder}

The following sections explore the physical properties of incompressible quantum liquids in which $\pi_{d-1}(S^{d-1})$ topological defects are abundant and mobile. Section \ref{secFract} introduces fractionalization of the intrinsic particles' quantum numbers. We will show that the topological Lagrangian term tends to bind a rationally quantized amount of elementary charge or spin to a mobile topological defect, and analyze how this fractionalization holds up to perturbations that spoil the conservation laws. We will find that stable topological orders can be characterized by ``filling factors'' associated with monopoles and hedgehogs, in analogy to fractional quantum Hall states. Section \ref{secTopDeg} presents a calculation of the topological ground state degeneracy on non simply connected manifolds. Section \ref{secBraid} contains a basic discussion of braiding operations, and Section \ref{secMicroPart} considers restrictions imposed on topological orders by microscopic properties of electrons. Section \ref{secEqMot} discusses soft boundary modes, and Section \ref{secME} analyzes response to certain external perturbations.

\subsection{Fractional quasiparticles}\label{secFract}

Here we consider fractionalization in an incompressible quantum liquid whose effective imaginary time Lagrangian contains the topological term (\ref{top-term-2}). Fractionalization is revealed by the kinematic relationship between the currents of charge and topological defects. We will find this relationship by converting the effective theory to real time and deriving the stationary action condition from the variations of the Abelian gauge field $A_\mu$. Apart from the topological term, the relevant parts of the real time Lagrangian density that contain $A_\mu$ are collected from (\ref{CM1}) and (\ref{L3b}), with a substitution $\kappa_1 = |\psi|^2$ to fix the units:
\begin{eqnarray}\label{Lrt}
\mathcal{L} &=& \frac{|\psi|^2}{2}(j_\mu+A_\mu)(j^\mu+A^\mu) - \frac{1}{4e_1^2} F_{\mu\nu}F^{\mu\nu} \nonumber \\
&& -\frac{\kappa'_2}{2} \left(\frac{F_{\mu\nu}}{2}-A_{\mu\nu}\right)\left(\frac{F^{\mu\nu}}{2}-A^{\mu\nu}\right) \nonumber \\
&& -K_d|\psi|^2 (j_\mu+A_\mu)\mathcal{J}^\mu + \cdots \ .
\end{eqnarray}
$J_\mu = |\psi|^2(j_\mu + A_\mu)$ is the gauge invariant charge current density, $F_{\mu\nu} = \partial_\mu A_\nu - \partial_\nu A_\mu$ is the electromagnetic field tensor, and $\mathcal{J}_\mu$ is the current density of topological defects. This is similar to the recently proposed BF theory \cite{Cho2010, Vishwanath2013, Chan2015, Fradkin2017} in $d=3$, without the linking term. Stationary variations $\delta\mathcal{L}=0$ with respect to $A_\mu$ produce the following field equation
\begin{equation}\label{eq-mot-1}
J^{\mu} = \frac{1}{e_{1}^{2}}\partial_{\nu}F^{\mu\nu}
  + \kappa'_2\partial_{\nu}\!\left(\frac{F^{\mu\nu}}{2}-A^{\mu\nu}\right)
  + K_{d}|\psi|^2\mathcal{J}^\mu
\end{equation}
if $\mathcal{J}_\mu$ is just the current of hedgehogs and hence independent of $A_\mu$. The case of monopoles is more complicated and we will revisit it later, even though much of this discussion applies to monopoles as well. Let us focus on the purely kinematic effect:
\begin{equation}\label{pf-attachment}
J^{\mu} \to K_{d}|\psi|^2\mathcal{J}^\mu \ .
\end{equation}
This describes the binding between particle charge and topological charge. Excitations must include the composites of particles and defects. Particle charge is microscopically quantized as an integer in the present formalism, and topological charge is quantized in the units of $q$ by the integral (\ref{top-charge-1}). In order for both particles and defects to be mobile in a uniform incompressible quantum liquid, they cannot propagate independently of each other -- the Heisenberg uncertainty principle necessarily localizes one or the other when they move relative to each other in the same space. These facts imply that a composite quasiparticle must be a bundle of an integer number $n$ of particles and an integer number $m$ of topological defects. Also, the composite quasiparticles must have hard-core repulsive interactions. Let us define a ``filling factor''
\begin{equation}
\nu = \frac{n}{m} \quad,\quad n,m\in\mathbb{Z}
\end{equation}
whose rational quantization is a \emph{necessary} condition for the stability of an incompressible quantum liquid. If we define a scalar ``magnetic field''
\begin{equation}
B = \mathcal{J}^{0} = \epsilon^{0\nu\lambda_{1}\cdots\lambda_{d-1}}\partial_{\nu}A_{\lambda_{1}\cdots\lambda_{d-1}} \ ,
\end{equation}
then we can express the number density of topological defects as $B/q$ according to (\ref{top-charge-2}). Since $|\psi|^2$ is the number density of particles, we can alternatively write the filling factor as
\begin{equation}
\nu = \frac{|\psi|^2}{B/q} = \frac{J^0}{\mathcal{J}^0/q} \ .
\end{equation}
Note that we use the non-relativistic charge current because the particle-hole symmetry is broken. Comparing with (\ref{pf-attachment}), we find $K_d = B^{-1}$, and hence:
\begin{equation}\label{pf-attachment-2}
J^{\mu} \to \frac{\nu}{q}\mathcal{J}^\mu \ .
\end{equation}
The topological Lagrangian for hedgehogs ($q=S_{d-1}$) can now be rewritten in real time as:
\begin{equation}\label{top-term-6}
\mathcal{L}_{\textrm{t}} = -(j_\mu+A_\mu) \times \frac{\nu}{q}\mathcal{J}^{\mu} \ .
\end{equation}

A composite bundle of $n$ particles and $m$ topological defects is not an elementary excitation of the incompressible quantum liquid. Since the topological defect number is conserved and quantized, one can apply an external field to trap a single topological defect in a small volume of the system. This perturbation does not by itself localize the particle charge. However, the charge fluid will dynamically redistribute to supply the amount $\nu$ of charge to the region where the defect is localized. This is described by the above field equation. The resulting charge-defect composite object is an elementary excitation, which can be also set to free motion. Charge fractionalization occurs at least when $\nu<1$. Similarly, one can localize a quantized conserved particle charge using an external field. It is evident even without an explicit derivation that the dynamics of particle and topological charges is self-dual: both charges are point-like and governed by the same symmetries and the duality-invariant topological term. In the dual description, a localized particle charge is quantized topologically. It attracts to itself a fractional amount $\nu^{-1}$ of fluid topological charge, forming a fractional quasiparticle. We would consider it an elementary excitation if $\nu>1$.

When particles and defects carry additional internal degrees of freedom (e.g. spin), these become fractionalized too. However, it is up to symmetries to conserve or not conserve these degrees of freedom. The precise conservation laws may look different in the two dual descriptions. For example, the topological charge conservation is guaranteed by topological protection in any local theory, while the particles' spin conservation is a matter of symmetry. Perturbations that break the relevant \emph{gauge} symmetry of the theory can modify the field equation (\ref{eq-mot-1}) and ruin the observable fractionalization condition (\ref{pf-attachment-2}) even when the gap of the topologically ordered state remains open. If such a perturbation is random (e.g. disorder), then the quantized fractionalization may still be asymptotically recovered in the long-wavelength limit -- for the fractional quasiparticles that experience the perturbation only on average.

Let us now analyze the relationship between charge and monopole currents, made complicated by an implicit dependence of the monopole current $\mathcal{J}_\mu$ on the gauge field $A_\mu$. To reveal this dependence, we must integrate out all gauge fields at ranks $n>1$. Substituting
\begin{equation}
A_{\lambda_1\cdots\lambda_n} = \frac{1}{n} \sum_{i=1}^{n}(-1)^{i-1}\partial_{\lambda_i}
    A_{\lambda_1\cdots\lambda_{i-1}\lambda_{i+1}\cdots\lambda_n} + \delta A_{\lambda_1\cdots\lambda_n}
\end{equation}
for all $n>1$ in (\ref{L3b}), (\ref{top-term-2}) and integrating out the gapped fluctuations of $\delta A_{\lambda_1\cdots\lambda_n}$ leads to a renormalization of the rank 1 Maxwell term and an effective substitution of $A_\mu$ in
\begin{equation}\label{Jmon}
\mathcal{J}^\mu = \epsilon^{\mu\nu\lambda_1\cdots\lambda_{d-1}} \partial_{\nu}\partial_{\lambda_{1}}\cdots\partial_{\lambda_{d-2}}A_{\lambda_{d-1}}
\end{equation}
consistent with the recursive inter-rank linking (\ref{singA1c}). As discussed before, the antisymmetrized derivatives have an effect only on singular functions. The full monopole topological part of the stationary action condition for (\ref{Lrt}) becomes:
\begin{eqnarray}
&& \frac{\partial\mathcal{L}_{t}}{\partial A_{\mu}} \propto -\mathcal{J}^{\mu} -A_\alpha \frac{\partial\mathcal{J}^\alpha}{\partial A_{\mu}} \\
&& ~~\to -\mathcal{J}^{\mu} -(-1)^{d-1}\epsilon^{\alpha\nu\lambda_{1}\cdots\lambda_{d-2}\mu}\partial_{\lambda_{d-2}}\cdots\partial_{\lambda_{1}}\partial_{\nu}A_{\alpha}
    \nonumber \\
&& ~~= -\mathcal{J}^\mu +(-1)^{d-1}\epsilon^{\mu\nu\lambda_{1}\cdots\lambda_{d-2}\lambda_{d-1}}\partial_{\lambda_{d-2}}\cdots\partial_{\lambda_{1}}\partial_{\nu}
    A_{\lambda_{d-1}} \nonumber \\
&& ~~= -\Bigl(1+(-1)^{\sigma+d}\Bigr)\mathcal{J}^{\mu} = -2 f_d \mathcal{J}^{\mu} \ . \nonumber
\end{eqnarray}
We substituted (\ref{Jmon}) and then carried out integration by parts (indicated by the arrow) to transfer all space-time derivatives onto $A_\alpha$. The index $\alpha$ was subsequently relabeled into $\lambda_{d-1}$. At the end, we reordered the indices $\lambda_{d-1},\dots,\lambda_1,\nu$ to reconstruct $\mathcal{J}^\mu$ given by (\ref{Jmon}), and this produced the sign $(-1)^\sigma$:
\begin{equation}
\sigma = (d-2)+(d-3)+\cdots+2+1=\frac{(d-2)(d-1)}{2} \ .
\end{equation}
The ensuing kinematic field equation that relates charge and monopole currents
\begin{equation}
J^{\mu} \to K_{d}|\psi|^2 2f_d \mathcal{J}^\mu
\end{equation}
is modified by the constant
\begin{equation}
f_d = \frac{1}{2} \Bigl(1+(-1)^{\sigma+d}\Bigr) = \begin{cases}
      1 & ,~\left\lbrack \frac{d}{2}\right\rbrack \;\textrm{is odd}\\
      0 & ,~\left\lbrack \frac{d}{2}\right\rbrack \;\textrm{is even}
    \end{cases} \ ,
\end{equation}
where $\lbrack x \rbrack$ is the integer part of $x$. This is merely a renormalization of the coupling $K_d$ in spatial dimensions $d\in \lbrace 2,3,6,7,10,11,\dots \rbrace$, so we can repeat the previous analysis to write
\begin{equation}\label{top-term-5}
J^{\mu} \to \frac{\nu}{q}\mathcal{J}^\mu \quad,\quad \mathcal{L}_{\textrm{t}} = -(j_\mu+A_\mu) \times \frac{\nu}{2q}\mathcal{J}^{\mu}
\end{equation}
with $q=2\pi$ for monopoles. Charge-monopole attachment in $d=3$ is often called Witten effect \cite{Witten1979}. However, if $d\in\lbrace 4,5, \dots \rbrace$ then $J^\mu \to 0$ does not mirror monopole currents and there is no binding of fractional charge to monopoles even though the ground state can be topologically ordered. This unusual pattern of dimensionality affecting response functions has been also found in \cite{Qi2008b}.

The full topological Lagrangian density of a generic system can govern both monopoles and hedgehogs with topological current densities $\mathcal{J}_\mu^{\textrm{m}}$ and $\mathcal{J}_\mu^{\textrm{h}}$ respectively. An incompressible quantum liquid is characterized by two independent ``filling factors'', $\nu^{\textrm{m}}$ for monopoles and $\nu^{\textrm{h}}$ for hedgehogs. Consider the topological Lagrangian density (\ref{top-term-4}) derived from a single spinor field $\psi$. In dimensions $d\in \lbrace 2,3,6,7,10,11,\dots \rbrace$, the coupling constant $K_d$ is related to the filling factors by:
\begin{eqnarray}
J^{0} &=& K_{d}|\psi|^2 \left(\mathcal{J}^{\textrm{h}0}-\frac{S_{d-1}}{2^{d-2}\pi S}\mathcal{J}^{\textrm{m}0}\right) \\
  &=& K_{d}|\psi|^2 \left(\frac{S_{d-1}}{\nu^{\textrm{h}}}-\frac{S_{d-1}}{2^{d-2}\pi S}\frac{2\pi}{\nu^{\textrm{m}}}\right) J^{0} \ , \nonumber
\end{eqnarray}
from which we conclude:
\begin{equation}
K_{d}|\psi|^2 = \left\lbrack S_{d-1}\left(\frac{1}{\nu^{\textrm{h}}}-\frac{1}{2^{d-3}S}\frac{1}{\nu^{\textrm{m}}}\right)\right\rbrack^{-1} \ .
\end{equation}
Therefore, the real time topological Lagrangian is
\begin{equation}
\mathcal{L}_{\textrm{t}} = -\frac{j_\mu+A_\mu}{S_{d-1}\left(\frac{1}{\nu^{\textrm{h}}}-\frac{1}{2^{d-3}S}\frac{1}{\nu^{\textrm{m}}}\right)}
  \left(\mathcal{J}^{\textrm{h}\mu}-\frac{S_{d-1}}{2^{d-1}\pi S}\mathcal{J}^{\textrm{m}\mu}\right) \ ,
\end{equation}
and its complete kinematic field equation is:
\begin{equation}
J^{\mu} = \frac{1}{S_{d-1}\left(\frac{1}{\nu^{\textrm{h}}}-\frac{1}{2^{d-3}S}\frac{1}{\nu^{\textrm{m}}}\right)}
  \left(\mathcal{J}^{\textrm{h}\mu}-\frac{S_{d-1}}{2^{d-2}\pi S}\mathcal{J}^{\textrm{m}\mu}\right) \ .
\end{equation}
Note that $\nu^{\textrm{m}}\to\infty$ or $\nu^{\textrm{h}}\to\infty$ correspond to confined monopoles or hedgehogs respectively. Additional degrees of freedom that support topologically protected defects could give rise to more filling factors.

The independence of monopole and hedgehog filling factors can be reduced by interactions that tend to bind monopoles to hedgehogs:
\begin{equation}
\mathcal{L}_{\textrm{Z}} \propto \Bigl( \mathcal{J}_\mu^{\textrm{h}}\mathcal{J}_\mu^{\textrm{h}}
  -\alpha \mathcal{J}_\mu^{\textrm{m}}\mathcal{J}_\mu^{\textrm{m}} \Bigr)^{2} \ .
\end{equation}
The physical origin of such interactions is the Zeeman coupling of magnetic moments to magnetic field. The number of monopoles bound to hedgehogs, and hence the ratio $\nu^{\textrm{m}}/\nu^{\textrm{h}}$, is determined dynamically through the strength $\alpha$ of the spin-orbit coupling. A linear coupling between the monopole and hedgehog currents is harder to justify -- it would enable a direct conversion of hedgehogs to monopoles, which is forbidden at least in $d=3$ by angular momentum conservation.

The coupling of spin currents to topological defects (\ref{top-term-alt}) is another route to incompressible quantum liquids, when symmetries allow. The ensuing field equation
\begin{equation}
J^{\mu,a_2\cdots a_{d-1}} = \frac{\nu^{a_2\cdots a_{d-1}}}{q}\mathcal{J}^{\mu}
\end{equation}
describes fractionalization of spin degrees of freedom. The value of the spin-related filling factor can be finite only if the Spin($d$) symmetry is broken and reduced to U(1). The residual symmetry still conserves one spin degree of freedom and enables its fractional quantization (assuming that no other perturbation spoils the spin conservation law). However, this symmetry reduction suppresses ordinary hedgehogs. Spin currents can still couple either to charge monopoles, or to the monopoles of the surviving spin U(1) degree of freedom in purely magnetic systems ($q=2\pi$). The latter gives rise to spin liquids.

\subsection{Topological ground state degeneracy on non-simply connected manifolds}\label{secTopDeg}

Topological order in $d\ge 2$ can be identified by the ground state degeneracy on a non-simply connected manifold $\mathcal{M}=S^{d-1}\times S^{1}$ consisting of a $d-1$ sphere swept along an orthogonal loop direction. For simplicity, the sphere and the loop have the same large radius $L$. Consider a vector field configuration $\hat{{\bf n}}(x_{1},\dots x_{d})$ where the coordinates $(x_{1},\cdots,x_{d-1})\in S^{d-1}$ live on the sphere and $x_{d}\in S^{1}$ lives on the loop. Since the possible local states of $\hat{{\bf n}}\in S^{d-1}$ span a sphere themselves, we can represent the field $\hat{{\bf n}}(x_{1},\dots,x_{d})$ by the spherical angles $(\theta_{1},\dots,\theta_{d-1})$ that depend on $(x_{1},\dots,x_{d})$. The topological invariant $N$ of the vector field is:
\begin{equation}
N=\frac{1}{S_{d-1}}\oint\limits_{S^{d-1}}\prod_{i=1}^{d-1}dx_{i}\,\epsilon_{j_{1}\cdots j_{d-1}}A_{j_{1}\cdots j_{d-1}} \ ,
\end{equation}
where
\begin{equation}
A_{j_1\cdots j_{d-1}} = \frac{\epsilon_{k_{1}\cdots k_{d-1}}}{(d-1)!} \prod_{i=1}^{d-1} (\sin\theta_{i})^{d-1-i}\, \partial_{j_{i}} \theta_{k_{i}} \ .
\end{equation}
is given by (\ref{singA2b}). For any finite $N$, the angles $\theta_{i}$ vary at most by $\Delta\theta_{i}\sim2\pi N$ on any closed loop of perimeter $\Delta x\sim L$ around the sphere, so $A_{\mu_{1}\cdots\mu_{d-1}}\propto L^{d-1}$. 
 
The procedure for finding the ground state degeneracy on $\mathcal{M}$ starts with identifying ``fundamental'' field configurations characterized by topological invariants. These are classical configurations that all cost vanishing energy in the thermodynamic limit. If quantum processes cause tunneling between these configurations, their classical infinite degeneracy is lifted down to a finite quantum degeneracy. Smooth local deformations of a fundamental configuration are gapped in an incompressible topological quantum liquid, and readily integrated out only to renormalize the Maxwell coupling constant. In that sense, we just need to study the spectrum that arises from the quantum dynamics of the lowest energy fundamental configurations. We can even introduce weak perturbations that adjust the form of the fundamental configurations to our liking, as long as the topological gap is not closed.

So, let the fundamental configurations $\hat{{\bf n}}(x_{1},\dots,x_{d})$ be undistorted hedgehogs given by (\ref{n-vs-theta}), with $(\theta_{1}^{\phantom{,}}, \dots, \theta_{d-2}^{\phantom{,}}, \theta_{d-1}^{\phantom{,}}) \!\!=\!\! (\theta'_{1},\dots,\theta'_{d-2}, N\theta'_{d-1})$ determined only by the spherical angles $\theta'_i$ of the points on $S^{d-1}$. The resulting gauge field is constant:
\begin{equation}
A_{j_{1}\cdots j_{d-1}}(x_{1},\dots,x_{d}) = \frac{N}{(d-1)!\,L^{d-1}}\epsilon_{j_{1}\cdots j_{d-1}} \ .
\end{equation}
We now extend the definition of the gauge field to allow its indices $\mu_{i}$ to represent all space-time directions:
\begin{equation}
A_{\mu_{1}\cdots\mu_{d-1}}(x_{0},x_{1},\dots,x_{d})\equiv\frac{N}{(d-1)!\,L^{d-1}}\epsilon_{0\mu_{1}\cdots\mu_{d-1}d} \ .
\end{equation}
This is the correct extension because it implements flux conservation by satisfying a Faraday law:
\begin{equation}
\partial_{\mu}\mathcal{J}_{\mu}=0 \ ,
\end{equation}
where $\mathcal{J}_{\mu} = \epsilon_{\mu\nu\lambda_{1}\cdots\lambda_{d-1}}\partial_{\nu}A_{\lambda_{1}\cdots\lambda_{d-1}}$ is the topological defect current density (i.e. flux). Local quantum processes on $\mathcal{M}$ are not able to change the protected topological invariant $N$, but global tunneling (instantons) will introduce some quantum dynamics for $N$.

\begin{figure*}[!]
\raisebox{0.65in}{(a)}
  \includegraphics[height=1.4in]{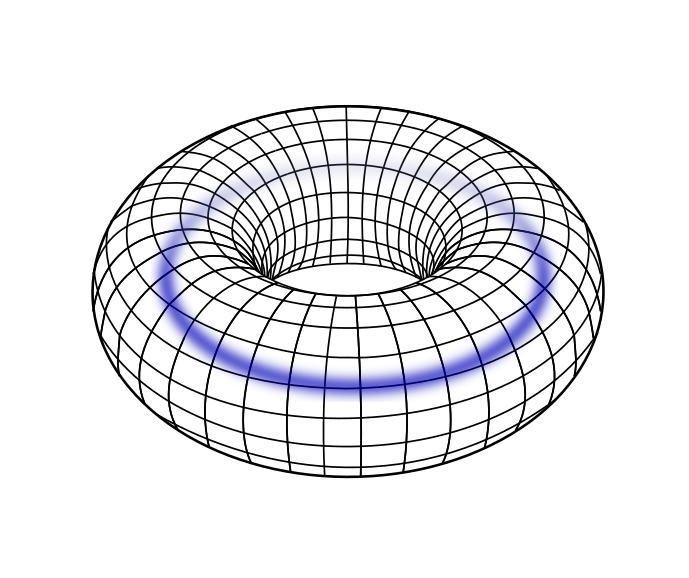}
  \includegraphics[height=1.4in]{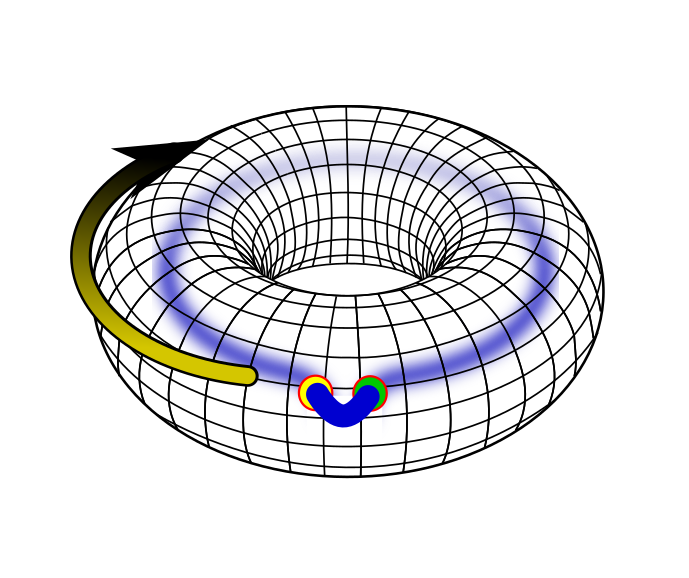}
  \includegraphics[height=1.4in]{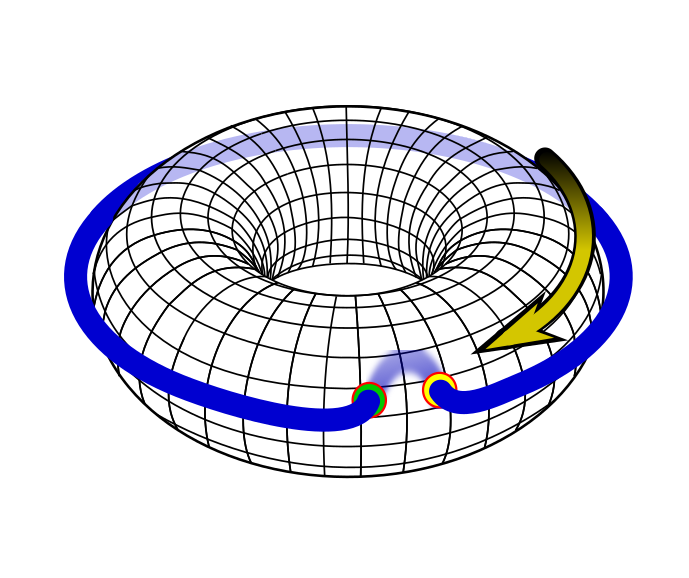}
  \includegraphics[height=1.4in]{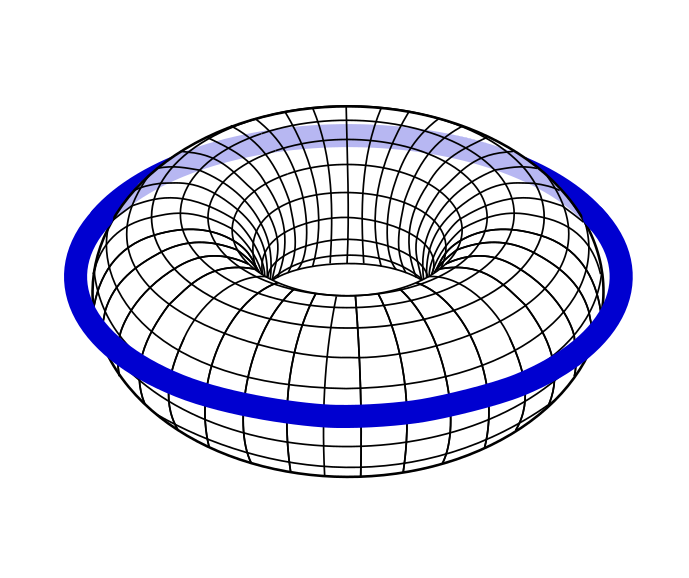}
\vskip -0.2in
\raisebox{0.65in}{(b)}
  \includegraphics[height=1.4in]{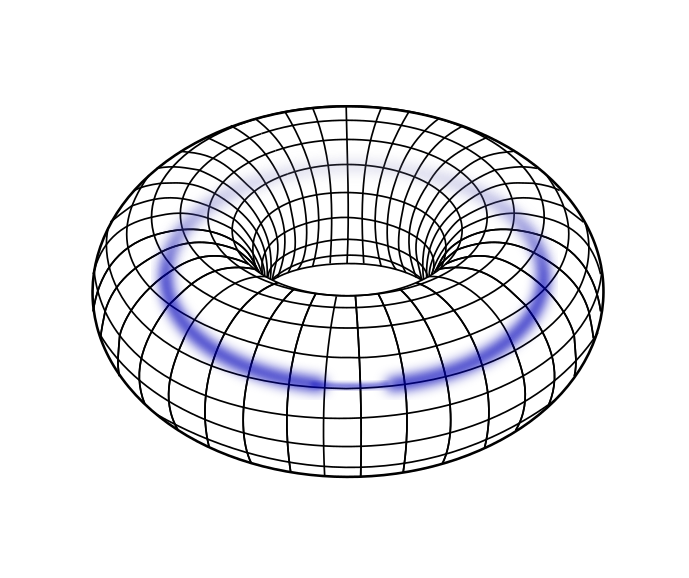}
  \includegraphics[height=1.4in]{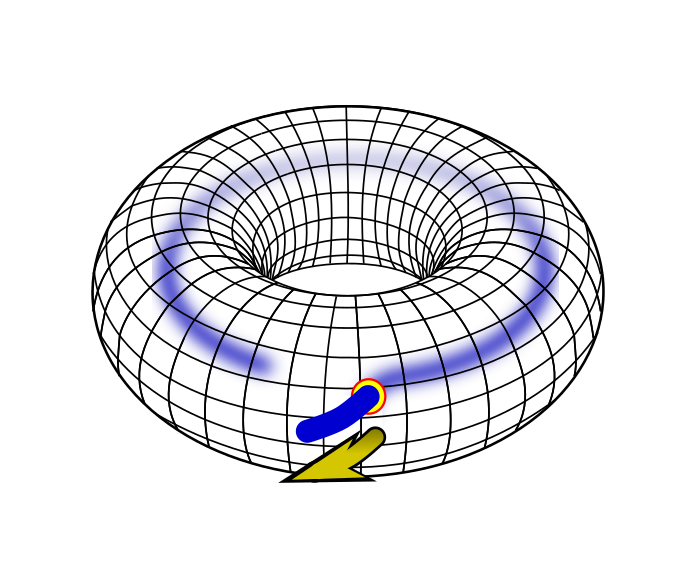}
  \includegraphics[height=1.4in]{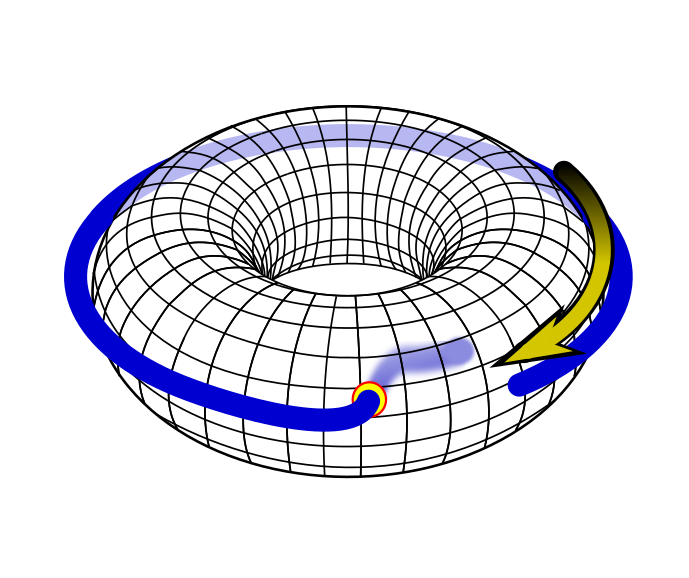}
  \includegraphics[height=1.4in]{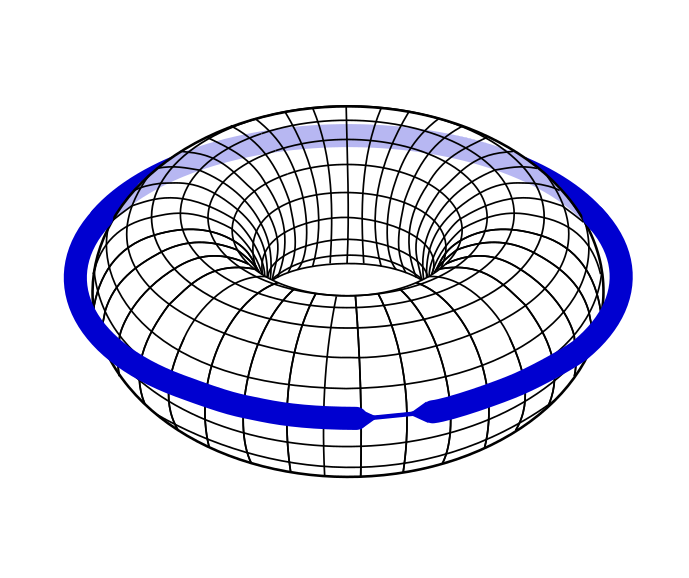}
\caption{\label{local-flux-change}Processes that change the Abelian rank 1 flux $2\pi N'$ on a 2-torus representation of space. (a) Conserved flux: instantaneously create a vortex-antivortex pair, drag the vortex around the torus (non-local), then annihilate it with the antivortex. (b) Non-conserved flux -- a flux line can tear and terminate with a monopole: create a vortex, \emph{afterwards} create an antivortex, then annihilate the pair (all local). The transformations of the shown fictitious flux line (thick blue) that do not take place on the torus manifold are not physical processes that need to involve some energy or time.}
\end{figure*}

Next, let us similarly set up the U(1) gauge field of charge currents to:
\begin{equation}
A_{\mu}(x_{0},x_{1},\dots,x_{d})\equiv\frac{N'}{L}\delta_{\mu,d} \ .
\end{equation}
This threads a U(1) flux $2\pi N'$ through the opening of the $S^{1}$ loop. $N'$ is quantized as an integer because $A_{\mu}=\partial_{\mu}\phi$ was obtained in a singular gauge transformation from the single-valued U(1) phase $\phi$. No local processes can change $N'$ if the flux of $A_\mu$ is conserved. However, if the Abelian rank 1 flux is not conserved, then there are local processes shown in Fig.\ref{local-flux-change} that can change $N'$. This possibility in incompressible $d>2$ quantum liquids, e.g. with deconfined monopoles, was discussed in Section \ref{secDynPhaseDiag}. We will first construct a Hamiltonian that describes only global flux tunneling, and eventually patch it to take into account any locally-caused $N'$ fluctuations.

In $d>2$, the topological Lagrangian term (\ref{top-term-2}) takes the form:
\begin{eqnarray}
\mathcal{L}_{t} &=& i\,K_{d}|\psi|^{2}(\partial_{\mu}\theta+A_{\mu})\epsilon_{\mu\nu\lambda_{1}\cdots\lambda_{d-1}}\partial_{\nu}A_{\lambda_{1}\cdots\lambda_{d-1}}
        \nonumber \\
&\to& i\,K_{d}|\psi|^{2}\epsilon_{\mu\nu\lambda_{1}\cdots\lambda_{d-1}}A_{\mu}\partial_{\nu}A_{\lambda_{1}\cdots\lambda_{d-1}} \nonumber \\
&=&   \frac{i\,K_{d}|\psi|^{2}}{L^{d}(d-1)!}(-1)^{d}\epsilon_{\nu\lambda_{1}\cdots\lambda_{d-1}d}\epsilon_{0\lambda_{1}\cdots\lambda_{d-1}d} N'\partial_{\nu}N  
        \nonumber \\
&=&\frac{i\,K_{d}|\psi|^{2}}{L^{d}}(-1)^{d}N'\partial_{0}N
\end{eqnarray}
after integrating out the residual phase $\theta$. At this point, taking the time derivative of an integer $N$ is sensible because the actual quantization $N/L^d$ vanishes in the thermodynamic limit $L\to\infty$. If we substitute the gauge fields in the rest of the Lagrangian density
\begin{eqnarray}
\mathcal{L} &=&\frac{1}{(d-1)!}\frac{1}{2e_{1}^{2}}\Bigl(\epsilon_{\mu_{1}\cdots\mu_{d-1}\nu\lambda}\partial_{\nu}A_{\lambda}\Bigr)^{2} \\
&& +\frac{1}{2e_{d-1}^{2}}\Bigl\lbrack\epsilon_{\mu\nu\lambda_{1}\cdots\lambda_{d-1}}\partial_{\nu}A_{\lambda_{1}\cdots\lambda_{d-1}}\Bigr\rbrack^{2}+\mathcal{L}_{t}
  \nonumber \ ,
\end{eqnarray}
and integrate out the spatial coordinates on $\mathcal{M}$, we obtain the Lagrangian in which $N$ and $N'$ are canonical coordinates:
\begin{eqnarray}
L &=& \int d^{d}x\,\mathcal{L} = 2\pi S_{d-1}\biggl\lbrack L^{d-2}\frac{1}{2e_{1}^{2}}(\partial_{0}N')^{2} \\
&& +\frac{1}{L^{d-2}}\frac{1}{2e_{d-1}^{2}}(\partial_{0}N)^{2} + i(-1)^{d}K_{d}|\psi|^{2}N'\partial_{0}N\biggr\rbrack \nonumber \ .
\end{eqnarray}
Here, the quantization of $N,N'$ becomes a significant feature. This theory can be understood only as the ``continuum limit'' of a more accurate compact theory that we will construct in the end. From the corresponding real time Lagrangian:
\begin{eqnarray}
L &=& 2\pi S_{d-1}\biggl\lbrack L^{d-2}\frac{1}{2e_{1}^{2}}(\partial_{0}N')^{2}+\frac{1}{L^{d-2}}\frac{1}{2e_{d-1}^{2}}(\partial_{0}N)^{2} \nonumber \\
&& -(-1)^{d}K_{d}|\psi|^{2}N'\partial_{0}N\biggr\rbrack \nonumber 
\end{eqnarray}
we can obtain the canonical momenta:
\begin{eqnarray}
P &=& \frac{\delta L}{\delta\partial_{0}N}=2\pi S_{d-1}\left\lbrack \frac{\partial_{0}N}{L^{d-2}e_{d-1}^{2}}-(-1)^{d}K_{d}|\psi|^{2}N'\right\rbrack \nonumber \\
P' &=& \frac{\delta L}{\delta\partial_{0}N'}=2\pi S_{d-1}\frac{L^{d-2}}{e_{1}^{2}}(\partial_{0}N') \ ,
\end{eqnarray}
and the Hamiltonian
\begin{equation}
H = P(\partial_{0}N)+P'(\partial_{0}N')-L = \frac{(P-\alpha N')^{2}}{2M}+\frac{P'^{2}}{2M'} \ , \nonumber
\end{equation}
where the ``masses'' are:
\begin{eqnarray}
M &=& \frac{2\pi S_{d-1}}{e_{d-1}^{2}}\frac{1}{L^{d-2}}\xrightarrow[d>2]{L\to\infty}0 \\
M' &=& \frac{2\pi S_{d-1}}{e_{1}^{2}}L^{d-2}\xrightarrow[d>2]{L\to\infty}\infty \nonumber \ .
\end{eqnarray}
The canonical momentum shift coefficient
\begin{equation}
\alpha = -(-1)^{d}2\pi S_{d-1}K_{d}|\psi|^{2}
\end{equation}
is determined by the hedgehog filling factor:
\begin{equation}
\frac{\alpha}{2\pi} = -(-1)^{d}S_{d-1}\frac{\nu}{q} = -(-1)^{d}\nu
\end{equation}
based on the discussion in Section \ref{secFract}. We interpret the Hamiltonian $H$ as a quantum-mechanical operator whose spectrum determines the topological ground state degeneracy of incompressible quantum liquids on $\mathcal{M}$. The canonical coordinate operators have eigenvalues $N,N'\in\mathbb{Z}$, so their canonical conjugate operators have eigenvalues $P,P'\in(-\pi,\pi)$. This means that the Hamiltonian must be properly regularized at high energies into a compact form, to treat $P$ and $P+2\pi$ as the same state:
\begin{equation}
H=t+t'-t\cos(P-\alpha N')-t'\cos(P') \ .
\end{equation}
A vortex ``line'' $N'\neq0$ threaded through the $S^{1}$ opening of $\mathcal{M}=S^{d-1}\times S^{1}$ can occasionally drift through $S^{1}$. Such instanton events connect different classical ground states and affect the kinetic energy $E\sim P'^{2}$ in the Hamiltonian $H$. Since the Maxwell coupling constants $e_{n}$ are not allowed to depend on the system size, the ``masses'' for $d>2$ have extreme behaviors $t\sim M^{-1} \to \infty$ and $t'\sim M'^{-1} \to 0$ in the thermodynamic limit $L\to\infty$. However, this picture includes only the global instanton processes. If the Abelian rank 1 flux is not conserved in the incompressible quantum liquid, then local tunneling processes introduce $N'$ fluctuations. It is clear from the symmetries that such processes merely renormalize $t'$ to a finite value. Still, $t\gg t'$ allows us to diagonalize $H$ perturbatively.

First consider the unperturbed problem $t'=0$, which is also physically relevant when the Abelian rank 1 flux is conserved. The unperturbed ground state energy does not depend on the eigenvalue of $P'$. Any superposition of $P'$ eigenstates is a Hamiltonian eigenstate, including the eigenstates of $N'$. Fixing $N'$ also completely determines the eigenvalue of $P$ in the ground state because $t\to\infty$. The ground state degeneracy is infinite given that the smallest possible ground state energy $E_{0}=0$ is obtained with $P=\alpha N'\;(\textrm{mod}\;2\pi)$ for every $N'\in\mathbb{Z}$. If $\alpha$ is quantized as:
\begin{equation}
\frac{\alpha}{2\pi}=\pm\nu=\frac{p}{q}\quad,\quad p\in\mathbb{Z}\quad,\quad q\in\mathbb{Z}/\lbrace0\rbrace \ ,
\end{equation}
then $P$ takes one of the $q$ possible discrete values in any degenerate ground state. Otherwise, if $\alpha/2\pi$ is irrational, the values of $P$ span the entire continuous interval $(-\pi,\pi)$ across all ground states. The latter indicates frustration, which can be illustrated in a quantum phase transition from a Mott insulator (with non-conserved topological charges) to an incompressible quantum liquid characterized by quantum numbers $P$ and $N'$. The system must pick some values of $P$ and $N'$ at the transition. If the system enters a state with an arbitrary eigenvalue of $P$, it also needs to find the matching $N'$ in order to not pay energy that scales as $L^{d-2}$ with the system size $L$. The matching $N'$ for an irrational $\alpha/2\pi$ is generally infinite and arbitrarily far away from the established $N'$ in the system's original state. Fine-tuning $P$ is equally difficult. Without a dynamics that can make either $P$ or $N'$ fluctuate ($t'\to0$), the system is stuck in a metastable state with high energy. There is probably a better conventional state that resolves metastability and has lower energy. On the other hand, a rational $\alpha/2\pi=p/q$ leaves behind only a finite set of $P$ eigenvalues in the degenerate ground states, each of which corresponds to an infinite set of matching $N'$ values spaced by $q$. This is much less frustration, especially if $q$ is small, and gives the system a good chance to enter a topologically ordered state across the quantum phase transition.

Now consider a perturbation $0<t'\ll t$ in the absence of the Abelian rank 1 flux conservation. Its main effect is to lift the macroscopic degeneracy of states labeled by $N'$. The instanton Hamiltonian turns into a Hofstadter problem, with a finite $q$-fold ground state degeneracy for $\alpha/2\pi = p/q$. The Hofstadter gaps are of the order of $t'$, and the bandwidths are dominated by $t$. Note that the instanton spectrum is completely discrete in $d>2$: the upper cutoff for $N$ is of the order of $L/a$, where $a$ is the lattice spacing of the microscopic crystal, so that the quantization of $P$ is of the order of $2\pi a/L$ and the quantization of instanton energy levels inside a Hofstadter band is roughly $t\times 2\pi a/L \sim L^{d-3}$.

Certain aspects of this topological order have classical character. First of all, the Abelian flux conservation with $t'=0$ produces infinite ground state degeneracy -- a separate topological sector is defined for every classical configuration of vortices threaded through the manifold opening. Even if $t'>0$ in the absence of flux conservation, the scaling $t\sim L^{d-2}$ creates macroscopic energy barriers in the instanton spectrum $E(P)$ between the remaining $q$-fold degenerate ground states. Instanton and local fluctuations at finite temperatures have energy $\delta E \sim T \ll t$ that is not sufficient to change a topological sector. Therefore, the $d>2$ topological order survives at finite temperatures, until the rank $d-1$ flux conservation fails.

It is instructive to compare the classical topological order for $\alpha/2\pi=p/q$ in $d>2$ with the $d=2$ topological order of quantum Hall states. When the latter is analyzed on a $S^{1}\times S^{1}=T^{2}$ torus, one finds two integer canonical variables $N\equiv N_{1}$ and $N'\equiv N_{2}$ of the same type, having the same \emph{finite} mass $M$. The corresponding canonical momenta are similarly shifted in the Hamiltonian:
\begin{eqnarray}
H &=& \frac{(P-\alpha N')^{2}}{2M}+\frac{(P'+\alpha N)^{2}}{2M} \\
  &\to& t-t\sum_{i=1,2}\cos(P_{i}-\alpha\epsilon_{ij}N_{j}) \ , \nonumber
\end{eqnarray}
but, since there is only one finite mass, the Hamiltonian always describes motion of a particle on a lattice in the presence of an external magnetic flux $\alpha$. The ensuing Hofstadter problem produces a finite $q$-fold ground state degeneracy, with finite energy barriers between the ground states. Therefore, the $d=2$ topological order is fundamentally shaped by quantum processes and does not survive at any finite temperature.
  
Monopoles can establish the same $\pi_{d-1}(S^{d-1})$ topological orders as hedgehogs on the manifold $\mathcal{M}=S^{d-1}\times S^{1}$, but their deconfinement implies the Abelian flux non-conservation at rank 1. If the monopole gauge field configuration $A_{\mu_1\cdots\mu_{d-1}}$ carries a non-trivial topological charge inside the opening of $S^{d-1}$
\begin{equation}
N=\frac{1}{2\pi}\oint\limits_{S^{d-1}}\prod_{i=1}^{d-1}dx_{i}\,\epsilon_{\mu_{1}\cdots\mu_{d-1}}A_{\mu_{1}\cdots\mu_{d-1}} \ ,
\end{equation}
then the field configuration on $\mathcal{M}$ cannot be smoothly deformed to change $N$. Any attempt to smoothly change $N$ would have to start with a local deformation of the fields at some $x\in S^{1}$ that creates a non-zero flux divergence $\partial_{\mu}\mathcal{J}_{\mu}\neq 0$ across the $S^{d-1}$ submanifold at $x$. This monopole front at $x$ would need to be gradually swept across the entire $S^{1}$ subspace (by changing $x$) in order to bring the desired monopole charge difference $\delta N$ from infinity into the interior opening of $S^{d-1}$ where the existing charge $N$ sits. The entire procedure is prohibited because the dynamics of incompressible quantum liquid maintains $\partial_{\mu}\mathcal{J}_{\mu}=0$ and the monopole front would cost infinite energy. Nevertheless, $N$ is quantized because the monopole must bring its unobservable Dirac string through $\mathcal{M}$.

The dynamics of monopole topological sectors is analogous to that of hedgehogs. Repeating the above derivation in $d\in \lbrace 2,3,6,7,10,11,\dots \rbrace$ dimensions for monopoles and hedgehogs combined leads to the instanton Hamiltonian on $\mathcal{M}=S^{d-1} \times S^1$ up to a constant:
\begin{equation}
H = - t^{\textrm{h}} \cos(P^{\textrm{h}}-\alpha^{\textrm{h}} N') - t^{\textrm{m}} \cos(P^{\textrm{m}}-\alpha^{\textrm{m}}N') - t'\cos(P') \nonumber
\end{equation}
with:
\begin{equation}
\frac{\alpha^{\textrm{h}}}{2\pi} = -(-1)^d \nu^{\textrm{h}} \quad,\quad \frac{\alpha^{\textrm{m}}}{\pi} = -(-1)^d \nu^{\textrm{m}} \ .
\end{equation}
Both $\nu^{\textrm{h}}$ and $\nu^{\textrm{m}}$ must be quantized as rational numbers in stable topologically ordered phases.

The topological degeneracy of the ground states on a torus $T^d = (S^1)^d$ can be obtained similarly because the topological invariants of monopoles and hedgehogs are protected on $(S^1)^{d-1}$ just as well as on $S^{d-1}$. Instanton dynamics is captured by the Hamiltonian:
\begin{eqnarray}
H &=& - \sum_i \Bigl\lbrack
    t^{\textrm{h}} \cos(P_i^{\textrm{h}}-\alpha^{\textrm{h}} N'_i) + t^{\textrm{m}} \cos(P_i^{\textrm{m}}-\alpha^{\textrm{m}}N'_i) \nonumber \\
&& \qquad + t'\cos(P'_i) \Bigr\rbrack \nonumber \ ,
\end{eqnarray}
where a set of canonical coordinates $N'_i, N^{\textrm{h}}_i, N^{\textrm{m}}_i \in \mathbb{Z}$ and the corresponding canonical momenta $P'_i, P^{\textrm{h}}_i, P^{\textrm{m}}_i \in (-\pi,\pi)$ is defined for every torus direction $i \in \lbrace 1,\dots,d \rbrace$.

Finally, we comment on the topological ground state degeneracy in pure spin systems without charge degrees of freedom. The spin current and the non-Abelian gauge field coupled to it cannot be topologically quantized in $d>2$ because every $\hat{{\bf n}}$ configuration of the vector field on $S^{1}$ is smoothly deformable to any other configuration (the integer quantization of $N'$ in the instanton Hamiltonian is lost). The spin-only topological Lagrangian density (\ref{top-term-alt}) ``knows'' this, and anticipates dynamics in which the spin symmetry is reduced to U(1) -- hence enabling the topological quantization of the remaining conserved current, and monopole-like topological orders.

\subsection{Quantum entanglement and braiding}\label{secBraid}

A defining feature of topological order is long-range quantum entanglement that makes it impossible to smoothly deform the ground state into an unentangled product state. Here we discuss a few aspects of entanglement in the $\pi_{d-1}(S^{d-1})$ topological orders of monopoles and hedgehogs, mostly focusing on $d=3$ dimensions. Some entanglement manifestations are \emph{topologically protected} and thus immune to all smooth gap-preserving deformations of the Hamiltonian. They sharply characterize the topological order in all circumstances. Other entanglement manifestations may be only \emph{dynamically protected} by the finite energy gap of the topologically ordered state. They can depend on the symmetries and Hamiltonian details, can be altered by the presence of gapped excitations, and may be prone to having non-quantized read-out values. Nevertheless, they still characterize topological order under certain conditions.

Ground state degeneracy on non-simply connected manifolds, unrelated to symmetry breaking, is the basic manifestation of entanglement in every topologically ordered phase. A Hamiltonian with a gapped topologically degenerate spectrum cannot be smoothly transformed (without crossing a quantum phase transition) into a topologically trivial Hamiltonian $H_0 = H_1 + H_2 + \cdots$ whose ground state $|0\rangle = |1\rangle \otimes |2\rangle \otimes \cdots$ is unentangled in terms of microscopic degrees of freedom $1,2,\dots$ and hence non-degenerate (every factor $|n\rangle$ can independently minimize its corresponding Hamiltonian's $H_n$ energy). The basic characterization of topological entanglement is a map $N(\mathcal{M})$ which evaluates to the ground state degeneracy on any given manifold $\mathcal{M}$.

A topological quantum computer can implement qubits with isolated topologically ordered systems whose ground state degeneracy is $N$. The established topological sector  $q\in\lbrace 1,2,\dots,N\rbrace$ of a qubit's quantum state or thermal ensemble in $d \ge 3$ is the stored quantum information that enjoys topological protection. The state of a qubit can be changed externally by global braiding processes of the kind shown in Fig.\ref{local-flux-change}. Multiple qubits can be entangled on purpose by performing correlated braiding operations on them.

Beyond these basics, a topological order may be characterized by additional topologically protected features. Quantum Hall liquids in $d=2$ dimensions are sharply characterized by the ``exchange statistics'' of fractional quasiparticles (anyons), or more accurately by the U(1) or higher internal-symmetry-group rotations of the many-body quantum state in braiding operations that exchange quasiparticles' positions. Exchanging identical particles in $d>2$ dimensions is topologically protected only within two homotopy classes of braiding trajectories, allowing a sharp distinction between only bosonic and fermionic exchange statistics \cite{WenQFT2004}. We will show later that various aspects of $d>2$ braiding are still dynamically protected and possibly interesting for quantum computing. But first we wish to identify the topologically protected braiding.

Generally, a braiding of excitations in a quantum state $|\Psi(v)\rangle$ is generated by some external time-dependent perturbation $G(t)$ to the Hamiltonian $H$. The excitations may have some internal degrees of freedom specified by a finite complex vector $v$. We can express the effect of braiding using the matrix $W(t)$ acting on $v$ for which
\begin{eqnarray}
v' &=& Wv \\
\langle\Psi(v')| U |\Psi(v)\rangle &\equiv& \langle\Psi| \bar{W}^\dagger U |\Psi\rangle = 1 \nonumber
\end{eqnarray}
holds, where $U = \prod_t e^{-i(H+G)dt}$ is the time evolution operator. In the case of Abelian braiding $\bar{W}=e^{i\varphi}$, we have:
\begin{equation}\label{EntanglementPhase}
\langle\Psi| U |\Psi\rangle = e^{i\varphi} \ .
\end{equation}
The calculation of $\varphi$, or $W$ in general, proceeds by the standard construction of a path integral where the time parameter $t$ is broken up into infinitesimal increments $dt$. Any aspect of the action of $G$ that involves many degrees of freedom can be evaluated classically using the saddle-point approximation. The simplest braiding involves pushing the point quasiparticles on paths ${\bf x}(t)$ with $G(t) = {\bf p}\, d{\bf x}(t)/dt$, where the canonical momentum operator ${\bf p}$ generates movement. In the adiabatic limit, only the gauge field part of ${\bf p}$ matters and produces the Aharonov-Bohm phase ${\bf A} d{\bf x}$ in every braiding step. Similar movement of multi-dimensional excitations gives rise to generalized Aharonov-Bohm phases at higher ranks of the gauge theory. The formalism for this is derived in Appendix \ref{appMultiBraiding}.

The generalization of anyon braiding to $d>2$ is the topologically protected braiding of point quasiparticles $p$ with $d-2$ dimensional excitations (``loops'') $l$. All variations of such braiding are topologically equivalent to the enclosure of $p$ inside the closed $d-1$ dimensional braiding trajectory $S^{d-1}$ of $l$ relative to $p$. The many-body quantum state collects the entire quantized Aharonov-Bohm phase at rank $d-1$ from the quantized defect bound to $p$. However, it turns out that only monopoles can produce non-trivial Aharonov-Bohm effects that characterize topological order. The topological orders of hedgehogs are scrambled, and possibly characterized by some other data.

How do the hedgehogs scramble their topological orders? The $\pi_{d-1}(S^{d-1})$ topological orders studied in this paper attach the rank 1 charge to the gauge flux at a different rank $d-1$ in $d>2$ dimensions. Therefore, non-trivial Aharonov-Bohm effect is possible only if the gauge fields at different ranks are coherently linked. Indeed, rigid linking (\ref{link1}) is established within the Abelian gauge field hierarchy that describes monopoles. For example, $A_\mu$ in $d=3$ is coerced into $\frac{1}{2} (\partial_\mu A_\nu - \partial_\nu A_\mu) = A_{\mu\nu}$ by linking, and couples the charge current to the monopole $A_{\mu\nu} \neq 0$ via a $j_\mu A_\mu$ Lagrangian density term. In contrast, inter-rank links within the non-Abelian hedgehog hierarchy (\ref{link2}) involve an additional field $\hat{\bf n}$ of spins. Abundant spin fluctuations in incompressible quantum liquids spoil the correlations between the gauge fields at different ranks (assuming the full spin symmetry). Specifically, $A_\mu^a$ in $d=3$ does not have extended correlations on any large loop. Therefore, hedgehogs do not produce a coherent spin Aharonov-Bohm effect, and similarly they lack an intrinsic correlation with the Abelian gauge fields for the charge Aharonov-Bohm effect. Hedgehogs can still experience non-trivial braiding indirectly if they bind monopoles.

\begin{figure}
\subfigure[{}]{\raisebox{0.25in}{\includegraphics[height=0.4in]{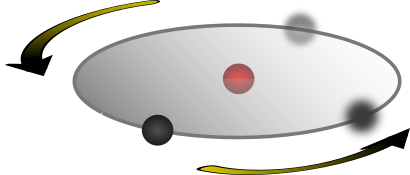}}}
\subfigure[{}]{\includegraphics[height=1.0in]{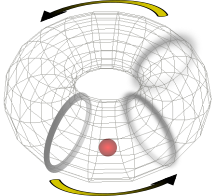}}
\subfigure[{}]{\includegraphics[height=1.0in]{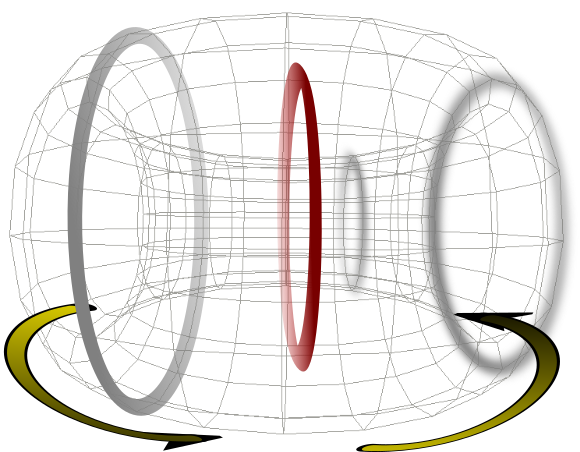}}
\caption{\label{Braiding3D}Braiding of elementary excitations in $d=3$ dimensions: (a) particle-particle, (b) particle-loop, (c) loop-loop. Only (b) is topologically protected. One excitation (red) is kept fixed, while the other one sweeps a closed trajectory. Simulated exchange of two particles is a half of the process depicted in (a).}
\end{figure}

At least, we can make progress by characterizing braiding in monopole quantum liquids. In $d=3$ dimensions, the elementary excitations amenable to braiding are particles and loops. The former are tied to rank 1 currents $j_\mu$ and arise from Witten effect \cite{Witten1979}, while the latter are tied to rank 2 currents $j_{\mu\nu}$ and shaped by magnetoelectric effect (see Section \ref{secME}). Braiding is Abelian, and the braiding phase (\ref{EntanglementPhase}) for the particle-particle, particle-loop and loop-loop braiding shown in Fig.\ref{Braiding3D} is calculated in Appendix \ref{appBraiding}, assuming monopole filling factors $\nu = 1/n$:
\begin{eqnarray}\label{D3braiding}
\textrm{particle-particle ``exchange''} &\quad\cdots\quad& \varphi \overset{*}{=} \frac{1}{2}\pi\nu \nonumber \\
\textrm{particle-loop braiding} &\quad\cdots\quad& \varphi = 2\pi\nu \nonumber \\
\textrm{loop-loop braiding} &\quad\cdots\quad& \varphi \overset{*}{=} 0 \nonumber \ .
\end{eqnarray}
An asterisk above the equality sign indicates the absence of topological protection. Only the particle-loop braiding is topologically protected with $\varphi = 2\pi\nu$. Generalizing to hierarchical states at other filling factors is straight-forward within the $K$-matrix formalism discussed in Section \ref{secMicroPart}. One would hope that non-Abelian particle-loop braiding is also possible in appropriately modified effective theories, since the monopole quantum liquids have much in common with quantum Hall liquids and the latter have non-Abelian varieties \cite{Moore1991}. Non-trivial topologically protected braiding of two and three loops has been explored in various models \cite{Wang2014a, Jiang2014, Wang2014b, Juan2014, Wang2015}, and some discussions of non-Abelian quasiparticle braiding have also appeared in the literature \cite{Teo2010, Swingle2011b}.

The remainder of this section analyzes the dynamically protected braiding. One can simulate an exchange of two identical fractional quasiparticles by driving them in a plane on semi-circular paths about their center of mass. In the simplest case, the braiding phase is $\varphi = \frac{1}{2}\pi\nu$ for $d=3$ monopole liquids with $\nu=1/n$. It is calculated from the ``field-induced'' corrections (caused by the topological Lagrangian term) to the total rank 1 Aharonov-Bohm phase $\pi\nu$ of the two particles. This comes with a problem -- the rank 1 Aharonov-Bohm phase due to a monopole is affected by the attached Dirac string, and hence gauge dependent. This issue is resolved by the emergent spin of fractional quasiparticles, which we discuss later. Identifying $\varphi$ with an exchange statistical phase does not make sense for multiple reasons. Mainly, $\varphi$ is not topologically protected (topological protection is hidden only in the amount of gauge flux attached to a particle). Modifications of the braiding particle trajectories can change the value of $\varphi$, even smoothly.

But, let us assume that we can put the particles on ``fixed rails'' and drive them on accurate rigid trajectories. Now $\varphi$ appears topologically quantized, unless some perturbations or excitations cause bending of the gauge field-lines created by the braided particle's charges and monopoles. This can alter the gauge fluxes through the loops formed by particle trajectories and modify the Aharonov-Bohm phase. Nevertheless, there is a dynamical protection mechanism against this: if the Hamiltonian is perfectly symmetric, then the field-line bending costs finite energy -- all excitations are gapped. Random disorder and neutral fluctuations will average out and have a hard time spoiling the braiding phase if the braiding trajectories are large enough.

Braiding operations are also not unique, and can be state-dependent. Different braiding operations generated by $G_i$, $i=1,2,\dots$ in (\ref{EntanglementPhase}) can transform a \emph{given} initial state of two identical quasiparticles into the same final state (up to a braiding phase), but generally produce different final states from an \emph{arbitrary} initial one. Consider $\pi$-rotations of two quasiparticles generated by $\int dt\, G = \pi {\bf L} \hat{\bf n}$, where ${\bf L}$ is the total angular momentum operator. Rotations about the center of mass always exchange the locations of two quasiparticles, but produce different final states depending on the rotation axis $\hat{\bf n}$ if the quasiparticles have spin. Specifically in this case, one can limit the $\pi$-rotation to unentangled angular momentum eigenstates of two identical particles: the lack of entanglement makes the particles independent, and the restriction to an eigenstate of ${\bf L} \hat{\bf n}$ is guaranteed to modify the state only by a phase. Since the unentangled state of two spins has the maximum possible eigenvalue of ${\bf L} \hat{\bf n}$, it is easy to see that the braiding (\ref{EntanglementPhase}) by such a $\pi$-rotation reproduces the statistical phase $\varphi$ of microscopic particles according to the spin-statistics theorem. In the case of fractional quasiparticles, the $\pi$ rotation will also pick the field-corrected Aharonov-Bohm phase due to attached monopoles.

Differences between exchange-simulating braiding operations are another reason to not identify braiding with exchange. However, these differences, and dynamically protected braiding in general, can characterize topological order in a given fixed set of dynamical conditions. For example, if one ensures certain symmetries, a temperature well below the gap scale, good isolation from the environment, etc, then it is possible to probe topological order by dynamically protected braiding. Hedgehog and monopole quantum liquids have different ``exchange'' braiding due to different Aharonov-Bohm effects. The difference between rotational and irrotational braiding can reveal the role of spin in the quantum entanglement.

A charge-monopole bound state, or a dyon, has emergent spin. The spin angular momentum is stored in the electromagnetic field of the charge-monopole pair, as we review in Appendix \ref{appZwanziger}. The quantum dynamics of this spin resolves the naive gauge dependence of the monopole-induced Aharonov-Bohm effect -- the spin fixes the gauge. We show in Appendix \ref{appDyonExchange} that one must calculate the braiding Aharonov-Bohm phase using the rank 1 gauge field that obtains when the monopole's Dirac string is oriented in the same/opposite direction as its spin -- so that the charge currents that try to screen-out the string generate a magnetic dipole moment consistent with the spin. This directly applies to unfractionalized dyons, which contain one unit of electric and magnetic charge, and carry the elementary unit $S=\frac{1}{2}$ of spin. In the case of fractional dyons, one needs to solve a difficult Schrodinger equation for an entire set of entangled dyon partons in order to determine how their internal degrees of freedom form the total properly quantized angular momentum.

Can the hedgehog quasiparticles experience fractional braiding given that they lack an Aharonov-Bohm phase? While we do not have a rigorous answer here, we cannot find obvious obstacles to fractional and even non-Abelian braiding. A necessary ingredient are internal degrees of freedom. Suppose we inject an electron into a topologically ordered liquid. When this electron breaks up into $n$ fractional quasiparticles, the partons must remain entangled in the total angular momentum state identical to the original electron's spin, even as they fly far apart. Clearly the partons must have some internal degrees of freedom. The long-range entanglement is necessarily protected because the physical dynamics can change the total charge and spin of the system only locally in integer amounts. The outcome of braiding operations (\ref{EntanglementPhase}) can depend on the joint state of all quasiparticles that the braided ones are entangled with, and a state-dependent braiding can be non-Abelian. Everything said so far applies equally to fractional dyons because they have emergent spin.

How can this enable fractional braiding? Suppose that the ground state breaks time reversal but respects rotational symmetry. If the $\pi$-rotation about an axis $\hat{\bf n}$ induces a braiding phase $\varphi(\hat{\bf n})$, then the time reversal of that braiding induces $\varphi' = -\varphi(\hat{\bf n})$. The latter is equivalent to the rotation about $-\hat{\bf n}$, i.e. $\varphi' = \varphi(-\hat{\bf n})$. Broken time reversal allows $-\varphi(\hat{\bf n}) \neq \varphi(\hat{\bf n}) \; (\textrm{mod }2\pi)$, but the ensuing $\varphi(-\hat{\bf n}) \neq \varphi(\hat{\bf n}) \; (\textrm{mod }2\pi)$ is inconsistent with rotational symmetry for generic fractional phases $\varphi$. The rotational bias needed for fractional braiding can exist in the entangled (spin) state of excited quasiparticles.

\bigskip

\subsection{Microscopic particle operators and hierarchical states}\label{secMicroPart}

The effective theory $\mathcal{L}$ of an incompressible quantum liquid must be able to produce an excitation with the characteristics of an electron. This introduces additional restrictions on the hedgehog $\nu^{\textrm{h}}$ and monopole $\nu^{\textrm{m}}$ filling factors, beyond their rational quantization that enables stable topological orders. It also constrains long-range quantum entanglement and correlation among fractional quasiparticles.

In the absence of monopoles, a fractional quasiparticle attaches charge $\nu^{\textrm{h}}$ to a hedgehog quantum. A Laughlin-like fractionalization $\nu^{\textrm{h}}=1/n$ with $n\in\mathbb{Z}$ reproduces a physical particle as a composite of $n$ fractional quasiparticles. The gauge fields of hedgehogs carry no angular momentum and do not generate any Aharonov-Bohm phases. Therefore, the simplest composite particle of charge $e=1$ is a spinless boson. In order to reconstruct spin $S=\frac{1}{2}$, we must have two flavors $\uparrow$ and $\downarrow$ of fractional quasiparticles. The causality of the microscopic Lorentz-invariant dynamics requires that the particles with spin $S=(2n+1)/2$ be fermions and those with $S=n$ bosons. Since we cannot obtain fermionic statistics from Aharonov-Bohm phases, we must impose it at the operator level -- the current operators
\begin{equation}
J_\mu = -\frac{i}{2}\Bigl\lbrack \psi^\dagger (\partial_\mu \psi) - (\partial_\mu \psi^\dagger) \psi \Bigr\rbrack + A_\mu \psi^\dagger \psi
\end{equation}
in the effective theory must be constructed using complex fields $\psi$ in the case of bosons, and Grassmann fields in the case of fermions. The fractional quasiparticles can themselves be fermions whose composites become physical fermionic particles. Then, $\nu^{\textrm{h}}=1/n$ is restricted to odd $n$ in the case of fermions, and even $n$ in the case of bosons (at least when the fractional quasiparticles are represented as fermions). Other rationally-quantized values of $\nu^{\textrm{h}}$ must be obtained by hierarchical constructions.

The field theory of incompressible quantum liquids admits hierarchical states with multiple flavors of fractional quasiparticles. We can express the dynamics of such states using the formalism developed for quantum Hall liquids. Let us introduce the fractional quasiparticle field operators $\psi_n$ and charge currents
\begin{equation}
j_{n,\mu}^{\phantom{,}} = -\frac{i}{2}\Bigl\lbrack \psi^\dagger_n (\partial_\mu^{\phantom{,}} \psi_n^{\phantom{,}})
  - (\partial_\mu^{\phantom{,}} \psi^\dagger_n) \psi_n^{\phantom{,}} \Bigr\rbrack
\end{equation}
for each flavor $n=1,\dots,N_{\textrm{f}}$. The operators $\psi^\dagger_n, \psi_n^{\phantom{,}}$, which create and annihilate a fractional quasiparticle, must be either complex or Grassmann as required by their compatibility with the exchange statistics of microscopic particles. They can exist only in local combinations that protect the quantization of charge and spin. The currents $j_{n,\mu}$ are minimally coupled to emergent rank 1 gauge fields $a_{n,\mu}$. Every flavor has its own hierarchy of singular configurations, ultimately leading to a gauge field $a_{n,\lambda_1\cdots\lambda_{d-1}}$ of monopoles or hedgehogs at rank $d-1$. The effective Lagrangian density of fractional quasiparticles contains Maxwell and linking terms for the flavor gauge fields at all ranks, and a generalized topological term that takes the following real-time form in the case of hedgehogs (replace $S_{d-1}$ with $4\pi$ for monopoles):
\begin{eqnarray}\label{effFractTheory}
&& \mathcal{L}'_{\textrm{t}} =
      \frac{1}{S_{d-1}}\epsilon^{\mu\nu\lambda_{1}\cdots\lambda_{d-1}} \Biggl\lbrack
         \sum_{n,m} K_{nm} a_{n,\mu}\partial_{\nu} a_{m,\lambda_{1}\cdots\lambda_{d-1}} \nonumber \\
&& \qquad + \sum_m \,q_{m} A_{\mu}\partial_{\nu} a_{m,\lambda_{1}\cdots\lambda_{d-1}}
     + \sum_n \,q_{n} a_{n,\mu}\partial_{\nu} A_{\lambda_{1}\cdots\lambda_{d-1}} \Biggr\rbrack \nonumber \\
&& \qquad -\sum_n j_{n,\mu}a_{n,\mu} \ .
\end{eqnarray}
Integrating out all flavor gauge fields reproduces the Lagrangian density (\ref{top-term-6}) with an emphasized gauge coupling of the physical charge current:
\begin{equation}
\mathcal{L}_{\textrm{t}} = -\frac{\nu}{S_{d-1}}\epsilon^{\mu\nu\lambda_{1}\cdots\lambda_{d-1}}A_{\mu}\partial_{\nu}A_{\lambda_{1}\cdots\lambda_{d-1}}+A_{\mu}J_{\mu} \ .
\end{equation}
If we collect all coefficients $K_{nm}\in\mathbb{Z}$ into a matrix $K$, $q_n\in\mathbb{Z}$ into a vector $q$ and $j_{n,\mu}\in\mathbb{Z}$ into a vector $j_\mu$, then the filling factor $\nu$ and physical charge current are
\begin{eqnarray}\label{KJ}
\nu=q^{T}K^{-1}q \quad,\quad J_{\mu}=q^{T}K^{-1}j_{\mu} \ .
\end{eqnarray}
The $K$ matrix and ``charge'' vector $q$ specify a hierarchical incompressible quantum liquid. Elementary quasiparticles correspond to integer quanta of the flavor currents. Setting $j_{n,0}$ to a combination of integers and calculating $J_0$ reveals the fractional charges carried by quasiparticles.

Capturing the electron's spin in an effective theory requires a basic hierarchical construction. The simplest fractionalization of a spin $S=1/2$ electron via hedgehogs is represented by:
\begin{equation}
K = \left(\begin{array}{cc} m & 0\\ 0 & m \end{array}\right) \quad,\quad q=\left(\begin{array}{c} 1\\ 1 \end{array}\right)
\end{equation}
with an odd $m$ and fermionic fractional quasiparticles. All operators are allowed to change charge and spin only by an integer (multiple of $e$ and $\hbar$ respectively), so that a group of $m$ fractional quasiparticles created from a single electron retains long-range entanglement. We can specify the nature of spin entanglement with an additional ``spin'' vector $s$ that defines the spin of fractional quasiparticles:
\begin{eqnarray}
S=\frac{1}{2}s^{T}K^{-1}j_{0} \ .
\end{eqnarray}
For example, consider
\begin{equation}
s=\left(\begin{array}{c} m \\ -m \end{array}\right) \quad,\quad s'=\left(\begin{array}{c} 1\\ -1 \end{array}\right) \ .
\end{equation}
The first case $s$ describes fractional quasiparticles that individually carry spin $S=1/2$ and keep a group entangled state with total spin $S=1/2$. This is compatible with a fermionic quasiparticle statistics. The second case $s'$ corresponds to a fractionalized spin -- still permissible since the usual spin-statistics theorem, deduced from local causality, cannot be justified for non-locally entangled fractional quasiparticles. The description of braiding statistics involving internal degrees of freedom will be postponed for future study. This will require additional data to specify the braiding operation details, because such statistics is not topologically protected. The general structure of braiding operations discussed in Section \ref{secBraid} admits a non-Abelian statistics.

The fractionalization by monopoles in $d=3$ dimensions is more complicated than the fractionalization by hedgehogs. When point-like dyons with electric and magnetic charges $(e_i,m_i)$, $i=1,\dots,N$ are treated classically (Appendix \ref{appZwanziger}), the quantization of electromagnetic angular momentum reduces to Schwinger-Zwanziger condition:
\begin{equation}\label{Zwanziger}
(\forall i, j)\qquad e_{i}m_{j}-e_{j}m_{i}\in\mathbb{Z} \ .
\end{equation}
Incompressible quantum liquids with Laughlin-like monopole filling factors $\nu^{\textrm{m}} = 1/n$, $n\in\mathbb{Z}$ are consistent with (\ref{Zwanziger}). These liquids break time-reversal symmetry. An electron-like object made from $n$ fractional dyons is itself a dyon that carries a unit-charge and $n$ monopoles. For odd $n$, the total angular momentum of the composite dyon and its electromagnetic field can be $L=1/2$, depending on its internal state. It has been argued that such dyons behave as fermions under exchange in agreement with the spin-statistics theorem \cite{Goldhaber1976, Goldhaber1989}, although the presence of monopoles makes them non-local objects and complicates the causality-based relationship between their spin and exchange statistics.

A different possible electron fractionalization involves fractional quasiparticles with electric and monopole charges $(e,m)\in\lbrace(\frac{1}{2},1),(\frac{1}{2},-1),(-\frac{1}{2},1),(-\frac{1}{2},-1)\rbrace$. This obtains from
\begin{equation}\label{DyonDipoles}
K = \left(\begin{array}{cc} 2 & 0\\ 0 & 2 \end{array}\right) \quad,\quad q = \left(\begin{array}{c} 1\\ -1 \end{array}\right)
\end{equation}
at $\nu^{\textrm{m}}=1$, and satisfies (\ref{Zwanziger}). Consider a composite particle $(\frac{1}{2},1)+(\frac{1}{2},-1)$ made from a flavor-1 quasiparticle and flavor-2 quasihole. This is a charge $e=1$ object with no net monopole charge that could produce unconventional Aharonov-Bohm phases. The monopole and antimonopole can form a magnetic dipole that carries a non-zero magnetic moment. Furthermore, if the electric charge is displaced from the monopoles, the composite will carry a quantized spin angular momentum. A composite particle $(\frac{1}{2},1)+(\frac{1}{2},-1)$ has two charges $e_{1}=e_{2}=\frac{1}{2}$ and two monopoles $m_{1}=-m_{2}=1$. If we arrange $e_{1},m_{1}$ to sit at the position ${\bf r}/2$ and $e_{2},m_{2}$ to sit at $-{\bf r}/2$, then the total angular momentum is:
\begin{equation}
{\bf L} = \frac{1}{2}\hat{{\bf r}} \ . \nonumber
\end{equation}
This follows from the classical derivation in Appendix \ref{appZwanziger}, but we shall assume that it holds quantum-mechanically as well. The composite particle effectively carries spin $S=\frac{1}{2}$, and hence behaves as a fermion under any $\pi$-rotation exchange. The composite particle is also a magnetic dipole, with dipole moment ${\bf m}=2\pi{\bf r}$ obtained from the total magnetic field of the two monopoles:
\begin{equation}
{\bf B}({\bf x}) = 2\pi\left(\frac{{\bf x}-\frac{{\bf r}}{2}}{\left\vert {\bf x}-\frac{{\bf r}}{2}\right\vert^{3}}
     -\frac{{\bf x}+\frac{{\bf r}}{2}}{\left\vert {\bf x}+\frac{{\bf r}}{2}\right\vert^{3}}\right)
  \approx 2\pi\frac{3\hat{{\bf x}}(\hat{{\bf x}}{\bf r})-{\bf r}}{|{\bf x}|^{3}}
\end{equation}
for $|{\bf r}|\ll|{\bf x}|$. A relativistic fermion of mass $m_e$ has magnetic moment $|{\bf m}| = 1/2m_e$ according to the Dirac equation. This sets the average distance $|{\bf r}| = 1/4\pi m_e$ between the two monopoles of the composite to roughly a half of the Compton length. A physical electron in vacuum cannot be modeled this way because its Compton length $\lambda_{\textrm{C}} \sim 10^{-12}\;\textrm{m}$ is much larger than its size. However, an effective electron in a correlated solid state material has renormalized properties in addition to being localized and sized with uncertainty of at least one lattice constant $a>10^{-10}\;\textrm{m}$. Therefore, this model can provide an adequate construction of an electron operator in the effective theory.

The states like (\ref{DyonDipoles}) with $S=\frac{1}{2}$ dipoles can be fractionalized into even smaller partons. One approach involves monopole clustering. Consider
\begin{equation}\label{ZnSpinLiquid}
K=\left(\begin{array}{cc} 2n^{2} & 0\\ 0 & 2n^{2} \end{array}\right) \quad,\quad q=\left(\begin{array}{c} 1\\ -1 \end{array}\right)
\end{equation}
with $n\in\mathbb{Z}$ and $\nu^{\textrm{m}}=1/n^2$. This state breaks time-reversal symmetry, unless monopoles are strictly bound into $n$-tuplets by some force \cite{Fradkin2017}. A generally time-reversal-invariant variation with $\nu^{\textrm{m}}=0$ is:
\begin{equation}
K=\left(\begin{array}{cc} 2n^{2} & 0\\ 0 & -2n^{2} \end{array}\right) \quad,\quad q=\left(\begin{array}{c} 1\\ 1 \end{array}\right) \ .
\end{equation}
Schwinger-Zwanziger condition (\ref{Zwanziger}) imposes restrictions on the actual elementary dyons that could make up an electron. Assume the existence of a composite fractional quasiparticle $(e,m)=(1/n,0)$ in the spectrum. Then, an elementary dyon must be a bound state of $n$ fundamental quasiparticles, i.e. $(e,m)=(1/2n,n)$. All elementary dyons are compatible with one another, but must not be fractionalized into fundamental quasiparticles. The quasiparticle $(e,m)=(1/n,0)=(1/2n,n)+(1/2n,-n)$ is a bound state dipole of two elementary dyons, and hence is compatible with them. Combining $n$ such dipoles together, with odd $n$, can reconstitute an electron-like particle with total electric charge 1, spin $S=\frac{1}{2}$ and finite dipole moment. As explained earlier, this object has fermionic statistics under exchange. Similarly, even $n$ describes best the fractionalization of bosonic spin-singlets or magnons into fermionic spinons. Note that the electromagnetic response captured by the axion term $\theta = 2\pi\left(\nu^{\textrm{m}} + \frac{1}{2}\right)$ with $\nu^{\textrm{m}} = 1/n^2$ (the $1/2$ shift comes from a quantum anomaly, see Section \ref{secME}) is equivalent to $\theta = \pi/n^2$ for odd $n$, and $\theta=0$ for even $n$, within the periodicity $\Delta\theta = 2\pi/n^2$ due to monopole clustering \cite{Fradkin2017}. 

An interesting and probably more stable fractionalization of (\ref{DyonDipoles}) into smaller partons is:
\begin{equation}\label{ZnSpinLiquid2}
K=\left(\begin{array}{cc} 2n^{2} & 0\\ 0 & 2n^{2} \end{array}\right) \quad,\quad q=\left(\begin{array}{c} n\\ -n \end{array}\right)
\end{equation}
with $n\in\mathbb{Z}$ and $\nu^{\textrm{m}}=1$. There is no need for a separate mechanism to bind the fundamental fractional partons -- the microscopic particles are bound into $n$-tuplets prior to fractionalization. The elementary quasiparticle charge is $\pm 1/2n$ just like in the monopole clustered states (\ref{ZnSpinLiquid}).

Spin systems without charge degrees of freedom can host topologically ordered phases if the Spin($d$) symmetry is reduced to U(1). This precludes the formation of hedgehogs, but allows the formation of spin-monopoles in the residual U(1) order parameter. The dynamics of the low-energy spin U(1) degree of freedom and its monopoles is necessarily controlled by the same type of effective theory that we analyzed in the context of charge dynamics. Therefore, we can apply the same constructions of monopole topological orders to gapped spin liquids.

The simplest gapped spin liquid is the resonant-valence-bond (RVB) state with Z$_2$ topological order \cite{senthil00}. It obtains when electrons localized on a lattice form spin-singlets with their neighbors, and the singlets fail to crystallize due to quantum fluctuations. Breaking a singlet creates a particle-antiparticle pair of two neutral $S=\frac{1}{2}$ spinons, which can drift far apart at a finite energy cost. None of the topological orders considered so far, applied to quantum paramagnets, reproduce exactly these excitations. The states (\ref{ZnSpinLiquid}) and (\ref{ZnSpinLiquid2}) contain elementary dyons with a fraction of the $S=\frac{1}{2}$ angular momentum unit for every $n$, and reproduce spinons only as magnetic dipole bound states of multiple dyons. Also, they have two independent flavors of spinons and their antiparticles instead of one (multi-flavored spinons occur in some frustrated magnets, e.g. on the pyrochlore lattice). We can alternatively construct general single-flavor topological orders $K=(n)$, $q=(l)$ with monopole clusterization into $c$-tuplets. The elementary quasiparticles $(e,m)=(lc/n,c)$ carry spin $S=em/2 = \frac{1}{2}$ if $n=lc^2$, at the filling factor $\nu^{\textrm{m}}=l^2/n=l/c^2$ ($l,c\in\mathbb{N}$,  $n=2k$, $k\ge 1$). This is still not a basic RVB spin liquid -- the spinon is a dyon, a source of gauge flux (a 2D equivalent would be a spinon-vison bound state). Therefore the gapped spin liquids obtained here are fundamentally different from the spin liquids of short-range singlets -- they are made, instead, from magnetic moments that remain well-defined at some finite coarse-graining length scales.

\subsection{Transverse response and boundary states}\label{secEqMot}

The topological Lagrangian term describes a steady-state response of rank $1$ charge currents to rank $d-1$ gauge fields. The response is linear even though the bulk is insulating, so it implies the existence of soft boundary modes. This is the natural generalization of the quantum Hall effect to higher dimensions. Let us focus on the kinematic field equations (\ref{pf-attachment-2}) in real time:
\begin{equation}
J^{\mu} = \frac{\nu}{q}\mathcal{J}^{\mu} = \frac{\nu}{q}\,\epsilon^{\mu\nu\lambda_{1}\cdots\lambda_{d-1}}\partial_{\nu}A_{\lambda_{1}\cdots\lambda_{d-1}} \ .
\end{equation}
We can rewrite this as
\begin{equation}\label{transverse-response-1}
J^{\mu} = \frac{\nu}{q}\frac{e_{d-1}^{2}}{d!}\epsilon^{\mu\lambda_{1}\cdots\lambda_{d}}E_{\lambda_{1}\cdots\lambda_{d}}
\end{equation}
using the ``electromagnetic'' field tensor $E_{\lambda_{1}\cdots\lambda_{d}}$, which is the canonical conjugate momentum to the rank $d-1$ gauge field (\ref{electric}) derived in Appendix \ref{appCanonical}:
\begin{equation}
E_{\lambda_{1}\cdots\lambda_{d}} = \frac{(d-1)!}{e_{d-1}^{2}}\sum_{i=1}^{d}(-1)^{i-1}\partial_{\lambda_{i}}
  A_{\lambda_{1}\cdots\lambda_{i-1}\lambda_{i+1}\cdots\lambda_{d}} \ .
\end{equation}
The system already has an implanted scalar ``magnetic'' field $B = \mathcal{J}^0$, and we can similarly define an ``electric'' field $E_{j_{1}\cdots j_{d-1}}$:
\begin{eqnarray}
&& B = \epsilon^{0\nu\lambda_{1}\cdots\lambda_{d-1}}\partial_{\nu}A_{\lambda_{1}\cdots\lambda_{d-1}}
  = \frac{e_{d-1}^{2}}{d!}\epsilon^{0\lambda_{1}\cdots\lambda_{d}}E_{\lambda_{1}\cdots\lambda_{d}} \nonumber \\
&& E_{j_{1}\cdots j_{d-1}} = \partial_{0}A_{j_{1}\cdots j_{d-1}}-\sum_{i=1}^{d-1}\partial_{j_{i}}A_{j_{1}\cdots j_{i-1}0j_{i+1}\cdots j_{d-1}} \nonumber \\[-0.1in]
   && \qquad\qquad = \frac{e_{d-1}^{2}}{(d-1)!} E_{0j_{1}\cdots j_{d-1}} \nonumber \ .
\end{eqnarray}
It can be shown that the ``electric'' field $E_{j_{1}\cdots j_{d-1}}$ accelerates the particles which carry rank $d-1$ currents with canonical momentum $\pi_{j_{1}\cdots j_{d-1}}$. We can further obtain the spatial components of the topological current (\ref{top-current}) $\mathcal{J}^{i} = \epsilon^{i0j_{1}\cdots j_{d-1}} E_{j_{1}\cdots j_{d-1}}$, and rewrite (\ref{transverse-response-1}) as:
\begin{equation}\label{transverse-response-2}
J^{0} = \frac{\nu}{q}B \quad,\quad J^{i} = \frac{\nu}{q}\epsilon^{i0j_{1}\cdots j_{d-1}}E_{j_{1}\cdots j_{d-1}} \ .
\end{equation}
This response is characterized by a fractionally quantized transverse charge conductivity:
\begin{equation}\label{transverse-response-2a}
\sigma^{ij_{1}\cdots j_{d-1}} = \frac{J^{i}}{E_{j_{1}\cdots j_{d-1}}} = \frac{\nu}{q}\, \epsilon^{0ij_{1}\cdots j_{d-1}} \ .
\end{equation}
Analogous consideration of the spin-current topological Lagrangian (\ref{top-term-alt}) leads to a transverse spin conductivity:
\begin{equation}\label{transverse-response-2b}
\sigma^{ij_{1}\cdots j_{d-1};a_{1}\cdots a_{d-2}} = \frac{J^{i;a_{1}\cdots a_{d-2}}}{E_{j_{1}\cdots j_{d-1}}}
  = \frac{\nu^{a_{1}\cdots a_{d-2}}}{q}\, \epsilon^{0ij_{1}\cdots j_{d-1}} \ ,
\end{equation}
whose quantization has the same symmetry-dependent fate as the spin fractionalization discussed in Section \ref{secFract}. 

Linear response conductivities (\ref{transverse-response-2a}) and (\ref{transverse-response-2b}) indicate the presence of soft modes in the spectrum. Such modes can live only at the boundary of a perfectly homogeneous system when its bulk is gapped. A boundary $\mathcal{B}$ always corresponds to a violation of translational symmetry. The density $\pi_{0\lambda_{1}\cdots\lambda_{n-1}}$ of a rank $n$ matter field $\theta_{\lambda_{1}\cdots\lambda_{n-1}}$ will generally be inhomogeneous at the system boundary and hence introduce electric fields $E_{j_{1}\cdots j_{n}}$ at the boundary in order to satisfy the Gauss law (\ref{electric}). This also holds at the highest rank $n=d-1$. If we assume for simplicity that the spatial inhomogeneity is expressed only in the direction $b\perp\mathcal{B}$ perpendicular to the boundary, we find that $\partial_{j}E^{jk_{1}\cdots k_{n-1}}\neq0$ has a solution $E^{bk_{1}\cdots k_{n-1}}\neq0$ near the boundary. Then, (\ref{transverse-response-2a}) implies charge currents $J^{i}\neq0$ near the boundary in all directions $i\parallel\mathcal{B}$ parallel to the boundary. These currents exist in equilibrium and changing them infinitesimally requires only an infinitesimal change of the ``electric'' or ``magnetic'' field, which costs arbitrarily small amount of energy. Hence, some gapless boundary states must be available to carry these currents. A detailed study of these boundary states in fractional incompressible quantum liquids is left for future work, and below we give only some qualitative remarks about their properties.

The transverse response has a definite sense of chirality, so it prevents the back-scattering of currents on the system boundary. The boundary spectrum is gapless in the absence of perturbations that break the gauge symmetry. Note that an ordinary (rank 1) electric field, either externally applied or internally generated by disorder, cannot by itself cause back-scattering and open a gap. The equilibrium electric field is pinned to zero in the bulk of an incompressible quantum liquid, by screening via the mobile boundary charges. Otherwise, the transverse response equation would predict the existence of bulk equilibrium charge currents that can be infinitesimally changed by infinitesimal perturbations, which would require gapless bulk excitations.

The gapless boundary modes that produce a Lorentz-invariant linear response (\ref{transverse-response-2}) or (\ref{transverse-response-2a}) necessarily have a relativistic spectrum. Therefore, the boundary spectrum contains chiral relativistic Dirac points.

\subsection{Electromagnetic response in three dimensions}\label{secME}

Incompressible quantum liquids in $d=3$ dimensions have unconventional electromagnetic properties when monopoles proliferate and bind charge. Here we briefly explore the fractional magneto-electric effect and Faraday/Kerr effect. These properties arise from the Abelian part of the Lagrangian density (\ref{L3b}) combined with the topological term (\ref{top-term-5}) in real time:
\begin{eqnarray}\label{EMLG}
\mathcal{L} &=& \frac{|\psi|^2}{2}(j_{\mu}+A_{\mu})(j^{\mu}+A^{\mu}) - \frac{1}{16\pi \bar{e}_{1}^{2}} F_{\mu\nu}F^{\mu\nu} \nonumber \\
&& -\frac{\kappa'_{2}}{2}\left(\frac{F_{\mu\nu}}{2}-A_{\mu\nu}\right)
                           \left(\frac{F^{\mu\nu}}{2}-A^{\mu\nu}\right) \nonumber \\
&& -\frac{1}{8\pi \bar{e}_{2}^{2}} (\epsilon^{\mu\nu\lambda_{1}\lambda_{2}}\partial_{\nu}A_{\lambda_{1}\lambda_{2}})
                           (\epsilon_{\mu\alpha\beta_{1}\beta_{2}}\partial^{\alpha}A^{\beta_{1}\beta_{2}}) \nonumber \\
&& -\frac{\nu^{\textrm{m}}}{4\pi}A_{\mu}\epsilon^{\mu\nu\lambda_{1}\lambda_{2}}\partial_{\nu}A_{\lambda_{1}\lambda_{2}} \ .
\end{eqnarray}
We redefined the Maxwell couplings $e_n^2 = 4\pi\bar{e}_n^2$ in order to facilitate the switch to the commonly used Gaussian units. Integrating out $A_{\mu\nu}$ renormalizes the rank 1 Maxwell term and carries out the replacement
\begin{equation}
A_{\mu\nu} \to \frac{1}{2} F_{\mu\nu} = \frac{1}{2} (\partial_{\mu}A_{\nu}-\partial_{\nu}A_{\mu})
\end{equation}
in the topological Lagrangian density:
\begin{eqnarray}\label{top-term-ME}
\mathcal{L}_{\textrm{t}} &=& -\frac{\nu^{\textrm{m}}}{4\pi} \epsilon^{\mu\nu\alpha\beta}(\partial_{\mu}\theta+A_{\mu})\partial_{\nu}A_{\alpha\beta}
  \nonumber \\
&\to& \frac{\nu^{\textrm{m}}}{8\pi}\epsilon^{\mu\nu\alpha\beta}(\partial_{\nu}A_{\mu})F_{\alpha\beta}
    = -\frac{\nu^{\textrm{m}}}{16\pi}\epsilon^{\mu\nu\alpha\beta}F_{\mu\nu}F_{\alpha\beta} \nonumber \\
&=& -\frac{\nu^{\textrm{m}}}{8\pi}\widetilde{F}^{\mu\nu}F_{\mu\nu} \ . 
\end{eqnarray}
The arrow indicates integration by parts, and
\begin{equation}
\widetilde{F}^{\mu\nu}=\frac{1}{2}\epsilon^{\mu\nu\alpha\beta}F_{\alpha\beta}
\end{equation}
is the dual electromagnetic field tensor.

We have previously emphasized the \emph{emergent} gauge fields that collect topological defects from the matter fields through singular gauge transformations. However, the rank 1 gauge field also includes the physical electromagnetic U(1) gauge field. The derivations in Sections \ref{secQH} and \ref{secDynCharge} indicate that the emergent gauge field $A_\mu$ should be simply absorbed into the physical gauge field $A_\mu^{\textrm{em}}$, and the combined gauge field should appear in all terms of the effective theory. The coupling of the final renormalized Maxwell term, which involves the combined field, defines the physical charge unit $e$. When we change the path integral variables to relabel the combined gauge field into $A_\mu^{\phantom{x}}-e A_\mu^{\textrm{em}} \to -e A_\mu^{\phantom{x}}$, we obtain the usual Gaussian form of the effective Lagrangian density:
\begin{equation}
\mathcal{L} = \frac{|\psi|^2}{2}(j_{\mu} - eA_{\mu})(j^{\mu} - e A^{\mu})
  - \frac{1}{16\pi} F_{\mu\nu}F^{\mu\nu} + \mathcal{L}_{\textrm{t}} \nonumber
\vspace{-0.18in}
\end{equation}
\begin{equation}
\mathcal{L}_{\textrm{t}} = -\frac{e^2 \nu^{\textrm{m}}}{8\pi}\widetilde{F}^{\mu\nu}F_{\mu\nu} = \frac{e^2 \nu^{\textrm{m}}}{2\pi}\,{\bf E}{\bf B} \ .
\end{equation}
The electric $\bf E$ and magnetic $\bf B$ fields
\begin{equation}
F^{i0} = E^{i}\quad,\quad F^{ij} = -\epsilon^{ijk}B^{k}
\end{equation}
are the physical fields shifted by the emergent fields of the incompressible quantum liquid.

Let us first explore the low-energy electrodynamics in the bulk of an incompressible quantum liquid. If we rewrite the field equation (\ref{eq-mot-1}) using the charge density and current density components of $J^\mu = e(\rho,{\bf j})$, we obtain in Gaussian units
\begin{eqnarray}\label{ME-total-1}
\rho &=& \frac{1}{4\pi} \boldsymbol{\nabla}{\bf E}+\frac{\alpha\nu^{\textrm{m}}}{2\pi}\boldsymbol{\nabla}{\bf B} \\
{\bf j} &=& \frac{1}{4\pi} \left(\boldsymbol{\nabla}\times{\bf B}-\frac{\partial{\bf E}}{\partial t}\right)
         -\frac{\alpha\nu^{\textrm{m}}}{2\pi}\left(\boldsymbol{\nabla}\times{\bf E}+\frac{\partial{\bf B}}{\partial t}\right) \nonumber \ ,
\end{eqnarray}
where $\alpha = e^2/\hbar c$ ($\hbar=c=1$) is the fine-structure constant. The usual conservation law is satisfied for the total currents:
\begin{equation}\label{ME-total-2}
\frac{\partial\rho}{\partial t}+\boldsymbol{\nabla}{\bf j} = 0 \ .
\end{equation}
The equations (\ref{ME-total-1}) include both the emergent and physical gauge fields, and likewise the background and induced charge currents. Note, however, that the corrections proportional to $\nu^{\textrm{m}}$ come only from the \emph{emergent} gauge field with compact regularization, because the physical gauge field obeys Ampere and Faraday laws. When no external electromagnetic fields are applied, the background current $(\rho_0^{\phantom{'}}, {\bf j}_0^{\phantom{'}}) = \left(\frac{\alpha\nu^{\textrm{m}}}{2\pi}\boldsymbol{\nabla}{\bf B}, 0\right)$ is related only to the emergent gauge field. We will subtract this background and reinterpret (\ref{ME-total-1}) below as a relationship between the \emph{induced} currents and perturbed electromagnetic fields. Even though field perturbations are driven externally, they still include the contribution of the emergent gauge fields due to fractional charge-monopole attachment.

If there is no \emph{induced} current flow (${\bf j}=0$) or charge density ($\rho=0$) in the bulk, we find:
\begin{eqnarray}\label{eq-mot-ME}
&& \qquad\qquad\quad {\bf E} = -2\alpha\nu^{\textrm{m}} {\bf B} + \boldsymbol{\nabla}\times\boldsymbol{\alpha} \\
&& \left\lbrack 1+\left(2\alpha\nu^{\textrm{m}}\right)^{2}\right\rbrack {\bf B}-2\alpha\nu^{\textrm{m}}
           \boldsymbol{\nabla}\times\boldsymbol{\alpha}=\frac{\partial\boldsymbol{\alpha}}{\partial t}+\boldsymbol{\nabla}\alpha^{0} \nonumber
\end{eqnarray}
The parameter $\alpha^\mu = (\alpha^0, \boldsymbol{\alpha})$ can be considered a ``dual'' gauge field. When $\boldsymbol{\nabla}\times\boldsymbol{\alpha}=0$, the electric field has no curl and becomes proportional to the magnetic field, with a fractionally quantized proportionality constant $2\alpha\nu^{\textrm{m}}$. This is magnetoelectric effect: an applied electric field will induce magnetization, and an applied magnetic field will induce polarization. The induced magnetization and polarization are captured here through a bulk emergent electromagnetic field, but actually originate from the physical response at the system boundary. Most generally, the absence of currents $\rho=0$, ${\bf j}=0$ relates the full electromagnetic field to the ``dual'' gauge field $\alpha^\mu$
\begin{eqnarray}
{\bf E} &=& \frac{1}{1+\left(2\alpha\nu^{\textrm{m}}\right)^{2}}\left\lbrack (\boldsymbol{\nabla}\times\boldsymbol{\alpha})
    -2\alpha\nu^{\textrm{m}} \left(\frac{\partial\boldsymbol{\alpha}}{\partial t}+\boldsymbol{\nabla}\alpha_{0}\right)\right\rbrack \nonumber \\
{\bf B} &=&\frac{1}{1+\left(2\alpha\nu^{\textrm{m}}\right)^{2}}\left\lbrack 2\alpha\nu^{\textrm{m}}
    (\boldsymbol{\nabla}\times\boldsymbol{\alpha})+\left(\frac{\partial\boldsymbol{\alpha}}{\partial t}+\boldsymbol{\nabla}\alpha_{0}\right)\right\rbrack \nonumber
\end{eqnarray}
An electromagnetic wave $\boldsymbol{\alpha} \propto e^{i({\bf k}{\bf x}-\omega t)}$, $\boldsymbol{\alpha}\perp{\bf k}$, $\alpha^0=0$, 
\begin{eqnarray}
-i{\bf E} &=& \frac{{\bf k}\times\boldsymbol{\alpha}+2\alpha\nu^{\textrm{m}}\omega\boldsymbol{\alpha}}{1+\left(2\alpha\nu^{\textrm{m}}\right)^{2}} \\[0.07in]
-i{\bf B} &=&\frac{2\alpha\nu^{\textrm{m}}({\bf k}\times\boldsymbol{\alpha})-\omega\boldsymbol{\alpha}}{1+\left(2\alpha\nu^{\textrm{m}}\right)^{2}} \nonumber
\end{eqnarray}
only has a Faraday-rotated polarization inside the system by the angle $\arctan(2\alpha\nu^{\textrm{m}})$, but otherwise propagates conventionally with dispersion $k^2=\omega^2$ (polarization effects due to bound charges are not included in this analysis). In the presence of static current flow, we obtain an anomalous Ampere law:
\begin{equation}
4\pi {\bf j}=\left\lbrack 1+\left(2\alpha\nu^{\textrm{m}}\right)^{2}\right\rbrack (\boldsymbol{\nabla}\times{\bf B}) \ .
\end{equation}

It should be pointed out that the response derived here is a \emph{classical} approximation that works best in the limit $|\nu^{\textrm{m}}|\ll 1$. Otherwise, quantum correction are significant. The part of the action $S_{\textrm{t}} = \int dt d^3x \, \mathcal{L}_{\textrm{t}}$ obtained from (\ref{top-term-ME}) is topologically quantized \cite{Qi2009, Vazifeh2010}, so that all aspects of electromagnetic response are ultimately functions of $\exp(2\pi i \nu^{\textrm{m}})$.

Relationships and phenomena analogous to the magnetoelectric effect can also be derived for spin currents in the cases of topological orders governed by (\ref{top-term-alt}). However, such phenomena involve the emergent non-Abelian gauge field at rank 1, without an external physical counterpart that could be manipulated experimentally. At least, the monopole-related magnetoelectric effect can arise from the spin-orbit coupling when monopoles are bound to hedgehogs via Zeeman effect.

Now consider the contributions to electrodynamics from higher-energy degrees of freedom, which are beyond reach of the fractionalization effective theory. This response may be subject to a quantum anomaly, depending on the details of the microscopic particle dispersion. A quantum anomaly represents the absence of a Lagrangian density symmetry in the regularized action's integration measure. Therefore, the response due to a quantum anomaly cannot be obtained from a stationary action condition and must be extracted by integrating out fields in the path integral. The quantum anomaly of three-dimensional topological systems can lead to a correction of the topological term
\begin{equation}
\Delta \mathcal{L}_{\textrm{t}} = \frac{\theta}{(2\pi)^2}\,{\bf E}{\bf B} \ ,
\end{equation}
which effectively yields the shift $\nu^{\textrm{m}} \to \nu^{\textrm{m}} + \theta/2\pi$ in the response field equations. However, all aspects of topological order, such as fractionalization, are shaped at low energies and determined by the original unshifted filling factor $\nu^{\textrm{m}}$ because the quantum anomaly is a high-energy regularization feature.

\section{Conclusions, possible physical realizations and future directions}\label{secConclusions}

In this paper, we have established the existence of stable incompressible quantum liquids with topological order in general $d$-dimensional systems of spinors fields. Independent topological orders can be driven by the fluctuations of hedgehogs and monopoles. They generalize fractional quantum Hall states in many ways, but also have novel properties in $d\ge 3$ dimensions. We calculated the topological ground state degeneracy, and showed that it survives all sufficiently weak perturbations even when all other common signatures of topological order become washed out. The topological orders in $d\ge 3$ are sharply defined phases at low finite temperatures, in contrast to quantum Hall liquids in $d=2$ dimensions. Free topological defects can bind fractional amounts of charge or spin, but the ultimate sharp quantization of fractional quantum numbers also depends on symmetries and can be spoiled. We presented a preliminary discussion of the long-range entanglement in these topologically ordered phases, and considered fractional braiding operations. We also briefly explored the characteristic topological responses of these unconventional states, including the anticipated fractional magnetoelectric and Kerr effects.

There are many important questions left for future research. First of all, the phenomenology of quantum entanglement in $d\ge 3$ dimensions should be explored in great detail. The present analysis did not identify topologically protected braiding data that characterize the scrambled topological orders of hedgehogs. We also argued in Sections \ref{secBraid} and \ref{secMicroPart} that the non-local entanglement of electron's spin among multiple fractional quasiparticles can give rise to non-trivial \emph{dynamically protected} braiding operations -- possibly non-Abelian despite the fact that the topological orders discussed here are Abelian (the \emph{topologically protected} particle-loop braiding is Abelian). This conclusion was based on general considerations regardless of the dimensionality $d\ge 2$, and awaits a mathematical description in concrete terms (e.g. a classification of non-trivial spin entanglement and operations due to the quantum motion of hedgehogs in fractionalized $d=3$ chiral magnets). Non-Abelian dynamically protected braiding of spin could perhaps provide a fertile platform for universal topological quantum computing given the ability to externally manipulate and measure magnetic moments, and the three-dimensional space to perform braiding operations.

Another obvious subject left for future research is the nature of soft boundary modes in $d\ge 3$ topologically ordered phases. We have only rudimentarily established the existence of such modes and their Dirac-like spectra, but it would be extremely interesting to fully understand the consequences of bulk fractionalization on the boundary dynamics. Among anticipated phenomena \cite{Cho2010, Metlitski2013, Bonderson2013, Wang2014c, Wang2015b, Cappelli2016, Fradkin2017, Fradkin2017b} are surface topological order, fractional parity anomaly, etc. This is also important for practical reasons, because the boundary modes are accessible to experiments. A deeper analysis of unconventional bulk responses to external probes is also important, and new ideas are needed to envision unambiguous methods to detect topological order and measure its properties.

Undoubtedly, it would be interesting to construct microscopic models that realize some of the topological orders considered here. Solving such models would be difficult in $d>2$ dimensions, especially in the continuum limit. Lattice models will be more tractable, especially in the context of monopoles: $d=3$ Hamiltonians analogous to the Hofstadter problem in $d=2$ can be readily constructed and perhaps analyzed numerically in the strongly interacting regime. Such models, however, are not realistic and would serve mainly as a proof of concept. More realistic models would need to feature frustrated spin dynamics, and would be much harder to solve. At least, such models can be constructed by the requirement that their continuum limit reduce to the theory considered here. It should be pointed out that a lattice formulation of dynamics introduces its own constraints on topological orders by limiting the topological charge that can be stored and preserved in a finite volume.

This paper was focused on the essential phenomenology of topological order. Forthcoming sequel papers will focus on making connections between the fundamental picture presented here and concrete topological materials. It will be shown that the highest rank gauge flux of the hedgehog and monopole gauge fields is directly related to a Berry flux in momentum space. In other words, there is a generalization of the famous Thouless-Kohmoto-Nightingale-den Nijs (TKNN) formula \cite{Thouless1982} to all homotopy classes $\pi_n(S^n)$, and the present theory provides a universal real-space description of the known $d=3$ topological materials. Another forthcoming study will explore the dynamics of spins in the presence of spin-orbit coupling and possible mobile electrons. Its purpose is to provide the bridge to microscopic models of magnetic topological materials, and lay down a more concrete foundation for the phenomenological picture of hedgehog dynamics presented here.

The candidate materials that might realize some of the fractional states we discussed include chiral magnets and strongly correlated topological insulators or semimetals. Specifically, chiral magnets can be related to incompressible quantum liquids of hedgehogs in the same manner as superconductors are related to fractional quantum Hall liquids. A type-2 superconductor can host gapped localized vortex defects, which form an Abrikosov lattice in an externally applied magnetic field. If quantum fluctuations melt this lattice, e.g. because vortices become almost as dense as particles, then the ensuing quantum vortex liquid state is naturally a fractional quantum Hall liquid \cite{Cooper2001}. By analogy, the topological defects of a $d=3$ magnet are hedgehogs, which can form a lattice in a chiral magnetically ordered phase \cite{Fujishiro2019}. Quantum melting of such a hedgehog lattice could produce a topologically ordered quantum liquid of the kind analyzed in this paper. Note that even a truly periodic magnetic order with hedgehogs in a \emph{microscopic crystal} is interesting for this scenario -- its U(1) analogue would be the quantum melting of a commensurate vortex lattice in a crystal, which produces a Hofstadter spectrum and still enables quantum Hall or Chern insulator states. Related structures of magnetic topological defects are skyrmion lattices in $d=2$ dimensions \cite{Tokura2013}, although they are not topologically protected against quantum fluctuations. There are currently no known chiral magnets that approach the quantum limit, but some may be close.

Existing $d=3$ topological insulators are already incompressible quantum liquids of hedgehog topological defects, albeit without topological order. In this sense, they are analogous to integer quantum Hall liquids. The topological states analyzed in this paper generally break the time-reversal and mirror symmetries, so additional provisions are required to implement the non-trivial topology with time-reversal symmetry. We anticipate that this requires a lattice formulation of the theory -- in the same sense as it is possible to construct a time-reversal-invariant quantum ``Hall'' state by pushing a magnetic $\pi$-flux through every lattice plaquette. Strong interactions are then also needed to stabilize topological order with fractional charged quasiparticles. Interactions are strong in quantum Hall liquids due to the flatness of the Landau-level bands. Approximate band flatness can also be achieved in $d=3$ systems, but this is not a necessary condition for topological order. Some interacting Weyl and Luttinger \cite{Moon2013} semimetals, characterized by similar Berry fluxes as topological insulators, can perhaps become unstable in the presence of interactions or additional degrees of freedom such as magnetic moments. If an energy gap opens in the energy vicinity of the semimetal's nodal points, then the Berry flux is not removed but only redistributed. Hence it is possible to obtain a topological insulator, perhaps even with topological order and fractional excitations.

Spin liquids are another system of interest in the context of this work, but this is not a new story \cite{Savary2016}. The theory presented here regards spin liquids as phases of purely magnetic degrees of freedom with U(1) symmetry, and hence sharply distinguishes them from the similar phases that (also) involve charge. Topological order involves attaching spin to the monopoles of an emergent U(1) gauge field, and the simplest fractional quasiparticle is a fermionic spinon. The U(1) spin liquid in $d=3$ dimensions \cite{Hermele2004a, Savary2016}, with gapped matter and monopoles but gapless photons, is the Abelian version of the phase we labeled $C_1H_2$ in Section \ref{secDynPhaseDiag}. If one identified topological order by the conservation of abundant topological defects, then the U(1) spin liquid would not have it. A $d=3$ topological order is present only in the phases we labeled $C_1T_2$, which are fully gapped and generalize quantum Hall liquids. These phases can either break or respect the time-reversal (TR) symmetry, and may be characterized as ``chiral'' and Z$_n$ \cite{Swingle2011, Fradkin2017} spin liquids respectively. Nevertheless, we found in Section \ref{secMicroPart} that spin-monopole attachment produces gapped spin liquids that have fundamentally different excitations than the resonant-valence-bond spin liquids. Also related to these are symmetry-protected topological (SPT) phases of bosonic degrees of freedom in magnets \cite{Vishwanath2013}.

Quantum spin-ice materials like Tb$_2$Ti$_2$O$_7$, Pr$_2$Sn$_2$O$_7$, Pr$_2$Zr$_2$O$_7$, NaCaNi$_2$F$_7$ are promising candidates for realizing U(1) spin liquids \cite{Gingras2014, Plumb2017}, and possibly also topological orders with spin-monopole attachment. The essential low-energy Hamiltonian describes their dynamics as a compact U(1) gauge theory \cite{Hermele2004a}. The deconfined phase of this model is a stable U(1) spin liquid. A more realistic model may need to include spin interactions generated by the spin-orbit coupling \cite{Thompson2011, Ross2011}. If such interactions are strong enough, topological order also becomes an option. The spin-orbit coupling is generally able to stimulate the appearance of point topological defects, even in purely spin systems (to be shown in a forthcoming paper). The role of matter field is played by the spins themselves in the spin-ice materials, so the gauge ``charge'' can be accumulated simply by applying an external magnetic field. When both ``charge'' and ``monopoles'' are present in the ground state, there is a chance that interactions may bind them into fractional quasiparticles. The ensuing topologically ordered phase could perhaps be experimentally seen as an incompressible magnetization plateau state in a non-saturating magnetic field, with gapless neutral modes at the crystal boundary.

\subsection{Charge fractionalization in quantum chromodynamics (QCD)}

At the end, we entertain the possibility that the fractional quantization of quarks' charge in atomic nuclei could be a glimpse of a topological order discussed in this paper. Such a view is different than the traditional standard model picture of quarks as elementary particles, but it has appealing features. We will assume that fundamental quarks are elementary fermions that have the SU(3) color ``charge'' and carry the same unit $e=1$ of the electric U(1) charge just like electrons. This fits the essential idea raised in this paper that gauge fields emerge from the dynamics of topological defects -- hence all elementary particle fields coupled to the same gauge field should carry the same charge. The integer charge assignment considered here is color-independent and different from the Han-Nambu assignment \cite{Nambu1965, Jaffe1981}.

How and why can the charge of fundamental quarks fractionalize into the observed amounts $2/3$ and $-1/3$ for ``up'' and ``down'' quarks respectively? For simplicity, we can work with the most basic QCD Lagrangian density \cite{Peskin1995}
\begin{equation}
\mathcal{L}=\bar{u}i\cancel{D}u+\bar{d}i\cancel{D}d = \bar{q}_{L}i\cancel{D}q_{L}+\bar{q}_{R}i\cancel{D}q_{R}
\end{equation}
of massless $u$ and $d$ quarks, where $\cancel{D} = \gamma^\mu (\partial_\mu - i \mathcal{A}_\mu)$ is the Dirac operator gauged with all the gauge fields that quarks couple to, and
\begin{equation}
q_{L}=\frac{1-\gamma^{5}}{2}q\quad,\quad q_{R}=\frac{1+\gamma^{5}}{2}q
\end{equation}
are the left-handed (L) and right-handed (R) isospin quark spinors $q^{\textrm{T}} = (u,d)$. The chirality $\gamma^5 = i \gamma^0 \gamma^1 \gamma^2 \gamma^3$ is a good quantum number for ungauged massless Dirac fermions, which leads to the conservation of isospin singlet and triplet chiral currents
\begin{equation}
j^{\mu5}=\bar{q}\gamma^{\mu}\gamma^{5}q \quad,\quad j^{\mu5a}=\bar{q}\gamma^{\mu}\gamma^{5}\tau^{a}q
\end{equation}
at the classical level. Here, $\tau^a$ are the isospin SU(2) generators ($a\in\lbrace 1,2,3 \rbrace$). However, the Adler-Bell-Jackiw quantum anomaly \cite{Adler1969, Bell1969} breaks the conservation of chiral currents in the gauge theory. Specifically, the ``up'' and ``down'' isospin flavors independently experience the same chiral anomaly:
\begin{equation}
\partial_{\mu}j^{\mu5a}=-\frac{e^{2}}{32\pi^{2}}\delta^{a,3}\epsilon^{\mu\nu\alpha\beta}F_{\mu\nu}F_{\alpha\beta} \ .
\end{equation}
This response can be equivalently reproduced from a topological Lagrangian term \cite{Qi2008b}. Therefore, a quantum anomaly is related to the topological Berry flux of a gapped state -- opening up a gap creates a topological insulator. Indeed, spontaneous and explicit chiral symmetry breaking in the full QCD gaps out all quarks, mesons and baryons. The momentum-space Berry flux corresponds to a non-zero background flux of a hedgehog/monopole rank 2 gauge field in real space. The topological insulator of Dirac fermions is invariant under time-reversal, so one cannot collect a finite Berry flux in a band from the plain momentum-space spin texture of all states. A finite Berry flux is extracted from the spin helicity $\boldsymbol{\sigma} \hat{\bf p}$ of chiral currents instead of charge currents. Since the chirality and helicity of a massless positive-energy Dirac particle are identical, the chiral Berry flux (the difference of the Berry fluxes of left-handed and right-handed particles) is positive in the ``conduction'' band and each particle of positive chirality is matched by a spin hedgehog. The chirality of a negative-energy particle is opposite from its helicity, implying a negative chiral Berry flux in the ``valence'' band. But, an antiparticle is then again matched by a spin hedgehog (created by the removal of an antihedgehog).

\begin{figure}
\subfigure[{}]{\includegraphics[height=2.7in]{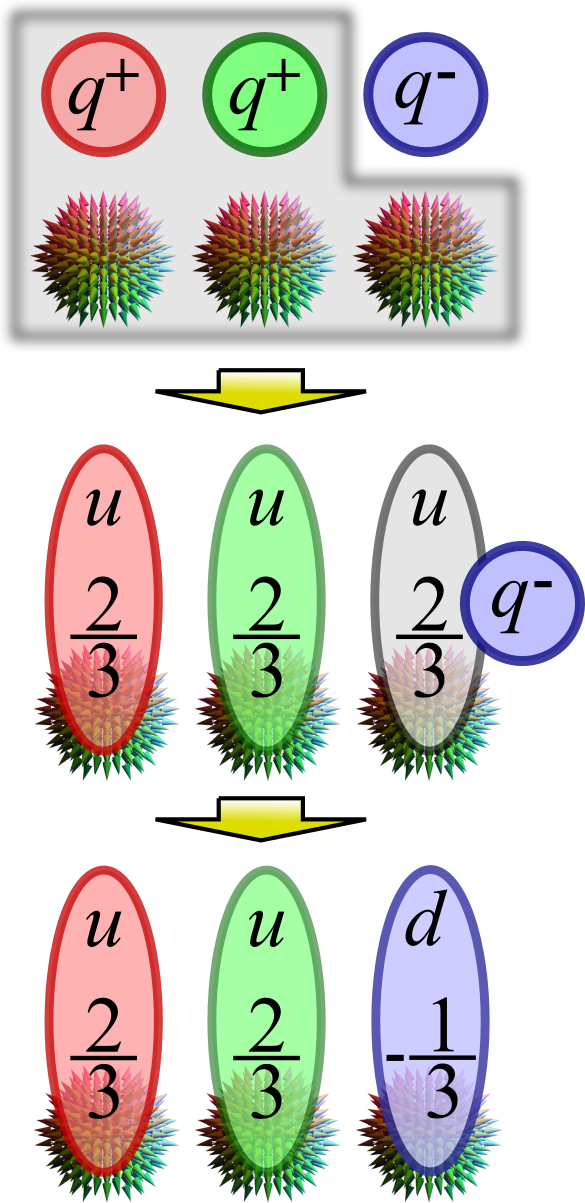}}\hspace{0.3in}
\subfigure[{}]{\includegraphics[height=2.7in]{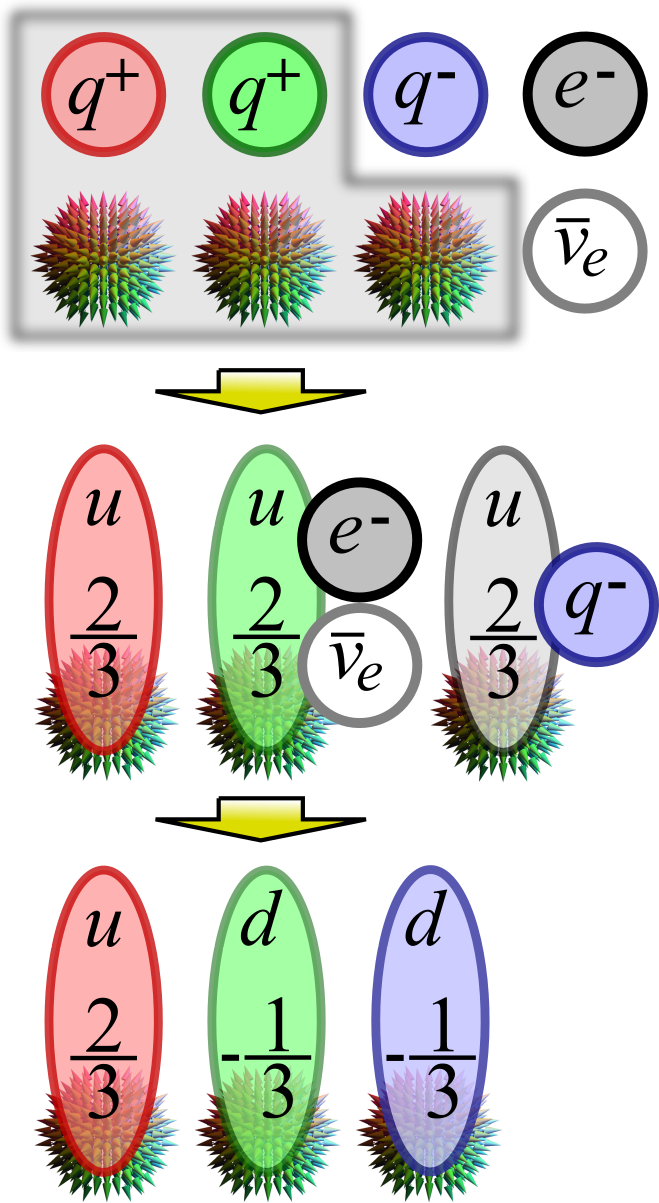}}
\caption{\label{NuclModel}Fractionalization toy models of a proton (a) and a neutron (b). The assumption is that quarks fundamentally have the same charge unit as electrons, but the strong nuclear force causes charge fractionalization by compressing quarks to relative distances smaller than the topological defect confinement length. The relevant topological defects are spin hedgehogs and U(1) monopoles bound to them. They emerge from the relativistic nature of Dirac fields and proliferate in the ground state due to the chiral/axial quantum anomaly. The actual $u$ and $d$ quarks are seen as fractional quasiparticles of an incompressible quantum liquid droplet where particle/antiparticle symmetry is locally broken.}
\end{figure}

We can now consider what might be happening inside a proton. As a color-neutral object, a proton must be made from three quarks. However, we assume that the fundamental quarks have electron's unit charge, and then we need two $e=+1$ quarks and one $e=-1$ antiquark. Being relativistic Dirac fermions, these particles bring three hedgehogs into the makeup of a proton. The hedgehogs intrinsically match the chiral currents, but since a fundamental quark carries both charge and chirality, we can equivalently associate charge to hedgehogs. We assume that the strong nuclear force compresses the quarks into such a small volume that the distance between them does not exceed the confinement length of hedgehog-antihedgehog pairs (which must be finite in the QCD ground state, or else quarks themselves would be deconfined). This is enough for a proton to become a droplet of an incompressible quantum liquid. Figure \ref{NuclModel} depicts an attachment of the two positive unit-charges to three hedgehogs, which creates three $u$ quarks as fractional quasiparticles with charge $2/3$. One of these $u$ quarks forms a tight bound state with the remaining fundamental antiquark, thus producing a $d$ quark with charge $-1/3$. This is a more complex composite object than the $u$ quark, hence naturally more massive and less stable. We can similarly envision the emergence of a neutron, starting from the stable particles that a free neutron decays into. First we fractionalize the positive charge and form three $u$ quarks just as in the case of a proton. Then, we convert two $u$ quarks into $d$ quarks by binding with the available negative unit charges. The schematic model does not do justice depicting the spin and color distribution among the nucleon constituents. Note that the hedgehogs associated with the electron and neutrino do not directly couple to the quarks and should not contribute to quark fractionalization.

Of course, more analysis is needed to verify this picture. For example, why should the hedgehogs attach to and fractionalize only the positive-charge fundamental quarks? This seems to involve a sort of Cooper pairing between positive fundamental quarks prior to fractionalization (mediated by the third quark?). The answer is hidden in the details of dynamics. Locally, the proton's matter breaks the particle-antiparticle symmetry, so a bias in the flux-particle attachment is not surprising -- and, it is even important for the stability of topological order. The effective topological Lagrangian $\mathcal{L}_{\textrm{t}}$ describing this situation in protons could have the form (\ref{effFractTheory}) with 
\begin{equation}
K = \left(\begin{array}{ccc} 3 & 0 & 0\\ 0 & 3 & 0\\ 0 & 0 & 3 \end{array}\right) \quad,\quad q=\left(\begin{array}{c} 2\\ 2\\ -1 \end{array}\right)
\end{equation}
and $\nu=3$. The fractional quasiparticle flavors correspond to three color states. The fractional charges in different colors are $(\frac{2}{3}, \frac{2}{3}, -\frac{1}{3})$ according to (\ref{KJ}), and $\nu=3$ reflects the degeneracy of color states (there is one Dirac fermion state per hedgehog in a ``band'', for each color and spin; a ``band'' is not fully populated at least due to spin degeneracy, so the incompressibility requires certain short-range repulsive interactions between fundamental quarks). Note that $\mathcal{L}_{\textrm{t}}$ written in (\ref{effFractTheory}) breaks parity P and time-reversal T, but stays invariant under PT, so it does not violate any symmetry of the standard model. A topological term that respects P and T can also be constructed by inserting a chirality $i \bar{\psi} \gamma^5 \psi$ factor in the definition of $\mathcal{L}_{\textrm{t}}$. As a matter of principle, if the fractionalization proposal is correct, it might be possible to obtain other charge fractions in different circumstances, constrained by color-neutrality. Quark-gluon plasmas might exist in many varieties of topologically ordered incompressible quantum liquids, perhaps as rich as fractional quantum Hall states.

\section{Acknowledgements}\label{secAck}

I am very grateful for eye-opening and inspiring discussions with T. Senthil and Yi Li. This research was partly supported by the Department of Energy, Basic Energy Sciences, Materials Sciences and Engineering Award DE-SC0019331. A part of this work was completed at the Aspen Center for Physics, which is supported by National Science Foundation grant PHY-1607611.

\newpage
\appendix

\section{Hedgehog gauge field}\label{appHedgehogs}

Here we derive (\ref{singA2b}) from (\ref{singA2a}) using the representation (\ref{n-vs-theta}) of the vector field $\hat{\bf n}({\bf x})$ in terms of angles $\theta_1,\dots,\theta_n$, where $n=d-1$ for notational convenience. The derivatives of $\hat{n}^{a}$ are:
\begin{equation}
\frac{\partial\hat{n}^{a}}{\partial\theta_{b}}=\hat{n}^{a}X_{a,b} \quad,\quad
  X_{a,b}=\cot\theta_{b}\,\delta_{b\le a}-\tan\theta_{b}\,\delta_{b,a+1} \ . \nonumber
\end{equation}
(here, the repeated index $a$ is not summed over) so that the antisymmetric gauge field tensor (\ref{singA2a}) is:
\begin{eqnarray}
A_{\mu_{1}\cdots\mu_{n}} &=& \frac{1}{n!}\epsilon_{a_{0}a_{1}\cdots a_{n}}\;\hat{n}^{a_{0}}\prod_{i=1}^{n}\frac{\partial\hat{n}^{a_{i}}}{\partial x_{\mu_{i}}} \\
  &=& \frac{\epsilon_{a_{0}a_{1}\cdots a_{n}}}{n!}\,
    \hat{n}^{a_{0}}\prod_{i=1}^{n}\frac{\partial\hat{n}^{a_{i}}}{\partial\theta_{b_{i}}}\frac{\partial\theta_{b_{i}}}{\partial x_{\mu_{i}}} \nonumber \\
  &=& A\,\epsilon_{a_{0}a_{1}\cdots a_{n}} \prod_{i=1}^n X_{a_i,b_i}\frac{\partial\theta_{b_{i}}}{\partial x_{\mu_{i}}} \ , \nonumber
\end{eqnarray}
where
\begin{equation}
A=\frac{1}{n!} \prod_{j=1}^{n}\cos\theta_{j}\,(\sin\theta_{j})^{n+1-j} \ .
\end{equation}
Since all indices $a_i$ are different, their values are all possible permutations $a_{i}=\mathcal{P}(i)$ of $(0,1,\dots,n)$. Similarly, $b_i=P(i)$ are all possible permutations of $(1,\dots,n)$. The parity $(-1)^{\mathcal{P}}$ of a permutation $\mathcal{P}$ is $\epsilon_{a_{0}a_{1}\cdots a_{n}}$, and the parity $(-1)^{P}$ of a permutation $P$ is $\epsilon_{b_{1}\cdots b_{n}}$. Then:
\begingroup
\allowdisplaybreaks
\begin{eqnarray}\label{appA1}
A_{\mu_{1}\cdots\mu_{n}} &=& A\sum_{\mathcal{P}}(-1)^{\mathcal{P}}\prod_{i=1}^{n}X_{\mathcal{P}(i),b_{i}}\frac{\partial\theta_{b_{i}}}{\partial x_{\mu_{i}}} \\
  &=& A\sum_{\mathcal{P}}(-1)^{\mathcal{P}}\sum_{P}\prod_{i=1}^{n}X_{\mathcal{P}(P^{-1}(i)),i}\frac{\partial\theta_{P(i)}}{\partial x_{\mu_{i}}} \nonumber \\
  &=& A\sum_{P}(-1)^{P}\sum_{\mathcal{P}-P}(-1)^{\mathcal{P}-P}\prod_{i=1}^{n}X_{(\mathcal{P}-P)(i),i}\frac{\partial\theta_{P(i)}}{\partial x_{\mu_{i}}} \nonumber \\
  &=& \left\lbrack A\sum_{\mathcal{P}'}(-1)^{\mathcal{P}'}\prod_{i=1}^{n}X_{\mathcal{P}'(i),i}\right\rbrack \epsilon_{b_{1}\cdots b_{n}} 
          \prod_{i=1}^{n}\frac{\partial\theta_{b_{i}}}{\partial x_{\mu_{i}}} \nonumber \\
  &=& A\det(X)\times\epsilon_{b_{1}\cdots b_{n}}\prod_{i=1}^{n}\frac{\partial\theta_{b_{i}}}{\partial x_{\mu_{i}}} \nonumber \ .
\end{eqnarray}
\endgroup
We also introduced a permutation $\mathcal{P}'\equiv\mathcal{P}-P$ of $(0,\dots,n)$ given by $\mathcal{P}'(i)=\mathcal{P}(P^{-1}(i))$ for $i\neq0$ and $\mathcal{P}'(0)=\mathcal{P}(0)$, whose parity is $(-1)^{\mathcal{P}'}=(-1)^{\mathcal{P}}(-1)^{P}$. Finally, we interpreted the expression in the square brackets as the determinant of the $n+1$ dimensional matrix $X$:
\begingroup
\allowdisplaybreaks
\begin{widetext}
\begin{eqnarray}
\det(X) &=& \left\vert \begin{array}{ccccccc}
1 & X_{0,1} & X_{0,2} & \cdots & X_{0,n-1} & X_{0,n}\\
1 & X_{1,1} & X_{1,2} & \cdots & X_{1,n-1} & X_{1,n}\\
1 & X_{2,1} & X_{2,2} & \cdots & X_{2,n-1} & X_{2,n}\\
1 & X_{3,1} & X_{3,2} & \cdots & X_{3,n-1} & X_{3,n}\\
\vdots & \vdots & \vdots & \ddots & \vdots & \vdots\\
1 & X_{n-1,1} & X_{n-1,2} & \cdots & X_{n-1,n-1} & X_{n-1,n}\\
1 & X_{n,1} & X_{n,2} & \cdots & X_{n,n-1} & X_{n,n}
\end{array}\right\vert
=
\left\vert \begin{array}{ccccccc}
1 & -\tan\theta_{1} & 0 & 0 & \cdots & 0 & 0\\
1 & \cot\theta_{1} & -\tan\theta_{2} & 0 & \cdots & 0 & 0\\
1 & \cot\theta_{1} & \cot\theta_{2} & -\tan\theta_{3} & \cdots & 0 & 0\\
1 & \cot\theta_{1} & \cot\theta_{2} & \cot\theta_{3} & \cdots & 0 & 0\\
\vdots & \vdots & \vdots & \vdots & \ddots & \vdots & \vdots\\
1 & \cot\theta_{1} & \cot\theta_{2} & \cot\theta_{3} & \cdots & \cot\theta_{n-1} & -\tan\theta_{n}\\
1 & \cot\theta_{1} & \cot\theta_{2} & \cot\theta_{3} & \cdots & \cot\theta_{n-1} & \cot\theta_{n}
\end{array}\right\vert \nonumber \\[0.1in]
&=&
\prod_{i=1}^{n}\frac{1}{\cos^{2}\theta_{i}}\left\vert \begin{array}{cccccccc}
1 & 0 & 0 & 0 & \cdots & 0 & 0 & 0\\
1 & \cot\theta_{1} & 0 & 0 & \cdots & 0 & 0 & 0\\
1 & \cot\theta_{1} & \cot\theta_{2} & 0 & \cdots & 0 & 0 & 0\\
1 & \cot\theta_{1} & \cot\theta_{2} & \cot\theta_{3} & \cdots & 0 & 0 & 0\\
\vdots & \vdots & \vdots & \vdots & \ddots & \vdots & \vdots & \vdots\\
1 & \cot\theta_{1} & \cot\theta_{2} & \cot\theta_{3} & \cdots & \cot\theta_{n-2} & \cot\theta_{n-1} & 0\\
1 & \cot\theta_{1} & \cot\theta_{2} & \cot\theta_{3} & \cdots & \cot\theta_{n-2} & \cot\theta_{n-1} & \cot\theta_{n}
\end{array}\right\vert 
=
\prod_{i=1}^{n}\frac{1}{\cos^{2}\theta_{i}}\cot\theta_{i}=\prod_{i=1}^{n}\frac{1}{\sin\theta_{i}\cos\theta_{i}} \ . \nonumber 
\end{eqnarray}
\end{widetext}
\endgroup
Substituting in (\ref{appA1}) yields the formula (\ref{singA2b}):
\begin{eqnarray}
A_{\mu_{1}\cdots\mu_{n}} &=& \prod_{j=1}^{n}\frac{\cos\theta_{j}\,(\sin\theta_{j})^{n+1-j}}{\sin\theta_{j}\cos\theta_{j}} \frac{1}{n!}
      \epsilon_{b_{1}\cdots b_{n}}\prod_{i=1}^{n}\frac{\partial\theta_{b_{i}}}{\partial x_{\mu_{i}}} \nonumber \\
  &=& \prod\limits _{j=1}^{n}(\sin\theta_{j})^{n-j} \frac{1}{n!}\epsilon_{b_{1}\cdots b_{n}}\prod_{i=1}^{n}\frac{\partial\theta_{b_{i}}}{\partial x_{\mu_{i}}} \ .
\end{eqnarray}

\section{Non-Abelian Maxwell terms in the effective Lagrangian}\label{appNonAbelianMaxwell}

The rank $n$ gauge field obtained from the vector field $\hat{\bf n}$ by a singular gauge transformation (\ref{singN1}) is:
\begin{equation}\label{NA-sing-gauge-field}
A_{\lambda_1\cdots\lambda_n}^{a_{n+1}\cdots a_{d-1}} = \frac{1}{n!} \epsilon_{a_0\cdots a_{d-1}}
  \hat{n}^{a_0}\prod_{i=1}^{n} (\partial_{\lambda_{i}}\hat{n}^{a_{i}}) \ .
\end{equation}
Smooth infinitesimal deformations
\begin{equation}
\hat{\bf n} \to \hat{\bf n} + \delta \hat{\bf n}
\end{equation}
generate gauge transformations of $A_{\lambda_1\cdots\lambda_n}^{a_{n+1}\cdots a_{d-1}}$ that must leave invariant the rank $n$ Maxwell term in the Lagrangian density:
\begin{equation}
\mathcal{L}_{\textrm{M}n} = \frac{1}{2g_n^2} \mathcal{J}_{\mu_{1}\cdots\mu_{d-n}}^{a_{n+1}\cdots a_{d-1}}
  \mathcal{J}_{\mu_{1}\cdots\mu_{d-n}}^{a_{n+1}\cdots a_{d-1}} \ .
\end{equation}
Here we determine the form of the gauge fluxes $\mathcal{J}_{\mu_{1}\cdots\mu_{d-n}}^{a_{n+1}\cdots a_{d-1}}$ required by gauge invariance. It is immediately apparent that the flux cannot be a plain curl of the gauge field, since it acquires a non-zero correction under a gauge transformation:
\begin{eqnarray}
&& \delta\left(\epsilon_{\mu_{1}\cdots\mu_{d-n}\lambda_{0}\lambda_{1}\cdots\lambda_{n}}^{\phantom{,}}\partial_{\lambda_{0}}^{\phantom{,}}
      A_{\lambda_{1}\cdots\lambda_{n}}^{a_{n+1}\cdots a_{d-1}}\right) \\
&& \quad = \frac{1}{n!}\epsilon_{a_{0}\cdots a_{d-1}}\epsilon_{\mu_{1}\cdots\mu_{d-n}\lambda_{0}\cdots\lambda_{n}}\,
      \delta\prod_{i=0}^{n}\partial_{\lambda_{i}}\hat{n}^{a_{i}} \nonumber \\
&& \quad = \frac{1}{n!}\epsilon_{a_{0}\cdots a_{d-1}}\epsilon_{\mu_{1}\cdots\mu_{d-n}\lambda_{0}\cdots\lambda_{n}}
      \sum_{j=0}^{n}(\partial_{\lambda_{j}}\delta\hat{n}^{a_{j}})\prod_{i\neq j}^{0\dots n}\partial_{\lambda_{i}}\hat{n}^{a_{i}} \ . \nonumber
\end{eqnarray}
The same correction can be obtained from a quadratic form of (\ref{NA-sing-gauge-field}), so the gauge invariant flux can be expressed as:
\begin{eqnarray}\label{rank-flux-1}
&& \mathcal{J}_{\mu_{1}\cdots\mu_{d-n}}^{a_{n+1}\cdots a_{d-1}} = \epsilon_{\mu_{1}\cdots\mu_{d-n}\lambda_{0}\cdots\lambda_{n}}^{\phantom{,}}
  \Biggl\lbrack \partial_{\lambda_{0}}^{\phantom{,}}A_{\lambda_{1}\cdots\lambda_{n}}^{a_{n+1}\cdots a_{d-1}} \nonumber \\
&& \quad -\sum_{k=0}^{\lbrack\frac{n-1}{2}\rbrack} f_{d,n,k}^{a_{n+1}\cdots a_{d-1}b_{1}\cdots b_{d-k-2}c_{1}\cdots c_{d-n+k-1}} \nonumber \\
&& \qquad\qquad \times A_{\lambda_{0}\cdots\lambda_{k}}^{b_{1}\cdots b_{d-k-2}}A_{\lambda_{k+1}\cdots\lambda_{n}}^{c_{1}\cdots c_{d-n+k-1}} \Biggr\rbrack
\end{eqnarray}
using a suitable set of structure constants $f$. The bounds for the integer $k$ exhaust all possibilities for the quadratic gauge field combinations without repetitions.  The structure constants must be antisymmetric at least under any exchange among the $a$ indices, the $b$ indices, and the $c$ indices. The total number of indices in $f$ is $3d-2n-4$. We require $\mathcal{J}_{\mu_{1}\cdots\mu_{d-n}}^{a_{n+1}\cdots a_{d-1}}=0$ at all non-singular points in space-time -- a singularity gauge field (\ref{NA-sing-gauge-field}) should have a non-vanishing flux only at the positions of singularities in the field configuration $\hat{\bf n}$. This is a necessary condition for gauge invariance. A sufficient condition is obtained by also requiring that the flux transform only as a locally rotating tensor under gauge transformations. This will be achieved by simply ensuring a proper tensor structure for the flux. The necessary condition implies the following $\delta \mathcal{J}_{\mu_{1}\cdots\mu_{d-n}}^{a_{n+1}\cdots a_{d-1}}=0$ behavior under gauge transformations at non-singular points:
\begingroup
\allowdisplaybreaks
\begin{eqnarray}
&& \delta \mathcal{J}_{\mu_{1}\cdots\mu_{d-n}}^{a_{n+1}\cdots a_{d-1}} = \epsilon_{\mu_{1}\cdots\mu_{d-n}\lambda_{0}\cdots\lambda_{n}}^{\phantom{,}}\Biggl\lbrace \frac{1}{n!}\epsilon_{a_{0}\cdots a_{d-1}}\, \delta\prod_{i=0}^{n}\partial_{\lambda_{i}}\hat{n}^{a_{i}} \nonumber \\
&& \quad -\sum_{k=0}^{\lbrack\frac{n-1}{2}\rbrack}\frac{f_{d,n,k}^{a_{n+1}\cdots a_{d-1}b_{1}\cdots b_{d-k-2}c_{1}\cdots c_{d-n+k-1}}}{(k+1)!(n-k)!} \nonumber \\
&& \qquad\qquad \times \delta\Biggl\lbrack \left(\epsilon_{b_{1}\cdots b_{d}}^{\phantom{x}}\hat{n}^{b_{d-k-1}}\prod_{i=0}^{k}
      \partial_{\lambda_{i}}\hat{n}^{b_{d-k+i}}\right) \nonumber \\
&& \qquad\qquad\quad \times \left(\epsilon_{c_{1}\cdots c_{d}}^{\phantom{x}}\hat{n}^{c_{d-n+k}}
      \prod_{i=k+1}^{n}\partial_{\lambda_{i}}\hat{n}^{c_{d-n+i}}\right)\Biggr\rbrack \Biggr\rbrace = 0 \nonumber
\end{eqnarray}
\endgroup
After some relabeling of upper indices:
\begingroup
\allowdisplaybreaks
\begin{eqnarray}
&& \delta \mathcal{J}_{\mu_{1}\cdots\mu_{d-n}}^{a_{n+1}\cdots a_{d-1}} = \epsilon_{\mu_{1}\cdots\mu_{d-n}\lambda_{0}\cdots\lambda_{n}}^{\phantom{,}}\Biggl\lbrace \frac{1}{n!}\epsilon_{a_{0}\cdots a_{d-1}}\, \delta\prod_{i=0}^{n}\partial_{\lambda_{i}}\hat{n}^{a_{i}} \nonumber \\
&& \quad -\sum_{k=0}^{\lbrack\frac{n-1}{2}\rbrack}\frac{f_{d,n,k}^{a_{n+1}\cdots a_{d-1}b_{1}\cdots b_{d-k-2}c_{1}\cdots c_{d-n+k-1}}}{(k+1)!(n-k)!} \nonumber \\
&& \qquad\qquad \times \frac{\epsilon_{b_{1}\cdots b_{d-k-1}a_{0}\cdots a_{k}} \epsilon_{c_{1}\cdots c_{d-n+k}a_{k+1}\cdots a_{n}}}{(k+1)!(n-k)!} \nonumber \\
&& \qquad\qquad \times \, \delta\left(\hat{n}^{b_{d-k-1}}\hat{n}^{c_{d-n+k}}\prod_{i=0}^{n}\partial_{\lambda_{i}}\hat{n}^{a_{i}}\right) \Biggr\rbrace = 0 \nonumber
\end{eqnarray}
\endgroup
The antisymmetry under any $\lambda_{i}\leftrightarrow\lambda_{j}$ exchange imposes antisymmetric exchanges among $a_{0}\cdots a_{n}$ in the structure constant $f$ term (any contributions symmetric under $a_i\leftrightarrow a_j$ cancel out). The indices $b_{d-k-1}$ and $c_{d-n+k}$ of the $\hat{{\bf n}}$ components without derivatives are not present in the structure constant. If $b_{d-k-1}\neq c_{d-n+k}$, then the two $\epsilon$ factors (which carry all possible vector indices) either make $b_{d-k-1}$ or $c_{d-n+k}$ equal to one of $a_{0}\cdots a_{n}$, or enforce $b_{d-k-1}\in\lbrace c_{1},\cdots,c_{d-n+k-1}\rbrace$ and $c_{d-n+k}\in\lbrace b_{1},\cdots,b_{d-k-2}\rbrace$. In the former case, one takes a derivative $\hat{n}^a\partial_\lambda\hat{n}^a = \frac{1}{2}\partial_{\lambda}|\hat{{\bf n}}|^2=0$ under $\delta(\cdots)$ and the $f$ term vanishes. In the latter case, we can make the $f$ term vanish by requiring that the structure constants be antisymmetric under exchange of any $b_{i}\leftrightarrow c_{j}$ when $b_{i}\neq c_{j}$. Note that some $b_{i}$ indices must be equal to some $c_{i}$ indices and $f$ should be symmetric under the exchange of those. At this point, we ensured that the the $f$ term could be non zero only if $b_{d-k-1}=c_{d-n+k}$, and then $\hat{n}^a\hat{n}^a = |\hat{\bf n}|^2 = 1$ under $\delta(\cdots)$ yields:
\begin{eqnarray}\label{struct-factor-constr}
&& \delta \mathcal{J}_{\mu_{1}\cdots\mu_{d-n}}^{a_{n+1}\cdots a_{d-1}} = \epsilon_{\mu_{1}\cdots\mu_{d-n}\lambda_{0}\cdots\lambda_{n}}^{\phantom{,}}
      \left( \delta\prod_{i=0}^{n}\partial_{\lambda_{i}}\hat{n}^{a_{i}} \right) \\
&& \times \Biggl\lbrace \frac{1}{n!}\epsilon_{a_{0}\cdots a_{d-1}}
      - \sum_{k=0}^{\lbrack\frac{n-1}{2}\rbrack} f_{d,n,k}^{a_{n+1}\cdots a_{d-1}b_{1}\cdots b_{d-k-2}c_{1}\cdots c_{d-n+k-1}} \nonumber \\
&& ~~~ \times (-1)^{n+1}\frac{P_{b_{1}\cdots b_{d-k-2}a_{0}\cdots a_{k},c_{1}\cdots c_{d-n+k-1}a_{k+1}\cdots a_{n}}}{(k+1)!(n-k)!} \Biggr\rbrace
      = 0 \nonumber
\end{eqnarray}
where we defined:
\begin{equation}
P_{b_{1}\cdots b_{d-1},c_{1}\cdots c_{d-1}}=\sum_{\mathcal{P}}^{1\cdots d-1}(-1)^{\mathcal{P}}\prod_{i=1}^{d-1}\delta_{b_{i}c_{\mathcal{P}(i)}}
\end{equation}
in terms of permutations $\mathcal{P}$ of $d-1$ elements, and used:
\begin{eqnarray}
&& \epsilon_{b_{1}\cdots b_{d-k-2} i a_{0}\cdots a_{k}}^{\phantom{,}}\epsilon_{c_{1}\cdots c_{d-n+k-1} i a_{k+1}\cdots a_{n}}^{\phantom{,}} 
  = (-1)^{d-k-2} \nonumber \\
&& \qquad \times (-1)^{d-n+k-1} \epsilon_{i b_{1}\cdots b_{d-k-2}a_{0}\cdots a_{k}}^{\phantom{,}}
      \epsilon_{i c_{1}\cdots c_{d-n+k-1}a_{k+1}\cdots a_{n}}^{\phantom{,}} \nonumber \\
&& \quad = (-1)^{n+1} P_{b_{1}\cdots b_{d-k-2}a_{0}\cdots a_{k},c_{1}\cdots c_{d-n+k-1}a_{k+1}\cdots a_{n}} \ . \nonumber
\end{eqnarray}
We find from (\ref{struct-factor-constr}) the condition that the structure constants must satisfy for every combination of $a_{n+1},\dots,a_{d-1}$:
\begin{eqnarray}\label{struct-factor-constr-2}
&& \mathcal{A}\sum_{k=0}^{\lbrack\frac{n-1}{2}\rbrack}f_{d,n,k}^{a_{n+1}\cdots a_{d-1}b_{1}\cdots b_{d-k-2}c_{1}\cdots c_{d-n+k-1}} \\ 
&& \qquad\quad \times \frac{P_{b_{1}\cdots b_{d-k-2}a_{0}\cdots a_{k},c_{1}\cdots c_{d-n+k-1}a_{k+1}\cdots a_{n}}}{(k+1)!(n-k)!} \nonumber \\
&& \quad = \frac{(-1)^{n+1}}{n!}\epsilon_{a_{0}\cdots a_{d-1}} \ , \nonumber
\end{eqnarray}
where $\mathcal{A}$ indicates antisymmetrization with respect to $a_{0},\dots a_{n}$ (only the antisymmetric part survives the summation over $\lambda_{i}$). Since the Levi-Civita symbol $\epsilon$ and the Kronecker symbol $\delta$ transform like tensors under rotations in Euclidean space-time, we will automatically satisfy the sufficient condition for gauge invariance by finding the structure constants $f$ that themselves transform like tensors. We can convert this into a system of linear equations for structure constants which automatically expels the non-antisymmetric terms on the left-hand-side. Multiplying (\ref{struct-factor-constr-2}) by $\epsilon_{a_{0}\cdots a_{d-1}}$ and summing over $a_{0},\dots a_{d-1}$ yields:
\begin{eqnarray}
&& \epsilon_{a_{0}\cdots a_{d-1}}\sum_{k=0}^{\lbrack\frac{n-1}{2}\rbrack}f_{d,n,k}^{a_{n+1}\cdots a_{d-1}b_{1}\cdots b_{d-k-2}c_{1}\cdots c_{d-n+k-1}} \\
&& \quad \times \frac{P_{b_{1}\cdots b_{d-k-2}a_{0}\cdots a_{k},c_{1}\cdots c_{d-n+k-1}a_{k+1}\cdots a_{n}}}{(k+1)!(n-k)!}=\frac{d!}{n!}(-1)^{n+1} \ . \nonumber
\end{eqnarray}

To summarize, we are looking for $f^{a\cdots b\cdots c\cdots}$ which is antisymmetric under exchanges among its $a$ indices, $b$ indices, $c$ indices, and $b\leftrightarrow c$ indices for $b\neq c$. Also, any $f$ factor that multiplies $\epsilon_{a_{0}\cdots a_{d-1}}$ but is not antisymmetric under exchanges among $a_{0},\dots a_{d-1}$ can be set to zero, because it does not contribute to the above equation. It is always possible to find structure constants under these constraints, and in some cases there is a unique solution. However, there are multiple solutions when the variable $k$ in the above sum can take multiple values, i.e. when $n\ge3$. These solutions for flux can be written as $(\epsilon\partial A-\epsilon AA)^{2}$ where the quadratic part $\epsilon AA$ takes different forms. This complexity reveals intricate interactions between the non-Abelian topological defects, but it should be kept in mind that the above procedure establishes only the necessary and not the sufficient condition for gauge invariance. The full gauge transformation of non-singular gauge fields discussed above may further restrict the form of Maxwell terms. Some examples are:
\begin{itemize}
\item $d=3$, $n=1$:
\begin{eqnarray}
3! &=& \epsilon_{a_{0}a_{1}a_{2}}f_{3,1,0}^{a_{2}b_{1}c_{1}}\,P_{b_{1}a_{0},c_{1}a_{1}} \nonumber \\
   &=& \epsilon_{a_{0}a_{1}a_{2}}f_{3,1,0}^{a_{2}b_{1}c_{1}}\,(\delta_{b_{1}c_{1}}\delta_{a_{0}a_{1}}-\delta_{b_{1}a_{1}}\delta_{a_{0}c_{1}}) \nonumber \\
   &=& \epsilon_{a_{0}a_{1}a_{2}}f_{3,1,0}^{a_{0}a_{1}a_{2}} \ . \nonumber
\end{eqnarray}
There is a unique solution:
\begin{equation}
f_{3,1,0}^{abc}=\epsilon_{abc} \ . \nonumber
\end{equation}
\item $d=4$, $n=2$:
\begingroup
\allowdisplaybreaks
\begin{eqnarray}
-\frac{4!}{2!} &=& \epsilon_{a_{0}a_{1}a_{2}a_{3}}f_{4,2,0}^{a_{3}b_{1}b_{2}c_{1}}\, \frac{P_{b_{1}b_{2}a_{0},c_{1}a_{1}a_{2}}}{2!} \nonumber \\
&=& \frac{1}{2!}\epsilon_{a_{0}a_{1}a_{2}a_{3}}f_{4,2,0}^{a_{3}b_{1}b_{2}c_{1}} \nonumber \\
&& \quad\times \bigl( \delta_{b_{1}c_{1}}\delta_{b_{2}a_{1}}\delta_{a_{0}a_{2}} -\delta_{b_{1}c_{1}}\delta_{b_{2}a_{2}}\delta_{a_{0}a_{1}} \nonumber \\
&& \quad             +\delta_{b_{1}a_{1}}\delta_{b_{2}a_{2}}\delta_{a_{0}c_{1}} -\delta_{b_{1}a_{1}}\delta_{b_{2}c_{1}}\delta_{a_{0}a_{2}} \nonumber \\
&& \quad             +\delta_{b_{1}a_{2}}\delta_{b_{2}c_{1}}\delta_{a_{0}a_{1}}-\delta_{b_{1}a_{2}}\delta_{b_{2}a_{1}}\delta_{a_{0}c_{1}} \bigr) \nonumber \\
&=&\frac{1}{2!}\epsilon_{a_{0}a_{1}a_{2}a_{3}}(f_{4,2,0}^{a_{3}a_{1}a_{2}a_{0}}-f_{4,2,0}^{a_{3}a_{2}a_{1}a_{0}}) \ . \nonumber
\end{eqnarray}
\endgroup
There is a unique solution:
\begin{equation}
f_{4,2,0}^{abcd} = \frac{1}{2}\epsilon_{abcd} \ . \nonumber
\end{equation}
\item $d=4$, $n=1$:
\begingroup
\allowdisplaybreaks
\begin{eqnarray}
4! &=& \epsilon_{a_{0}a_{1}a_{2}a_{3}}f_{4,1,0}^{a_{2}a_{3}b_{1}b_{2}c_{1}c_{2}}\,P_{b_{1}b_{2}a_{0},c_{1}c_{2}a_{1}} \nonumber \\
   &=& \epsilon_{a_{0}a_{1}a_{2}a_{3}}f_{4,1,0}^{a_{2}a_{3}b_{1}b_{2}c_{1}c_{2}} \nonumber \\
&& \quad\times \bigl( \delta_{b_{1}c_{1}}\delta_{b_{2}c_{2}}\delta_{a_{0}a_{1}} -\delta_{b_{1}c_{1}}\delta_{b_{2}a_{1}}\delta_{a_{0}c_{2}} \nonumber \\
&& \quad             +\delta_{b_{1}c_{2}}\delta_{b_{2}a_{1}}\delta_{a_{0}c_{1}} -\delta_{b_{1}c_{2}}\delta_{b_{2}c_{1}}\delta_{a_{0}a_{1}} \nonumber \\
&& \quad             +\delta_{b_{1}a_{1}}\delta_{b_{2}c_{1}}\delta_{a_{0}c_{2}}-\delta_{b_{1}a_{1}}\delta_{b_{2}c_{2}}\delta_{a_{0}c_{1}} \bigr) \nonumber \\
&=& \epsilon_{a_{0}a_{1}a_{2}a_{3}} \bigl(-f_{4,1,0}^{a_{2}a_{3}ba_{1}ba_{0}}+f_{4,1,0}^{a_{2}a_{3}ba_{1}a_{0}b} \nonumber \\
&& \quad             +f_{4,1,0}^{a_{2}a_{3}a_{1}bba_{0}}-f_{4,1,0}^{a_{2}a_{3}a_{1}ba_{0}b} \bigr) \nonumber \\
&=& 4\epsilon_{a_{0}a_{1}a_{2}a_{3}}f^{a_{0}a_{1}ba_{2}ba_{3}} \ . \nonumber
\end{eqnarray}
\endgroup
There is a unique symmetrized solution:
\begin{equation}
f_{4,1,0}^{abpcqd}=\frac{1}{8}(\epsilon_{abcd}\delta_{pq}-\epsilon_{abpd}\delta_{cq}+\epsilon_{abpq}\delta_{cd}-\epsilon_{abcq}\delta_{pd}) \nonumber
\end{equation}
\item $d=5$, $n=3$:
\begingroup
\allowdisplaybreaks
\begin{widetext}
\begin{eqnarray}
\frac{5!}{3!} &=& \epsilon_{a_{0}a_{1}a_{2}a_{3}a_{4}}\left\lbrack f_{5,3,0}^{a_{4}b_{1}b_{2}b_{3}c_{1}}\,
        \frac{P_{b_{1}b_{2}b_{3}a_{0},c_{1}a_{1}a_{2}a_{3}}}{3!}+f_{5,3,1}^{a_{4}b_{1}b_{2}c_{1}c_{2}}\,
        \frac{P_{b_{1}b_{2}a_{0}a_{1},c_{1}c_{2}a_{2}a_{3}}}{2!\cdot2!}\right\rbrack \nonumber \\
&=& \epsilon_{a_{0}a_{1}a_{2}a_{3}a_{4}}\Biggl\lbrack -f_{5,3,0}^{a_{4}b_{1}b_{2}b_{3}c_{1}}\,\frac{\delta_{a_{0}c_{1}}}{3!}
        \sum_{\mathcal{P}}^{1\cdots3}(-1)^{\mathcal{P}}\prod_{i=1}^{3}\delta_{b_{i}a_{\mathcal{P}(i)}} \nonumber \\
&& \qquad\qquad +f_{5,3,1}^{a_{4}b_{1}b_{2}c_{1}c_{2}}\,
        \frac{\delta_{b_{1}a_{2}}\delta_{b_{2}a_{3}}\delta_{c_{1}a_{0}}\delta_{c_{2}a_{1}}
             -\delta_{b_{1}a_{2}}\delta_{b_{2}a_{3}}\delta_{c_{1}a_{1}}\delta_{c_{2}a_{0}}
             +\delta_{b_{1}a_{3}}\delta_{b_{2}a_{2}}\delta_{c_{1}a_{1}}\delta_{c_{2}a_{0}}
             -\delta_{b_{1}a_{3}}\delta_{b_{2}a_{2}}\delta_{c_{1}a_{0}}\delta_{c_{2}a_{1}}}{2!\cdot2!}\Biggr\rbrack \nonumber \\
&=& \epsilon_{a_{0}a_{1}a_{2}a_{3}a_{4}}\left\lbrack -\frac{1}{3!}\sum_{\mathcal{P}}^{1\cdots3}(-1)^{\mathcal{P}}
       f_{5,3,0}^{a_{4}a_{\mathcal{P}(1)}a_{\mathcal{P}(2)}a_{\mathcal{P}(3)}a_{0}}+\frac{1}{2!\cdot2!}(
              f_{5,3,1}^{a_{4}a_{2}a_{3}a_{0}a_{1}}-f_{5,3,1}^{a_{4}a_{2}a_{3}a_{1}a_{0}}
             +f_{5,3,1}^{a_{4}a_{3}a_{2}a_{1}a_{0}}-f_{5,3,1}^{a_{4}a_{3}a_{2}a_{0}a_{1}})\right\rbrack \nonumber \\
&=& \epsilon_{a_{0}a_{1}a_{2}a_{3}a_{4}}(-f_{5,3,0}^{a_{4}a_{1}a_{2}a_{3}a_{0}}+f_{5,3,1}^{a_{4}a_{2}a_{3}a_{0}a_{1}})
   =\epsilon_{a_{0}a_{1}a_{2}a_{3}a_{4}}(f_{5,3,0}^{a_{0}a_{1}a_{2}a_{3}a_{4}}+f_{5,3,1}^{a_{0}a_{1}a_{2}a_{3}a_{4}}) \ . \nonumber
\end{eqnarray}
\end{widetext}
\endgroup
There are multiple solutions:
\begin{equation}
f_{5,3,0}^{abcde}+f_{5,3,1}^{abcde} = \frac{1}{6}\epsilon_{abcde} \ .
\end{equation}
\end{itemize}
\raggedbottom

\section{Singular gauge transformations and dynamical generation of higher ranks in the non-Abelian effective theory}\label{appNonAbelianranks}

Maxwell terms $\mathcal{L}_{\textrm{M}n}$ at rank $n$ in the effective Lagrangian arise from integrating out the smooth short-wavelength fluctuations of currents at the same rank. A further integration of singular currents at short length scales generates the current term at rank $n+1$, and gives it the form compatible to that of the rank $n$ Maxwell term. We will roughly sketch this process here. Let us write the rank $n$ flux as:
\begin{equation}\label{rank-flux-2}
\mathcal{J}_{\mu_{1}\cdots\mu_{d-n}}^{a_{n+1}\cdots a_{d-1}} =
  \epsilon_{\mu_{1}\cdots\mu_{d-n}\alpha_{0}\cdots\alpha_{n}}^{\phantom{,}} X_{\alpha_{0}\cdots\alpha_{n}}^{a_{n+1}\cdots a_{d-1}} \ .
\end{equation}
The formulas for $X$ were discussed and derived in Appendix \ref{appNonAbelianMaxwell}. The rank $n$ Maxwell term is:
\begin{eqnarray}
&& \mathcal{L}_{\textrm{M}n}^{\phantom{,}} \sim
  \mathcal{J}_{\mu_{1}\cdots\mu_{d-n}}^{a_{n+1}\cdots a_{d-1}} \mathcal{J}_{\mu_{1}\cdots\mu_{d-n}}^{a_{n+1}\cdots a_{d-1}} \\
&& ~~ = (d-n)!\sum_{\mathcal{P}}^{0\cdots n}(-1)^{\mathcal{P}} X_{\alpha_{0}\cdots\alpha_{n}}^{a_{n+1}\cdots a_{d-1}}
        X_{\alpha_{\mathcal{P}(0)}\cdots\alpha_{\mathcal{P}(n)}}^{a_{n+1}\cdots a_{d-1}} \nonumber \\
&& ~~ = \frac{(d-n)!}{(n+1)!}\left\lbrack \sum_{\mathcal{P}}^{0\cdots n}(-1)^{\mathcal{P}}
        X_{\alpha_{\mathcal{P}(0)}\cdots\alpha_{\mathcal{P}(n)}}^{a_{n+1}\cdots a_{d-1}}\right\rbrack ^{2} \nonumber \ ,
\end{eqnarray}
where $\mathcal{P}$ indicates a permutation and $(-1)^{\mathcal{P}}$ its parity. We will consider a fixed gauge in which singularity gauge fields have not yet been generated from the given matter field configuration $\hat{\bf n}$. In this gauge, the current at rank $n+1$ is:
\begin{equation}\label{Jcurrent}
J_{\lambda_{1}\cdots\lambda_{n+1}}^{a_{n+2}\cdots a_{d-1}} = \frac{1}{(n+1)!}
  \epsilon_{a_{0}\cdots a_{d-1}}^{\phantom{,}}\hat{n}^{a_{0}}\left(\prod_{i=1}^{n+1}\partial_{\lambda_{i}}\hat{n}^{a_{i}}\right)
\end{equation}
and the current term in the Lagrangian density is:
\begingroup
\allowdisplaybreaks
\begin{eqnarray}
&& \mathcal{L}_{\textrm{C},n+1}^{\phantom{,}} \sim \left(J_{\lambda_{1}\cdots\lambda_{n+1}}^{c_{n+2}\cdots c_{d-1}}\right)^{2} \\[0.1in]
&& ~~ = \frac{\epsilon_{a_{0}\cdots a_{n+1}c_{n+2}\cdots c_{d-1}}\epsilon_{b_{0}\cdots b_{n+1}c_{n+2}\cdots c_{d-1}}}{\lbrack(n+1)!\rbrack^{2}} \nonumber \\
&& \qquad\quad \times \, \hat{n}^{a_{0}}\left(\prod_{j=1}^{n+1}\partial_{\lambda_{j}}\hat{n}^{a_{j}}\right)
           \hat{n}^{b_{0}}\left(\prod_{k=1}^{n+1}\partial_{\lambda_{k}}\hat{n}^{b_{k}}\right) \nonumber \\
&& ~~ = \frac{(d-n-2)!}{\lbrack(n+1)!\rbrack^{2}}\,\sum_{\mathcal{P}}^{1\cdots n+1}(-1)^{\mathcal{P}}
           \left(\prod_{i=1}^{n+1}\delta_{a_{i}b_{\mathcal{P}(i)}}\right) \nonumber \\
&& \qquad\quad \times \left(\prod_{j=1}^{n+1}\partial_{\lambda_{j}}\hat{n}^{a_{j}}\right)
           \left(\prod_{k=1}^{n+1}\partial_{\lambda_{k}}\hat{n}^{b_{k}}\right) \nonumber \ .
\end{eqnarray}
\endgroup
The last line obtains for $a_0=b_0$, where the derivative-free factors $\hat{n}^a$ produce $\hat{n}^a\hat{n}^a = |\hat{\bf n}|=1$. The terms with $a_0\neq b_0$ imply $a_0 = b_i$, $i>0$ and hence vanish by $\partial_\lambda |\hat{\bf n}|^2 = 0$. Now we unpack the permutations:
\begin{eqnarray}
&& \mathcal{L}_{\textrm{C},n+1}^{\phantom{,}} \sim \left(J_{\lambda_{1}\cdots\lambda_{n+1}}^{c_{n+2}\cdots c_{d-1}}\right)^{2} \\[0.1in]
&& ~~ = \frac{\epsilon_{a_{1}\cdots a_{n+1}c_{n+2}\cdots c_{d}}\epsilon_{b_{1}\cdots b_{n+1}c_{n+2}\cdots c_{d}}}{\lbrack(n+1)!\rbrack^{2}(d-n-1)} \nonumber \\
&& \qquad\quad \times \left(\prod_{j=1}^{n+1}\partial_{\lambda_{j}}\hat{n}^{a_{j}}\right)\left(\prod_{k=1}^{n+1}\partial_{\lambda_{k}}\hat{n}^{b_{k}}\right) \nonumber \\
&& ~~ = \frac{1}{\lbrack(n+1)!\rbrack^{2}(d-n-1)} Y_{\lambda_{1}\cdots\lambda_{n+1}}^{c_{n+2}\cdots c_{d}}Y_{\lambda_{1}\cdots\lambda_{n+1}}^{c_{n+2}\cdots c_{d}}
   \ , \nonumber
\end{eqnarray}
and proceed working on the $Y$ factors:
\begin{eqnarray}
&& Y_{\lambda_{0}\cdots\lambda_{n}}^{c_{n+1}\cdots c_{d-1}} = \epsilon_{a_{0}\cdots a_{n}c_{n+1}\cdots c_{d-1}}^{\phantom{,}}
        \left(\prod_{j=0}^{n}\partial_{\lambda_{j}}\hat{n}^{a_{j}}\right) \nonumber \\
&& ~~ = \mathcal{A}\,\epsilon_{a_{0}\cdots a_{n}c_{n+1}\cdots c_{d-1}}^{\phantom{,}}
        \partial_{\lambda_{0}}\left(\hat{n}^{a_{0}}\prod_{j=1}^{n}\partial_{\lambda_{j}}\hat{n}^{a_{j}}\right) \nonumber \\
&& ~~ = n!\,\mathcal{A}\,\partial_{\lambda_{0}}^{\phantom{,}}J_{\lambda_{1}\cdots\lambda_{n}}^{c_{n+1}\cdots c_{d-1}}
      \propto \mathcal{A}\, X_{\lambda_{0}\cdots\lambda_{n}}^{c_{n+1}\cdots c_{d-1}}\ .
\end{eqnarray}
Here, $\mathcal{A}$ antisymmetrizes the space-time indices. The last proportionality follows from (\ref{rank-flux-1}) and (\ref{rank-flux-2}). Specifically in the chosen gauge, $Y$ is proportional to antisymmetrized $X$ without the non-Abelian part involving structure constants in (\ref{rank-flux-1}), where the rank $n$ gauge field emerges from the singularities of (\ref{Jcurrent}) upon a singular gauge transformation and coarse-graining. In conclusion, structurally:
\begin{equation}
\mathcal{L}_{\textrm{C},n+1} \sim (\mathcal{A} X)^2 \sim \mathcal{L}_{\textrm{M},n} \ . \nonumber
\end{equation}
The full coarse-graining procedure expands all terms into their full gauge-invariant forms featuring gauge fields. Given the above structural relationship, we anticipate that fluctuations indeed generate the Lagrangian terms at all ranks, starting from the fundamental ones at rank 1. The same effect was revealed in the case of Abelian charge dynamics.

\flushbottom

\section{Duality mapping of the Abelian compact gauge theory at rank $n$}\label{appDuality}

Here we consider the action:
\begin{eqnarray}\label{origTh}
&& S =  -k\sum_{\square}^{(n+1)}\cos\left(\epsilon_{\mu_{1}\cdots\mu_{d-n}\nu\lambda_{1}\cdots\lambda_{n}}\partial_{\nu}A_{\lambda_{1}\cdots\lambda_{n}}\right) \\
&& ~ -t\sum_{\square}^{(n)}\cos\left(\sum_{l=1}^{n}(-1)^{l-1}\partial_{\lambda_{l}}\theta_{\lambda_{1}\cdots\lambda_{l-1}\lambda_{l+1}\cdots\lambda_{n}}
      -A_{\lambda_{1}\cdots\lambda_{n}}\right) \nonumber
\end{eqnarray}
of a matter field $\theta$ coupled to a rank $n$ gauge field $A$ on a $d+1$ dimensional space-time cubic lattice. The sums run over $n$ dimensional hypercube ``plaquettes'' of the lattice, and the indices $\mu,\dots$ label independent lattice directions along lattice bonds. A quantity $f_{\mu_{1}\cdots\mu_{n}}$ with $n$ indices $\mu_{i} \in \lbrace 0,1, \dots, d \rbrace$ lives on an oriented $n$ dimensional plaquette of the space-time lattice. One specifies a plaquette by a lattice site $i$ and an ordered set of indices $(\mu_{1},\dots,\mu_{n})$ that indicate space-time directions of the plaquette edges that emanate from the site $i$. The plaquette orientation is given by the value of $\epsilon_{\mu_{1} \cdots \mu_{n} \alpha_{n+1} \cdots \alpha_{d+1}}$ that obtains from $\epsilon_{1,2,3, \cdots, d+1}$ with the minimum number of index permutations. A plaquette orientation can be changed either by an exchange of two indices or by a sign change of one index; $f_{\mu_{1}\cdots\mu_{n}}$ defined on a plaquette with positive orientation is equivalent to $-f_{\mu_{1}\cdots\mu_{n}}$ on the same plaquette with negative orientation, and consistent with $f_{\mu_{1}\cdots\mu_{n}}$ being an antisymmetric tensor. The lattice derivative $\partial_{\nu}$ is defined by $\partial_{\nu}f_{i}=f_{i+\hat{\nu}}-f_{i}$. The Maxwell term $k$ is specified on an $n+1$ dimensional plaquette spanned by $\nu\lambda_{1}\cdots\lambda_{n}$, and the remaining indices $\mu_{1}\cdots\mu_{d-n}$ are redundant in the compact formulation but kept for the sake of the continuum limit where they are contracted in a quadratic form. Note that the nature of lattice derivatives is such that we can apply integration by parts (in an infinite system):
\begin{eqnarray}
\sum_{i}f\partial_{\mu}g &=& \sum_{i}(f_{i}g_{i+\hat{\mu}}-f_{i}g_{i}) = \sum_{i}(f_{i-\hat{\mu}}g_{i}-f_{i}g_{i}) \nonumber \\
  &=& -\sum_{i}g_{i}(f_{i}-f_{i-\hat{\mu}})=-\sum_{i}g\partial_{\mu}f \nonumber \ .
\end{eqnarray}

The dual action is derived as follows. Decouple the cosines in $S$ with integer-valued antisymmetric tensor fields $J_{\lambda_{1}\cdots\lambda_{n}}$ and $\Phi_{\mu_{1}\cdots\mu_{d-n}}$ using Villain's approximation $\exp(-t\cos x) \approx \sum_m \exp(-T m^2 + imx)$:
\begin{eqnarray}
&& S = \sum_{\square}^{(n-1)}\Biggl\lbrack \frac{1}{2\tau}J_{\lambda_{1}\cdots\lambda_{n}}J_{\lambda_{1}\cdots\lambda_{n}}
       +iJ_{\lambda_{1}\cdots\lambda_{n}} \times \\
&& \qquad\quad \times \left(\sum_{l=1}^{n}(-1)^{l-1}\partial_{\lambda_{l}}\theta_{\lambda_{1}\cdots\lambda_{l-1}\lambda_{l+1}\cdots\lambda_{n}}
      -A_{\lambda_{1}\cdots\lambda_{n}}\right)\Biggr\rbrack \nonumber \\
&& ~~~~ +\sum_{\square}^{(n)}\Biggl\lbrack \frac{1}{2\kappa}\Phi_{\mu_{1}\cdots\mu_{d-n}}\Phi_{\mu_{1}\cdots\mu_{d-n}}
      +i\Phi_{\mu_{1}\cdots\mu_{d-n}}  \times \nonumber \\
&& \qquad\quad \times \left(\epsilon_{\mu_{1}\cdots\mu_{d-n}\nu\lambda_{1}\cdots\lambda_{n}}
      \partial_{\nu}A_{\lambda_{1}\cdots\lambda_{n}}\right)\Biggr\rbrack \nonumber \ .
\end{eqnarray}
Large values of $t$ ($k$) correspond to large values of $\tau$ ($\kappa$), and small values of $t$ ($k$) correspond to small values of $\tau$ ($\kappa$). Integrating out the angles $\theta$ and $A$ produces
\begin{equation}
S = \sum_{\square}^{(n)}\frac{1}{2\tau}J_{\lambda_{1}\cdots\lambda_{n}}J_{\lambda_{1}\cdots\lambda_{n}}
  +\sum_{\square}^{(n+1)}\frac{1}{2\kappa}\Phi_{\mu_{1}\cdots\mu_{d-n}}\Phi_{\mu_{1}\cdots\mu_{d-n}}
\end{equation}
with constraints on $J$ and $\Phi$:
\begin{equation}
(\forall l)\quad\partial_{\lambda_{l}}J_{\lambda_{1}\cdots\lambda_{n}} = 0
\end{equation}
\vskip -0.3in
\begin{equation}
\epsilon_{\mu_{1}\cdots\mu_{d-n}\nu\lambda_{1}\cdots\lambda_{n}}\partial_{\nu}\Phi_{\mu_{1}\cdots\mu_{d-n}}+J_{\lambda_{1}\cdots\lambda_{n}} = 0 \nonumber \ .
\end{equation}
The constraints can be solved by expressing $J$ and $\Phi$ in terms of new antisymmetric tensor fields $a$ and $\phi$:
\begin{equation}\label{dualConstr}
J_{\lambda_{1}\cdots\lambda_{n}} = \epsilon_{\mu_{1}\cdots\mu_{d-n}\nu\lambda_{1}\cdots\lambda_{n}}\partial_{\nu}a_{\mu_{1}\cdots\mu_{d-n}}
\end{equation}
\vskip -0.3in
\begin{equation}
\Phi_{\mu_{1}\cdots\mu_{d-n}} = \sum_{l=1}^{d-n}(-1)^{l-1}\partial_{\mu_{l}}\phi_{\mu_{1}\cdots\mu_{l-1}\mu_{l+1}\cdots\mu_{d-n}}-a_{\mu_{1}\cdots\mu_{d-n}} \nonumber
\end{equation}
The values of $a$ and $\phi$ can be real as long as $J$ and $\Phi$ are integer-valued. In fact, the integer-value constraint on $\Phi$ automatically makes $J$ integer-valued. We can soften the integer value requirement on $\Phi$ using a sine-Gordon term ($\lambda_\kappa$) without affecting the universality class of the theory. Substituting the above constraint solutions in the action yields:
\begin{widetext}
\begin{eqnarray}\label{dualTh}
S &=& \frac{1}{2\tau}\sum_{\square}^{(d-n+1)}\Bigl(\epsilon_{\mu_{1}\cdots\mu_{d-n}\nu\lambda_{1}\cdots\lambda_{n}}\partial_{\nu}a_{\mu_{1}\cdots\mu_{d-n}}\Bigr)^{2}
     +\frac{1}{2\kappa}\sum_{\square}^{(d-n)}\left(\sum_{l=1}^{d-n}(-1)^{l-1}\partial_{\mu_{l}}\phi_{\mu_{1}\cdots\mu_{l-1}\mu_{l+1}\cdots\mu_{d-n}}
       -a_{\mu_{1}\cdots\mu_{d-n}}\right)^{2} \nonumber \\
&& -\lambda_{\kappa}\sum_{\square}^{(d-n)}\cos\left(2\pi\sum_{l=1}^{d-n}(-1)^{l-1}\partial_{\mu_{l}}\phi_{\mu_{1}\cdots\mu_{l-1}\mu_{l+1}\cdots\mu_{d-n}}
      -2\pi a_{\mu_{1}\cdots\mu_{d-n}}\right) \ .
\end{eqnarray}
\end{widetext}
This is the dual theory, formulated on the dual cubic lattice. Note that an $n$ dimensional plaquette of the original lattice is dual to a $d-n$ dimensional plaquette of the dual lattice.

\subsubsection{Phase diagram}

The original theory of angle-valued fields (\ref{origTh}) is expected to have Higgs and Coulomb phases. The Higgs phase is a $\theta$ condensate with all excitations gapped by Higgs mechanism. At least in this phase, we may take the continuum limit of both compact terms because all field fluctuations are suppressed. The Coulomb phase must have a massless gauge boson due to gauge invariance. It features abundant field fluctuations, so it is not a priori clear that the compact terms can be expanded to quadratic order for the continuum limit. The surviving compactness means that the electric field $E$ (canonically conjugate to $A$) is integer-valued, and any charged sources of the frozen $E$ (with fluctuating $A$) are confined.

The phase diagram of the dual theory (\ref{dualTh}) must match the phase diagram of the original theory (\ref{origTh}). The condensation of $\theta$ in the original Higgs phase is consistent with abundant fluctuations of $J$ (due to the $iJ\partial_{\mu}\theta$ term in the intermediate action), implying correspondence to a dual Coulomb phase with abundant $a$ fluctuations. The suppression of $A$ in the original Higgs phase is similarly tied to the fluctuations of $\phi$. Conversely, abundant fluctuations of $\theta,A$ in the original Coulomb phase correspond to the suppressed fluctuations of $\phi,a$ in a dual Higgs-like phase. However, there is a problem: the original Coulomb phase is gapless, so the dual Higgs phase must be gapless despite the presence of the dual gauge field $a$. Similarly, the original Higgs phase is gapped, so the dual Coulomb phase must be gapped too despite the presence of the dual gauge field $a$. If the dual theory contained only the matter field $\phi$, then its superfluid (Higgs) and disordered (Coulomb) phases would correctly match the excitation spectra of the original theory. Another problem is how to obtain gapless phases from fields whose gauge-invariant configurations are discrete-valued and always classically cost a finite action.

The hint to solving the second problem is that the only gapless mode in the original theory is the $A$ photon of the Coulomb phase, and the corresponding gapless mode of the dual theory is the Goldstone mode of $\phi$. Indeed, the field $\phi$ is not required to be integer-valued, only the combination $\Phi$ in (\ref{dualConstr}) is. Let us separate the transverse and longitudinal modes of $a$ and absorb the longitudinal modes into $\partial\phi$ by a change of variables. This amounts to gauge fixing through some condition imposed on $\phi$ and $a$. However, neither $\phi$ nor $a$ can be integer-valued in a fixed state-independent gauge on the lattice. Hence, separating transverse from longitudinal gauge modes inevitably runs into geometric frustration on the lattice. A set of non-quantized transverse gauge modes $a$ can emerge from frustration and live at very long wavelengths with small amplitudes, as long as they are compensated by appropriate longitudinal $\phi$ modes to make $\Phi$
integer-valued. These long-wavelength $a$ modes cost arbitrarily small Maxwell energy, but the attached longitudinal modes are generally costly through the non-compact $\kappa^{-1}$ term of the dual action (\ref{dualTh}). For this reason, the presence of the $\kappa^{-1}$ term effectively gaps out the transverse $a$ modes, and we can safely integrate them out. The resulting effective theory may have a sine-Gordon term for the surviving longitudinal modes $\phi$, but the sine-Gordon coupling cannot be infinite because $\phi$ is fundamentally not quantized as a result of frustration (it cannot be even finite if it is a relevant perturbation that flows to infinity under renormalization group). The final dual theory for low-energy excitations is approximately a non-compact longitudinal model:
\begin{equation}\label{dualTh2}
S = \frac{1}{2\kappa'}\sum_{\square}^{(d-n)}\left(\sum_{l=1}^{d-n}(-1)^{l-1}\partial_{\mu_{l}}\phi_{\mu_{1}\cdots\mu_{l-1}\mu_{l+1}\cdots\mu_{d-n}}\right)^{2} \ .
\end{equation}
This model correctly matches the phases of the original theory and their excitation spectra. The analogous compact model
\begin{equation}
S' = -\lambda'\!\sum_{\square}^{(d-n)}\!\cos\left(2\pi\sum_{l=1}^{d-n}(-1)^{l-1}\partial_{\mu_{l}}\phi_{\mu_{1}\cdots\mu_{l-1}\mu_{l+1}\cdots\mu_{d-n}}\right)
  \nonumber
\end{equation}
also matches the phases of the original theory and almost all of its excitations. It only differs from the original theory in regard to its confinement of charges in the Coulomb phase. An excitation with quantized ``electric'' charge in the original theory corresponds to a quantized topological defect of the dual theory. The compact theory $S'$ does not confine its defects in the ordered phase because the frustration of $\phi$ due to separated quantized defects can be collected into a singular multi-dimensional ``fault line'' that terminates at the defects and costs no $\lambda'$ energy. Consequently, the non-compact theory (\ref{dualTh2}) correctly describes the spectrum of charged excitations in the original model's Coulomb phase.

\subsubsection{Without gauge fields in the original theory}

A special case of the original theory (\ref{origTh}) obtains in the $k\to\infty$ limit, which suppresses the gauge field $A$. The ensuing action contains only the matter field:
\begin{equation}
S = -t\sum_{\square}^{(n)}\cos\left(\sum_{l=1}^{n}(-1)^{l-1}\partial_{\lambda_{l}}\theta_{\lambda_{1}\cdots\lambda_{l-1}\lambda_{l+1}\cdots\lambda_{n}}\right) \ .
\end{equation}
Its phase diagram consists of a superfluid ordered phase with gapless Goldstone modes, and a disordered gapped phase. Following the same procedure as before, we obtain the dual theory
\begin{eqnarray}
S &=& \frac{1}{2\tau}\sum_{\square}^{(d-n+1)}\Bigl(\epsilon_{\mu_{1}\cdots\mu_{d-n}\nu\lambda_{1}\cdots\lambda_{n}}\partial_{\nu}a_{\mu_{1}\cdots\mu_{d-n}}\Bigr)^{2}
  \nonumber \\
&& -\lambda_{\kappa}\sum_{\square}^{(d-n)}\cos\Biggl(2\pi\sum_{l=1}^{d-n}(-1)^{l-1}\partial_{\mu_{l}}\phi_{\mu_{1}\cdots\mu_{l-1}\mu_{l+1}\cdots\mu_{d-n}} \nonumber \\
&& \qquad\qquad\qquad  -2\pi a_{\mu_{1}\cdots\mu_{d-n}}\Biggr) \ .
\end{eqnarray}
The main difference from (\ref{dualTh}) is the absence of the non-compact term with dual matter ($\kappa\to\infty$). Repeating the previous analysis, we find that the long-wavelength transverse modes of $a$ must still bind some longitudinal modes of $\phi$ in order to satisfy the integer constraint (\ref{dualConstr}) on $\Phi$, but these longitudinal modes now cost no energy in the absence of the non-compact matter term. Therefore, the gauge bosons are massless by gauge-invariance, unless a Higgs mechanism occurs. The Higgs phase of the dual model is fully gapped and corresponds to the disordered phase of the original model. The dual Coulomb phase has massless photons, which corresponds to the massless Goldstone modes of the original model's superfluid phase.

Note that we could also consider a fully compact Maxwell term in the dual theory $S'$. However, such a compact theory would confine its charges in the Coulomb phase, so the topological defects of the original model's ordered phase would need to be confined too. Defect confinement is not present in the original compact theory, so the correct dual theory indeed has a non-compact Maxwell term.

\subsubsection{Without matter fields in the original theory}

Another special case of the original theory is the limit $t\to0$:
\begin{equation}\label{origTh3}
S = -k\sum_{\square}^{(n+1)}\cos\left(\epsilon_{\mu_{1}\cdots\mu_{d-n}\nu\lambda_{1}\cdots\lambda_{n}}\partial_{\nu}A_{\lambda_{1}\cdots\lambda_{n}}\right) \ .
\end{equation}
This is a pure compact gauge theory at rank $n$, which describes the phase $\mathcal{G}_{d-1} = C_1\cdots C_{d-1}$ from Section \ref{secDynPhaseDiag} after all other gapped fields have been integrated out. The dual theory is derived using the same approach as before, but with $J=0$, $a=0$:
\begin{eqnarray}\label{dualTh3}
&& S = \frac{1}{2\kappa}\!\sum_{\square}^{(d-n)}\!\left(\sum_{l=1}^{d-n}(-1)^{l-1}\partial_{\mu_{l}}\phi_{\mu_{1}\cdots\mu_{l-1}\mu_{l+1}\cdots\mu_{d-n}}\right)^{\!2} \\
&& ~~~ -\lambda_{\kappa}\sum_{\square}^{(d-n)}\cos\left(2\pi\sum_{l=1}^{d-n}(-1)^{l-1}\partial_{\mu_{l}}\phi_{\mu_{1}\cdots\mu_{l-1}\mu_{l+1}\cdots\mu_{d-n}}\right) 
  \nonumber
\end{eqnarray}
This is a gauge theory at rank $d-n-1$ provided that $n<d-1$, because the Abelian current term of rank $d-n$ is equivalent to the Maxwell term at rank $d-n-1$.

The highest rank $n=d-1$ is special because the ``longitudinal'' field $\phi$ becomes a scalar without indices. The dual action (\ref{dualTh3}) reduces to sums over lattice links:
\begin{equation}
S = \frac{1}{2\kappa}\sum_{-}\left(\partial_{\mu}\phi\right)^{2}-\lambda_{\kappa}\sum_{-}\cos\left(2\pi\partial_{\mu}\phi\right) \ .
\end{equation}
It has been shown \cite{zheng89} that $\lambda_{\kappa}$ is always relevant here and flows to infinity under renormalization group (in $d=2$, but the argument naively extends to $d>2$), so this turns into a height model with $\phi\in\mathbb{Z}$ (an arbitrary real-valued uniform offset to $\phi$ is irrelevant). The final dual theory:
\begin{equation}\label{HeightModel}
S = \frac{1}{2\kappa}\sum_{-}\left(\partial_{\mu}\phi\right)^{2}\quad,\quad\phi\in\mathbb{Z}
\end{equation}
with integer-valued field is known to have \emph{only} an ordered ``smooth'' phase in $d=2$ dimensions \cite{sachdev02e}, and the ordered phase is only more stable in $d>2$. This ordered phase is evidently gapped and corresponds to the confining phase of the original pure-gauge theory (\ref{origTh3}).

A disordered phase of the dual theory would corresponds to the large $k$ phase of the original theory (\ref{origTh3}). At large $k$, the fluctuations of $A$ would be suppressed and we would naively be able to expand the cosine to quadratic order and take the continuum limit. The ensuing non-compact gauge theory would then have a massless photon mode. However, this is not actually accurate. What does this correspond to in the dual theory? In order to quantize the dual theory, we introduce a canonically conjugate observable $n$ to $\phi$, defined on dual lattice sites $i$ through $\lbrack n_{i},\phi_{j}\rbrack=i\delta_{ij}$. Since $\phi\in\mathbb{Z}$, we find that $n\in\lbrack0,2\pi)$ is a continuous variable. The dual Hamiltonian takes form
\begin{equation}
H = -\frac{u}{2}\sum_{i}\cos(n_{i})+\frac{1}{2\kappa}\sum_{\langle ij\rangle}(\phi_{i}-\phi_{j})^{2} \ .
\end{equation}
The $\phi$-disordered state is ordered in $n$, but the settled value of $\langle n_{i}\rangle=0$ in the ground state is not separated from other values $\langle n_{i} \rangle \neq 0$ by a finite gap and consequently $\kappa^{-1}$ is not a small perturbation. It has been shown that the actual spectrum is gapped in $d=2$. The dual height model is in its smooth phase for all $\kappa$, so that $\langle\phi\rangle$ is always well defined in the ground state. Conversely, the original compact gauge theory at the highest rank $n=d-1$ is always confined and gapped.

\section{Canonical formalism of the multi-rank Abelian gauge theory: energy-momentum tensor and angular momentum}\label{appCanonical}

The Abelian Lagrangian density (\ref{L3}) of gauge and matter fields at rank $n$ in $d$ dimensions is:
\begin{eqnarray}
\mathcal{L} &=& \frac{1}{(d-n)!}\frac{1}{2e_{n}^{2}}\left(\epsilon_{\alpha_{1}\cdots\alpha_{d-n}\mu\lambda_{1}\cdots\lambda_{n}}
  \partial_{\mu}A_{\lambda_{1}\cdots\lambda_{n}}\right)^{2} \nonumber \\
&& +\frac{\kappa_{n}}{2}\left(\sum_{i=1}^{n}(-1)^{i-1}\partial_{\lambda_{i}}\theta_{\lambda_{1}\cdots\lambda_{i-1}\lambda_{i+1}\cdots\lambda_{n}}
  +A_{\lambda_{1}\cdots\lambda_{n}}\right)^{2} \nonumber
\end{eqnarray}
The canonical conjugates to $A_{\lambda_{1}\cdots\lambda_{n}}$ and $\theta_{\lambda_{1}\cdots\lambda_{n-1}}$ are:
\begin{eqnarray}
E_{\mu\lambda_{1}\cdots\lambda_{n}} &=& \frac{\delta\mathcal{L}}{\delta\partial_{\mu}A_{\lambda_{1}\cdots\lambda_{n}}} \\
\pi_{\mu\lambda_{1}\cdots\lambda_{n-1}} &=& \frac{\delta\mathcal{L}}{\delta\partial_{\mu}\theta_{\lambda_{1}\cdots\lambda_{n-1}}} \nonumber
  = n\,\frac{\delta\mathcal{L}}{\delta A_{\mu\lambda_{1}\cdots\lambda_{n-1}}} \nonumber \\
&& = -n\,\partial_{\nu}E_{\mu\nu\lambda_{1}\cdots\lambda_{n-1}} \nonumber
\end{eqnarray}
$E_{\mu\lambda_{1}\cdots\lambda_{n}}$ is the generalized field tensor ($F_{\mu\nu}$ in electrodynamics). The formulas for $\pi_{\mu\lambda_{1}\cdots\lambda_{n-1}}$ follow from gauge-invariance and the Lagrange field equation for the gauge field. The explicit expressions are:
\begin{eqnarray}\label{electric}
&& E_{\lambda_{1}\cdots\lambda_{n+1}} \!=\! \frac{(\epsilon_{\alpha_{1}\cdots\alpha_{d-n}\rho\nu_{1}\cdots\nu_{n}}\partial_{\rho}A_{\nu_{1}\cdots\nu_{n}})
      \epsilon_{\alpha_{1}\cdots\alpha_{d-n}\lambda_{1}\cdots\lambda_{n+1}}}{(d-n)! e_{n}^{2}} \nonumber \\
   && \qquad = \frac{n!}{e_{n}^{2}}\sum_{i=1}^{n+1}(-1)^{i-1}\partial_{\lambda_{i}}A_{\lambda_{1}\cdots\lambda_{i-1}\lambda_{i+1}\cdots\lambda_{n+1}} \\
&& \pi_{\lambda_{1}\cdots\lambda_{n}} \!=\! n\kappa_{n}\left\lbrack \sum_{i=1}^{n}(-1)^{i-1}\partial_{\lambda_{i}}
       \theta_{\lambda_{1}\cdots\lambda_{i-1}\lambda_{i+1}\cdots\lambda_{n}} \!-\! A_{\lambda_{1}\cdots\lambda_{n}}\right\rbrack \nonumber
\end{eqnarray}

The symmetry under translations by $a$
\begin{eqnarray}
x_{\mu} &\to& x_{\mu}+\delta a\,\delta_{\alpha\mu} \nonumber \\
A_{\lambda_{1}\cdots\lambda_{n}} &\to& A_{\lambda_{1}\cdots\lambda_{n}}+\delta a\,\partial_{\alpha}A_{\lambda_{1}\cdots\lambda_{n}} \nonumber \\
\mathcal{L} &\to& \mathcal{L}+\partial_{\mu}(\delta a\,\delta_{\alpha\mu}\mathcal{L}) \nonumber
\end{eqnarray}
yields the conserved canonical energy-momentum tensor:
\begin{eqnarray}
T_{\alpha,\mu} &=& E_{\mu\lambda_{1}\cdots\lambda_{n}}\,\partial_{\alpha}A_{\lambda_{1}\cdots\lambda_{n}} \nonumber \\ 
  && +\pi_{\mu\lambda_{1}\cdots\lambda_{n-1}}\,\partial_{\alpha}\theta_{\lambda_{1}\cdots\lambda_{n-1}} -\delta_{\alpha\mu}\mathcal{L} \nonumber \ .
\end{eqnarray}
Similarly, rotations lead to the conservation of angular momentum. Under infinitesimal rotations by $\delta\theta$
\begin{eqnarray}
R_{\alpha\beta}(\delta\theta) &=& e^{-iM_{\alpha\beta}\delta\theta} \to 1-iM_{\alpha\beta}\delta\theta \nonumber \\
(M_{\alpha\beta})_{\mu\nu} &=& -i(\delta_{\alpha\mu}\delta_{\beta\nu}-\delta_{\alpha\nu}\delta_{\beta\mu}) \nonumber \ ,
\end{eqnarray}
the coordinates $x_{\mu}$ and tensor fields like $A_{\lambda_{1}\cdots\lambda_{n}}$ transform as:
\begin{eqnarray}
x_{\mu} &\to& x_{\mu}-\delta\theta(\delta_{\alpha\mu}x_{\beta}-\delta_{\beta\mu}x_{\alpha}) \nonumber \\
A_{\lambda_{1}\cdots\lambda_{n}} &\to& A_{\lambda_{1}\cdots\lambda_{n}}+\delta\theta(x_{\alpha}\partial_{\beta}-x_{\beta}\partial_{\alpha})
  A_{\lambda_{1}\cdots\lambda_{n}} \nonumber \\
&& \quad +\delta\theta\,\Sigma_{\alpha\beta;\lambda_{1}\cdots\lambda_{n},\gamma_{1}\cdots\gamma_{n}}A_{\gamma_{1}\cdots\gamma_{n}} \nonumber \ .
\end{eqnarray}
Rotations in the $\alpha\beta$ plane are obtained when both $\alpha$ and $\beta$ are spatial indices, otherwise we have the generalized Lorentz transformations in the imaginary (Euclidean) space-time. The ``spin matrix'' $\Sigma_{\alpha\beta}$ is responsible for rotating the internal degrees of freedom of the field. For the gauge fields in this theory, $R_{\alpha\beta}(\delta\theta)$ is separately applied on each index of $A_{\lambda_{1}\cdots\lambda_{d-1}}$ and only the lowest-order terms are kept (the sense of rotation is opposite to that of $x_{\mu}$):
\begin{eqnarray}
\Sigma_{\alpha\beta;\lambda_{1}\cdots\lambda_{n},\gamma_{1}\cdots\gamma_{n}} &=&
  \sum_{k=1}^{n}(M_{\alpha\beta})_{\lambda_{k}\gamma_{k}}\prod_{i\neq k}\delta_{\lambda_{i}\gamma_{i}} \nonumber \\
&=& \sum_{k=1}^{n}(\delta_{\alpha\lambda_{k}}\delta_{\beta\gamma_{k}}-\delta_{\alpha\gamma_{k}}\delta_{\beta\lambda_{k}})\prod_{i\neq k}\delta_{\lambda_{i}\gamma_{i}}
  \nonumber \ .
\end{eqnarray}
As a scalar, the imaginary-time (Euclidean) Lagrangian density transforms according to:
\begin{eqnarray}
\mathcal{L} &\to& \mathcal{L}+\delta\theta(x_{\alpha}\partial_{\beta}-x_{\beta}\partial_{\alpha})\mathcal{L} \nonumber \\
  &=& \mathcal{L}+\delta_{\mu}\Bigl\lbrack\delta\theta(x_{\alpha}\delta_{\beta\mu}-x_{\beta}\delta_{\alpha\mu})\mathcal{L}\Bigr\rbrack
     -\delta\theta(\delta_{\beta\alpha}-\delta_{\alpha\beta})\mathcal{L} \nonumber \\
  &=& \mathcal{L}+\partial_{\mu}W_{\mu}  \nonumber \ ,
\end{eqnarray}
where
\begin{equation}
W_{\mu}=\delta\theta\,(x_{\alpha}\delta_{\beta\mu}-x_{\beta}\delta_{\alpha\mu})\mathcal{L} \ . \nonumber
\end{equation}
Therefore, the canonical angular momentum current density is:
\begin{eqnarray}
J_{\alpha\beta,\mu} &\propto& E_{\mu\lambda_{1}\cdots\lambda_{n}}\delta A_{\lambda_{1}\cdots\lambda_{n}}
     +\pi_{\mu\lambda_{1}\cdots\lambda_{n-1}}\delta\theta_{\lambda_{1}\cdots\lambda_{n-1}}-W_{\mu} \nonumber \\
&=& x_{\alpha}T_{\beta,\mu}-x_{\beta}T_{\alpha,\mu} \nonumber \\
&& +E_{\mu\lambda_{1}\cdots\lambda_{d-1}}\Sigma_{\alpha\beta;\lambda_{1}\cdots\lambda_{d-1},\gamma_{1}\cdots\gamma_{d-1}}
     A_{\gamma_{1}\cdots\gamma_{d-1}} \nonumber \\
&& +\pi_{\mu\lambda_{1}\cdots\lambda_{n-1}}\Sigma_{\alpha\beta;\lambda_{1}\cdots\lambda_{n-1},\gamma_{1}\cdots\gamma_{n-1}}\theta_{\gamma_{1}\cdots\gamma_{n-1}}
     \nonumber \ .
\end{eqnarray}
The canonical construction of $T_{\alpha,\mu}$ and $J_{\alpha\beta,\mu}$ using Noether's theorem ensures the conservation laws $\partial_{\mu}T_{\alpha,\mu}=0$ and $\partial_{\mu}J_{\alpha\beta,\mu}=0$. However, the canonical tensors $T_{\alpha,\mu}$ and $J_{\alpha\beta,\mu}$ are not symmetric in their indices and do not look gauge-invariant. These issues are fixed by symmetrizing the energy-momentum tensor.

The symmetrized energy-momentum tensor and angular momentum currents (without commas separating their indices) are:
\begin{eqnarray}
T_{\mu\nu} &=& \frac{1}{n\,\kappa_{n}}\pi_{\mu\lambda_{1}\cdots\lambda_{n-1}}\pi_{\nu\lambda_{1}\cdots\lambda_{n-1}} \\
  && +\frac{e_{n}^{2}}{n!} E_{\mu\lambda_{1}\cdots\lambda_{n}}E_{\nu\lambda_{1}\cdots\lambda_{n}}-\delta_{\mu\nu}\mathcal{L} \nonumber \\[0.1in]
J_{\alpha\beta\mu} &=& x_{\alpha}T_{\beta\mu}-x_{\beta}T_{\alpha\mu} \nonumber \ .
\end{eqnarray}
They differ from the canonical ones only by total derivatives and hence describe the same bulk quantities.

\section{Braiding of multi-dimensional excitations}\label{appMultiBraiding}

A useful operation that generalizes from particle braiding is braiding of multi-dimensional objects. Consider creating a (quasi) particle-antiparticle pair at some point in space, driving the particle along a closed loop path, and eventually annihilating the pair. This operation leaves behind a Dirac string loop. Now consider creating adjacent Dirac loops that completely cover a sphere. If all loops have the same orientation relative to the local orientation of the sphere's surface, then the initial and final states are both free of charge or gauge flux and hence differ only by a phase. This phase can be fractionalized and it can characterize topological order. Let us begin with constructing the operator that creates a Dirac loop. First we want to create a particle-antiparticle pair in terms of charge. The operator:
\begin{equation}
\psi^{\dagger}({\bf x})=e^{-i\nu\theta({\bf x})}
\end{equation}
creates a point-like lump of charge $\nu$ at location ${\bf x}$ if $\theta$ is the angle operator conjugate to the integer-valued particle number, i.e. the usual U(1) phase. Creating a particle-antiparticle pair separated by a distance $\delta{\bf x}$ is accomplished with:
\begin{equation}
\psi({\bf x})\psi^{\dagger}({\bf x}+\delta{\bf x})\to e^{-i\nu(\partial_{\mu}\theta+A_{\mu})\delta x_{\mu}} \ .
\end{equation}
We have gauged this operation in order to create a physical state. Movement of a particle along a path is done by chaining similar pair-creation operations: in every infinitesimal movement step, one creates a new particle-antiparticle pair displaced by $\delta{\bf x}$ along the path in such a way that the antiparticle lands at the same position as the old ``driven'' particle and annihilates it, leaving behind just the new particle at a new position. It is easy to see that the loop creation operation is given by the operator:
\begin{eqnarray}
&& \prod_{\delta x_{\mu}\in\mathcal{P}}e^{-i\nu(\partial_{\mu}\theta+A_{\mu})\delta x_{\mu}}
  = \exp\left(-i\nu\oint\limits_{\mathcal{P}}dx_{\mu}(\partial_{\mu}\theta+A_{\mu})\right) \nonumber \\
&& \qquad = \exp\left(-i\nu\oint\limits_{\mathcal{P}}dx_{\mu}\,j_{\mu}\right) \ .
\end{eqnarray}
For the simplicity of notation, we work here with a renormalized charge current $j_{\mu}=\partial_{\mu}\theta+A_{\mu}$ ``per particle'' that does not contain the incompressible $|\psi|^{2}$ density factor. The quasiparticles of an incompressible quantum liquid combine a monopole with a lump $\nu$ of charge, so we also need to create a monopole/antimonopole pair and drive the monopole around the loop. Since the monopole is a point-particle, its normalized current $\mathcal{J}_{\mu} \sim \partial_{\mu} \phi$ can be similarly written in terms of an operator $e^{i\phi}$ that creates a monopole. By duality, we simply need to replace the charge current operator $j_{\mu}$ with the normalized defect current operator $\mathcal{J}_{\mu}$ in the above formula. The full Dirac loop operator is then:
\begin{equation}
L_{\mathcal{P}}=\exp\left(-i\oint\limits _{\mathcal{P}}dx_{\mu}(\nu j_{\mu}+q_{d}\mathcal{J}_{\mu})\right) \ .
\end{equation}
One must independently find the charge $q_{d}$. Since the operator $L_{\mathcal{P}}$ creates a Dirac loop, we can immediately extend it to an operator that moves the loop sweeping an open cylindrical surface $S$ from one end to another:
\begin{equation}
L_{\mathcal{S}} = \exp\left(-i\int\limits_{\mathcal{S}}da_{\mu}\,\epsilon_{\mu\nu\lambda}\partial_{\nu}(\nu j_{\lambda}+q_{d}\mathcal{J}_{\lambda})\right) \nonumber \ .
\end{equation}
Here, $da_{\mu}$ is a vector normal to the surface with a magnitude equal to the local surface area element. The surface is orientable, and its orientation in the integral corresponds to the direction of sweep. At this stage, we observe that the Dirac attachment links between different gauge field ranks require us to transport the matter and gauge field configurations associated with the quasiparticle at all intermediate ranks. The rank 2 matter field describes one-dimensional extended objects, which are actually quantized flux lines linked to Dirac strings. We must generate the appropriate rank 2 matter field, and the rank 2 gauge field it couples to, when we create and move a Dirac loop. Defining the gauge-invariant rank 2 current $j_{\mu\nu} = \partial_{\mu}\theta_{\nu} - \partial_{\nu}\theta_{\mu} - A_{\mu\nu}$, we have a correction:
\begin{equation}
L_{\mathcal{S}} = \exp\left(-i\int\limits_{\mathcal{S}}da_{\mu}\,\epsilon_{\mu\nu\lambda}\Bigl\lbrack
  \partial_{\nu}(\nu j_{\lambda}+q_{d}\mathcal{J}_{\lambda})+q_{2}j_{\nu\lambda}\Bigr\rbrack\right) \ .
\end{equation}
The charge constant $q_{2}$ for $j_{\mu\nu}$ needs to be separately determined. We have constructed the correction to $L_{S}$ purely on symmetry grounds and by formal analogy to the operator $L_{\mathcal{P}}$ that moves a point-like object along a one-dimensional path. More formally, $j_{\mu}$ is equivalent to the canonical momentum of the matter field per particle (the density factor $|\psi|^{2}$ is stripped away), and indeed generates translations. Similarly, $j_{\mu\nu}$ is the canonical momentum of the rank 2 matter field per particle per unit-length (given its units), so it can be used as above to generate translations of a line-shaped object. This would be all in $d=3$ ($\mathcal{J}_{\lambda}$ contains a derivative of $A_{\mu\nu}$), and we could continue in the same fashion to higher ranks in $d>3$ by defining operators that create $n-1$ dimensional objects and sweep them across $n$ dimensional manifolds $\mathcal{M}_{n}$:
\begin{eqnarray}
L_{\mathcal{M}_{n}} &=& \exp\biggl\lbrack -i\int\limits_{\mathcal{M}_{n}}\prod_{i=0}^{n}dx_{i}\;\epsilon_{\mu_{1}\cdots\mu_{n}}\,
  \biggl(q_{n}j_{\mu_{1}\cdots\mu_{n}} \\
&& +\frac{q_{n-1}}{n}\sum_{i=1}^{n}(-1)^{i-1}\partial_{\mu_{i}}j_{\mu_{1}\cdots\mu_{i-1}\mu_{i+1}\cdots\mu_{n}}
  +\cdots\biggr)\biggr\rbrack \ . \nonumber
\end{eqnarray}
If the manifold $\mathcal{M}_{n}$ is closed (without a boundary), then only the highest-rank gauge field can contribute:
\begin{equation}
L_{\mathcal{M}_{n}}=\exp\left(-iq_{n}\oint\limits_{\mathcal{M}_{n}}\prod_{i=0}^{n}dx_{i}\;\epsilon_{\mu_{1}\cdots\mu_{n}}\,A_{\mu_{1}\cdots\mu_{n}}\right) \ .
\end{equation}
So, creating a loop, sweeping it across a closed surface $S$, and then annihilating it in $d=3$ is achieved by:
\begin{eqnarray}\label{sweep2D}
L_{\mathcal{S}} &=& \exp\left(-i\oint\limits_{\mathcal{S}}da_{\mu}\,\epsilon_{\mu\nu\lambda}\Bigl\lbrack
  \partial_{\nu}(\nu j_{\lambda}+q_{3}\mathcal{J}_{\lambda})+q_{2}j_{\nu\lambda}\Bigr\rbrack\right) \nonumber \\
&=& \exp\left(-iq_{2}\oint\limits_{\mathcal{S}}da_{\mu}\,\epsilon_{\mu\nu\lambda}A_{\nu\lambda}\right) = e^{-2\pi iq_{2}N} \ ,
\end{eqnarray}
where $N$ is the total $\pi_{2}(S^{2})$ topological charge enclosed by $S$. The currents $j_{\mu}$ and $\mathcal{J}_{\mu}$ have no curl in an incompressible quantum liquid, so they dropped out.

\section{Fractional braiding statistics}\label{appBraiding}

Here we analyze the Aharonov-Bohm phase in both dynamically and topologically protected braiding operations in $d=3$ dimensions. The starting point is the following Lagrangian density of an incompressible quantum liquid with monopoles
\begin{eqnarray}\label{BrL1}
\mathcal{L} &=& -j_{\mu}A^{\mu} -j_{\mu\nu}A^{\mu\nu} - \kappa\left(\frac{F_{\mu\nu}}{2}-A_{\mu\nu}\right)\left(\frac{F^{\mu\nu}}{2}-A^{\mu\nu}\right) \nonumber \\
&&   -\frac{\nu}{4\pi}\epsilon^{\mu\nu\alpha\beta}A_{\mu}\partial_{\nu}A_{\alpha\beta} + \mathcal{L}_{\textrm{M}1} + \mathcal{L}_{\textrm{M}2} \ ,
\end{eqnarray}
where $F_{\mu\nu} = \partial_\mu A_\nu - \partial_\nu A_\mu$, $\mathcal{L}_{\textrm{M}n}$ are rank $n$ Maxwell terms, and the external currents $j_\mu$ and $j_{\mu\nu}$ describe inserted fractional excitations -- point-like quasiparticles at rank 1 and loops at rank 2 respectively.

Consider first the braiding of two point quasiparticles. The braiding outcome is non-trivial even though it is not topologically protected. The external currents of two quasiparticles at positions ${\bf x}_1(t)$ and ${\bf x}_2(t)$ are $j^{\mu\nu} = 0$ and $j^\mu = j^\mu_1 + j^\mu_2$:
\begin{eqnarray}
j_{1}^{0}({\bf x},t)=q_{1}^{\phantom{x}}\delta\Bigl({\bf x}-{\bf x}_{1}^{\phantom{x}}(t)\Bigr) &\;\;,\;\;&
   j_{1}^{i}({\bf x},t)=q_{1}^{\phantom{x}}\dot{x}_{1}^{i}\delta\Bigl({\bf x}-{\bf x}_{1}^{\phantom{x}}(t)\Bigr) \nonumber \\
j_{2}^{0}({\bf x},t)=q_{2}^{\phantom{x}}\delta\Bigl({\bf x}-{\bf x}_{2}^{\phantom{x}}(t)\Bigr) &\;\;,\;\;&
   j_{2}^{i}({\bf x},t)=q_{2}^{\phantom{x}}\dot{x}_{2}^{i}\delta\Bigl({\bf x}-{\bf x}_{2}^{\phantom{x}}(t)\Bigr) \nonumber 
\end{eqnarray}
Integrating out $\delta A_{\mu\nu}$ in $A_{\mu\nu}=\frac{1}{2}F_{\mu\nu}+\delta A_{\mu\nu}$ renormalizes the rank 1 Maxwell term and replaces the topological term with an axion term:
\begin{equation}
\mathcal{L} = -j_{\mu}A^{\mu}-\frac{\nu}{4\pi}\epsilon^{\mu\nu\alpha\beta}A_{\mu}\partial_{\nu}\partial_{\alpha}A_{\beta} + \mathcal{L}_{\textrm{M}1} \ .
\end{equation}
If we define the operator
\begin{equation}\label{BrC}
\mathcal{C}^{\mu\nu}=\frac{\nu}{2\pi}\epsilon^{\mu\nu\alpha\beta}\partial_{\alpha}\partial_{\beta} \ ,
\end{equation}
then integrating out $A_{\mu}$ yields:
\begin{equation}
\mathcal{L} = \frac{1}{2}j^{\mu}\mathcal{C}_{\mu\nu}^{-1}j^{\nu} + \cdots \ ,
\end{equation}
when we neglect the ``radiative'' interactions between charges and currents induced by electromagnetic field fluctuations (through $\mathcal{L}_{\textrm{M}1}$). It is important to note that (\ref{BrC}) is a symmetric operator (its anti-symmetric parts would not contribute the braiding action $S$). Namely, transposing $\mathcal{C}$ flips the sign through the exchange of external indices $\mu,\nu$, but this sign flip is canceled in the integration by parts $\epsilon^{\cdots\alpha\beta} \partial_\alpha \partial_\beta \to \epsilon^{\cdots\alpha\beta} (-\partial_\beta) (-\partial_\alpha)$. The effective braiding action
\begin{equation}\label{BrA}
S = \int d^4 x\; \mathcal{L} = \int d^4 x\; \frac{1}{2} \left( j_{1\mu}^{\phantom{,}}\mathcal{A}_{2}^{\mu}+j_{2\mu}^{\phantom{,}}\mathcal{A}_{1}^{\mu} \right)
\end{equation}
captures the Aharonov-Bohm phase associated with the braiding of two quasiparticles around each other. Each quasiparticle picks the Aharonov-Bohm phase only from the flux attached to the other quasiparticle, via
\begin{equation}
\mathcal{A}_{i\mu}^{\phantom{'}} = \mathcal{C}_{\mu\nu}^{-1}j_{i}^{\nu} \quad\Rightarrow\quad \mathcal{C}^{\mu\nu}\mathcal{A}_{i\nu}
  = \frac{\nu}{2\pi}\epsilon^{\mu\nu\alpha\beta}\partial_{\alpha}\partial_{\beta}\mathcal{A}_{i\nu} = j_{i}^{\mu} \ . \nonumber
\end{equation}
Substituting the above formulas for $j_n^\mu({\bf x},t)$ allows us to determine $\mathcal{A}_{n}^{\mu}$. For a quasiparticle at rest, $\mathcal{A}_{n}^{0}=0$ and:
\begin{equation}
\epsilon^{0ijk}\partial_{i}\partial_{j}\mathcal{A}_{nk} = \frac{2\pi q_n}{\nu} \delta({\bf x}-{\bf x}_{n}) \ .
\end{equation}
Therefore, $\mathcal{A}_n^\mu$ is the U(1) gauge field of a monopole at ${\bf x}_n^{\phantom{,}}$ with topological charge $2\pi q_n/\nu$. The fractional charge of a point quasiparticle in the Laughlin-like incompressible quantum liquid with the filling factor $\nu = 1/m$ is $q_n = \nu$. This endows $\mathcal{A}_n^\mu$ with a single $2\pi$ unit of monopole flux. Each quasiparticle of charge $q_n=\nu$ picks an Aharonov-Bohm phase from the monopole quantum attached to the other quasiparticle, but (\ref{BrA}) associates only a half of the total two-particle phase to the braiding operation. This result is analogous to the $d=2$ case of quantum Hall liquids \cite{WenQFT2004}, but different from the $d=3$ field-induced correction calculated in Ref.\cite{Goldhaber1989}

As an example, simulate the exchange of two identical quasiparticles by driving them on opposite semi-circular paths in a single plane about their center of mass. Each quasiparticle contributes $\frac{1}{2} \times \frac{\Omega\nu}{2}$ to the Aharonov-Bohm phase; $\Omega=2\pi$ is the solid angle that subtends a closed loop in the braiding plane, and the extra factor of $\frac{1}{2}$ is the patch for the semi-circular path. The braiding phase is a half of the total Aharonov-Bohm phase of both quasiparticles, $\varphi = \frac{1}{2} \nu \pi$. This phase is not topologically protected, but if one is able to precisely control the quasiparticle trajectories then the value of $\varphi$ is protected dynamically -- it can be arbitrarily changed only at the cost of exciting additional gapped excitations that distort the field lines of charges and monopoles. There is also another issue regarding the gauge-dependent choice for the unavoidable Dirac string, discussed in Appendix \ref{appDyonExchange}.

A few comments are in order. The above derivation utilizes antisymmetrized combinations of derivatives, which vanish when applied to any analytic function. Hence, we rely on singular field configurations at rank 1 in order to describe monopoles. As explained in the paper, quantized monopole singularities in the rank 1 gauge fields are made possible by the compact regularization of the field theory, and their mathematical structure is analogous to that of a U(1) phase $\theta$ in the description of quantized vortices. The antisymmetrized derivatives of gauge fields should be computed $\textrm{mod}\;2\pi$. If we naively annihilated them, we would never be able to capture any interaction between charge currents and monopoles at rank 1.

Now, let us calculate the braiding phase of a quasiparticle and a loop. Repeating the above procedure of integrating out $A_{\mu\nu}$ and then $A_\mu$ in (\ref{BrL1}) leads to the effective Lagrangian density
\begin{eqnarray}\label{BrL2}
\mathcal{L} &\to& -j_{\mu}A^{\mu} - \frac{F^{\mu\nu}}{2}j_{\mu\nu} - \frac{\nu}{8\pi}\epsilon^{\mu\nu\alpha\beta}A_{\mu}\partial_{\nu}F_{\alpha\beta}
   + \mathcal{L}_{\textrm{M}1} + \cdots \nonumber \\
&\to& \frac{1}{2}(j^{\mu}-j^{\mu\alpha}\partial_{\alpha}^{\phantom{x}})\mathcal{C}_{\mu\nu}^{-1}(j^{\nu}-\partial_{\beta}^{\phantom{x}}j^{\nu\beta}) + \cdots \ ,
\end{eqnarray}
with an abuse of notation: $\partial_\beta$ from the last bracket acts to the left on the objects outside the bracket. We are again neglecting radiative contributions to current-current interactions. The braiding action is
\begin{eqnarray}
S &=& \int d^{4}x\,\frac{1}{2}\Bigl\lbrack -j^{\mu\alpha}\partial_{\alpha}^{\phantom{x}}\mathcal{C}_{\mu\nu}^{-1}j^{\nu}
        -j^{\nu\beta}\partial_{\beta}^{\phantom{x}}\mathcal{C}_{\mu\nu}^{-1}j^{\mu}\Bigr\rbrack \nonumber \\
  &=& \int d^{4}x\,j^{\mu\nu}\mathcal{A}_{\mu\nu} \ ,
\end{eqnarray}
where
\begin{equation}
\mathcal{C}^{\mu\alpha}\mathcal{A}_{\mu\nu} = -\partial_{\nu}^{\phantom{x}}j^{\alpha} \nonumber \ .
\end{equation}
If we rewrite $\mathcal{A}_{\mu\nu} = \partial_\mu a_\nu - \partial_\nu a_\mu$, then
\begin{equation}
\epsilon^{\mu\nu\alpha\beta}\partial_{\alpha}\partial_{\beta}(\partial_{\mu}a_{\lambda}-\partial_{\lambda}a_{\mu}) 
  = -\frac{2\pi}{\nu} \partial_{\lambda}^{\phantom{x}}j^{\nu}
\end{equation}
yields a monopole solution
\begin{equation}
\epsilon^{\mu\nu\alpha\beta}\partial_{\alpha}\partial_{\beta}\partial_{\mu} a_{\lambda}=0 \quad,\quad
  \epsilon^{\mu\nu\alpha\beta}\partial_{\alpha}\partial_{\beta}a_{\mu} = \frac{2\pi}{\nu} j^{\nu}
\end{equation}
without higher-rank singularities. Since the fractional quasiparticle carries charge $j^0 \propto q_1 = \nu$, the ensuing rank 2 gauge field $\mathcal{A}_{\mu\nu}$ is simply the ``electromagnetic'' field tensor of a unit $2\pi$ monopole attached to the quasiparticle. The braiding phase is the full Aharonov-Bohm phase collected from a $2\pi$ monopole quantum at rank 2. We also need the loop's rank 2 charge $q_{2}$. The rank 2 ``electric charge'' density (per unit volume and loop's unit-length) $j^{0i}$ couples to $A_{0i}$ in the Lagrangian density, and similarly, the rank 2 ``magnetic flux'' density $j^{ij}$ couples to $A_{ij}$. The linking Lagrangian terms relate $A_{\mu\nu}$ to $\frac{1}{2}F_{\mu\nu}$, so the ratios of electric and magnetic charges are the same in the two descriptions. The magnetoelectric effect derived in Section \ref{secME} relates electric and magnetic fields as ${\bf E} = -2\alpha\nu{\bf B}$ where $\alpha = e^{2}/\hbar c \to 1/4\pi$ is the fine-structure constant in natural units. Therefore, a $2\pi$ unit of magnetic flux binds $\nu$ electric field units in the loop. The loop is a fractional excitation that carries rank 2 charge $q_2 = \nu$.

As an example, consider driving the point-particle on a closed path through the loop. Relative to the quasiparticle, the loop sweeps a torus-shaped surface that encloses the quasiparticle and collects all of its monopole's flux. The rank 2 Aharonov-Bohm phase (\ref{sweep2D}) is $2\pi\nu$, so the braiding phase is $\varphi = 2\pi\nu$. This is naively expected from the Aharonov-Bohm effect at rank 1, for a quasiparticle of charge $q_1=\nu$ that encircles a $2\pi$-quantized vortex loop, but now we have a confirmation that there are no special corrections from the topological Lagrangian term.

Lastly, we briefly show that loop-loop braiding is trivial in the presently considered topological orders. The braiding action for two loops described with currents $j^\mu=0$ and $j^{\mu\nu}=j^{\mu\nu}_1+j^{\mu\nu}_2$ is obtained from (\ref{BrL2}):
\begin{equation}
S = \int d^{4}x\,\frac{1}{2}\Bigl\lbrack-j_{1}^{\mu\alpha}\partial_{\alpha}^{\phantom{x}}\mathcal{C}_{\mu\nu}^{-1}\partial_{\beta}^{\phantom{x}}j_{2}^{\nu\beta}
  -j_{2}^{\mu\alpha}\partial_{\alpha}^{\phantom{x}}\mathcal{C}_{\mu\nu}^{-1}\partial_{\beta}^{\phantom{x}}j_{1}^{\nu\beta}\Bigr\rbrack \to 0 \nonumber
\end{equation}
after an integration by parts which transfers $\partial_\beta$ onto a target on its right. This action vanishes because the closed rank 2 loops satisfy $\partial_\mu j^{\mu\nu} = \partial_\nu j^{\mu\nu} = 0$. Physically, this braiding only moves electric and magnetic flux lines around one another, and hence does not generate Aharonov-Bohm phases. The situation would have been different if rank 1 electric or magnetic charge were attached to the loops \cite{Wang2014a, Jiang2014, Wang2014b, Juan2014, Wang2015}.

\section{The electromagnetic angular momentum of charges and monopoles}\label{appZwanziger}

Consider a system of point charges $e_{i}^{\phantom{,}}$ at locations ${\bf x}_{i}$ and point monopoles $m_{j}$ at positions ${\bf x}_{j}$. The angular momentum of the electromagnetic field contributed by the charge $e_{i}$ and all monopoles relative to ${\bf x}_{i}$ is:
\begin{eqnarray}
{\bf L}_{i} &=& \int d^{3}x\:({\bf x}-{\bf x}_{i})\times\Bigl({\bf E}({\bf x})\times{\bf B}({\bf x})\Bigr) \\
  &=& \frac{e_{i}}{4\pi}\int d^{3}x\:\frac{1}{|{\bf x}|}\left\lbrace \hat{{\bf x}}
      \Bigl\lbrack\hat{{\bf x}}\,{\bf B}({\bf x}+{\bf x}_{i})\Bigr\rbrack-{\bf B}({\bf x}+{\bf x}_{i})\right\rbrace \ . \nonumber
\end{eqnarray}
Using:
\begin{equation}
({\bf B}\boldsymbol{\nabla})\hat{{\bf x}} = \frac{{\bf B}-\hat{{\bf x}}(\hat{{\bf x}}{\bf B})}{|{\bf x}|}
\end{equation}
we obtain:
\begin{eqnarray}
{\bf L}_{i} &=& -\frac{e_{i}}{4\pi}\int d^{3}x\:\Bigl\lbrack{\bf B}({\bf x}+{\bf x}_{i})\boldsymbol{\nabla}\Bigr\rbrack\hat{{\bf x}} \\
  &\to& \frac{e_{i}}{4\pi}\int d^{3}x\:\Bigl\lbrack\boldsymbol{\nabla}{\bf B}({\bf x}+{\bf x}_{i})\Bigr\rbrack\hat{{\bf x}} \nonumber \\
  &=& \frac{2\pi}{4\pi}\sum_{j}e_{i}m_{j}\int d^{3}x\:\delta({\bf x}+{\bf x}_{i}-{\bf x}_{j})\hat{{\bf x}} \nonumber \\
  &=& \frac{1}{2}\sum_{j}e_{i}m_{j}\delta\hat{{\bf x}}_{ij} \nonumber \ .
\end{eqnarray}
The arrow indicates integration by parts, and $\delta{\bf x}_{ij}={\bf x}_{j}-{\bf x}_{i}=|\delta{\bf x}_{ij}|\delta\hat{{\bf x}}_{ij}$. Since the total momentum carried by the electromagnetic field is zero, the angular momentum does not depend on the anchor point. The total electromagnetic angular momentum of all charges and monopoles, with respect to any anchor point, is:
\begin{equation}\label{angmom}
{\bf L} = \sum_{i}{\bf L}_{i} = \frac{1}{2}\sum_{ij}e_{i}m_{j}\delta\hat{{\bf x}}_{ij} \ .
\end{equation}
This angular momentum was computed classically, and hence represents the expectation value $\langle\Psi|L|\Psi\rangle$ of the angular momentum operator $L$ in the quantum state $|\Psi\rangle$ of charges, monopoles and their electromagnetic field. If $|\Psi\rangle$ is an eigenstate of $L$ (along some direction), then the above expression must have a quantized magnitude in agreement with the quantization of angular momentum eigenvalues. We will specialize to such circumstances by aligning all $\delta\hat{{\bf x}}_{ij}$ in the same direction. 

Let us calculate the angular momentum produced by $N$ dyons whose sizes are negligible next to their separations. If $i^{\textrm{th}}$ dyon carries electric charge $e_{i}$ and magnetic charge $m_{i}$, then the total angular momentum is:
\begin{eqnarray}\label{totAngMom}
{\bf L} &=& \frac{1}{2}\sum_{i=1}^{N}\sum_{j=1}^{N}e_{i}m_{j}\delta\hat{{\bf x}}_{ij} \\
  &=& \frac{1}{2}\sum_{i}e_{i}m_{i}\delta\hat{{\bf x}}_{ii}+\frac{1}{2}\sum_{i<j}(e_{i}m_{j}-e_{j}m_{i})\delta\hat{{\bf x}}_{ij} \ . \nonumber
\end{eqnarray}
If all $\delta\hat{{\bf x}}_{ii}$ and $\delta\hat{{\bf x}}_{ij}=-\delta\hat{{\bf x}}_{ji}$ are independent variables, then the angular momentum quantization requires:
\begin{eqnarray}\label{DiracZwanziger}
(\forall i)\quad&&\quad e_{i}m_{i}\in\mathbb{Z} \nonumber \\
(\forall i\neq j)\quad&&\quad e_{i}m_{j}-e_{j}m_{i}\in\mathbb{Z} \ .
\end{eqnarray}
The former is Dirac's monopole charge quantization \cite{Dirac1931} and the latter is Schwinger-Zwanziger condition \cite{Schwinger1969, Zwanziger1968}. These conditions ensure that the sign changes of $\delta\hat{{\bf x}}_{ij}$, i.e. internal dyon rotations and pairwise dyon exchanges, do not violate angular momentum quantization or dyon exchange statistics. All alterations can change the total angular momentum only by an integer (multiple of $\hbar$). Note that an exchange of two identical dyons does not affect the total angular momentum. Violating either of these conditions due to a fractionalization of $e_{i}$ would need to be compensated by long-range correlations among $\delta\hat{{\bf x}}_{ij}$ in order to protect the quantization of $\bf L$.

The two conditions (\ref{DiracZwanziger}) together seem to prohibit charge fractionalization unless monopoles can exist only in multi-monopole clusters. However, there is a way out. Classically, a system of non-coinciding charges and monopoles that satisfy Dirac quantization can be transformed into a system of point-like dyons though a duality mapping ${\bf E}+i{\bf B} \to e^{i\theta}({\bf E}+i{\bf B})$ of electrodynamics with electric and magnetic charges. The resulting dyons satisfy Schwinger-Zwanziger condition. The total electromagnetic angular momentum is invariant under duality, and thus does not obtain contributions from the internal structure of dyons after duality, as if $\delta\hat{{\bf x}}_{ii} \equiv 0$ in (\ref{totAngMom}). This hints a path to the eventual quantum regularization, where neither electric nor magnetic charge are ever localized at a single point. At the least, it is more appropriate to write:
\begin{equation}
{\bf L}=\sum_{i}{\bf L}_{i}+\frac{1}{2}\sum_{i<j}(e_{i}m_{j}-e_{j}m_{i})\delta\hat{{\bf x}}_{ij} \ ,
\end{equation}
where ${\bf L}_{i}$ is the intrinsic dyon's quantum spin.

Next, we briefly review the quantum dynamics of dyon's emergent spin. Consider an elementary dyon made from a unit-charge $e=1$ and unit-monopole $m=1$. The electromagnetic angular momentum (\ref{angmom}) must be included in the angular momentum operator of the dyon. If the monopole is fixed at the origin, then the charge-particle's stationary state can be an eigenstate of the total angular momentum operators $L^2$ and $L_z$ by rotational symmetry. One can easily verify that the operators
\begin{equation}
L_i = -i\epsilon_{ijk}x_{j}(\partial_{k}-iA_{k}) - \frac{em}{2} \frac{x_{i}}{|{\bf x}|}
\end{equation}
satisfy the commutation relations $\lbrack L_i, L_j \rbrack = i\epsilon_{ijk} L_k$ for $i,j,k\in\lbrace x,y,z\rbrace$. The plain kinetic energy Hamiltonian of the charge-particle is found to be:
\begin{equation}
H = \frac{1}{2M r^2} \left\lbrack -\frac{\partial}{\partial r} \left(r^2 \frac{\partial}{\partial r}\right) + L^2 - \left(\frac{em}{2}\right)^2 \right\rbrack
\end{equation}
in spherical coordinates. Since the kinetic energy is positive, the minimum angular momentum magnitude is $em/2 \to 1/2$. Consequently, there are two degenerate lowest-energy eigenstates, corresponding to $L_z \to \pm 1/2$. The angular part of the wavefunctions is given by the monopole harmonics $Y_{\frac{1}{2}, l, m}$:
\begin{eqnarray}
Y_{\frac{1}{2},\frac{1}{2},\frac{1}{2}} &=& -\frac{1}{\sqrt{2\pi}} \sin\left(\frac{\theta}{2}\right) e^{i\phi} \\
Y_{\frac{1}{2},\frac{1}{2},-\frac{1}{2}} &=& -\frac{1}{\sqrt{2\pi}} \cos\left(\frac{\theta}{2}\right) \nonumber
\end{eqnarray}
in the gauge
\begin{equation}
{\bf A} = \frac{1}{2r} \frac{1-\cos\theta}{\sin\theta} \hat{\boldsymbol{\phi} }
\end{equation}
for the monopole's magnetic field
\begin{equation}
{\bf B} = \boldsymbol{\nabla} \times {\bf A} = \frac{\hat{\bf r}}{2r} \ .
\end{equation}
It is interesting to observe that the gauge-invariant current
\begin{eqnarray}\label{MonHarmonicCurrents}
J_\mu^{\phantom{*}} &\propto& -\frac{i}{2}\Bigl\lbrack Y^* (\partial_\mu Y) - (\partial_\mu Y^*) Y \Bigr\rbrack - |Y|^2 A_\mu \nonumber \\
  &=& \pm \frac{\sin\theta}{8\pi r}\,\hat{\phi}_\mu
\end{eqnarray}
contains vortex-like flow (concentrated near the monopole and with a proper core) that builds a magnetic moment consistent with the total angular momentum direction $L_z = \pm 1/2$.

\section{Dyon braiding}\label{appDyonExchange}

Here we attempt to analyze the braiding of dyons with a few thought experiments. We will initially assume that the Aharonov-Bohm phase completely determines the braiding statistics, and run into a paradox: the statistics is not symmetric and depends on the braiding path. We will then ``resolve'' the paradox by making a naive assumption that quantized Dirac strings have a physical effect on fractional charges. The ultimate correct understanding of Aharonov-Bohm phases will be obtained at the end from a closer scrutiny of dyon quantum mechanics.

Suppose that a dyon consists of a fractional charge $\nu$ bound to a monopole quantum. Driving a dyon on a loop $\mathcal{P}$ near another static dyon produces the Aharonov-Bohm phase
\begin{equation}\label{MSA}
\varphi = \nu\oint\limits_{\mathcal{P}}\delta{\bf y}\,{\bf A}({\bf y}) = \frac{\nu}{2} \Omega_{\mathcal{P}} \ ,
\end{equation}
where $\Omega_{\mathcal{P}}$ is the solid angle through which the loop is seen from the static dyon. Let us simulate an exchange of two identical dyons in $d=3$, by driving them in opposite directions on a circle centered at their center of mass. Relative to dyon 1, the dyon 2 completes a half of a twice larger circle which is seen through the solid angle $\Omega=2\pi$. This accumulates the phase $\varphi_2 = \nu\pi/2$ for the dyon 2. The same happens to the dyon 1, so the total Aharonov-Bohm phase of both dyons is $\varphi_1+\varphi_2 = \nu\pi$. The two-body wavefunction of dyons acquires a half of this phase as calculated in Appendix \ref{appBraiding}.

Since $\Omega_{\mathcal{P}}$ is well-defined only modulo $4\pi$, the Aharonov-Bohm phase $\varphi$ is well-defined modulo $2\pi\nu$. This is fine for integer $\nu$, but presents a problem when $\nu$ is fractional. Even worse, the simulated monopole-quasiparticle exchange is generally anisotropic: the time-reversal $\varphi = \nu\pi \to -\nu\pi \neq \varphi + 2\pi n$ has the same effect as the reorientation $\hat{\bf z}\to -\hat{\bf z}$ of the exchange rotation axis. Both issues could be easily resolved if the Dirac strings were detectable in the presence of fractional charge. A monopole source $A_{\mu\nu}$ is fundamentally isotropic, but the rank 1 gauge field linked to it by $A_{\mu\nu} \sim \frac{1}{2}(\partial_\mu A_\nu - \partial_\nu A_\mu)$ cannot be isotropic due to a necessary Dirac string. This inter-rank link is frustrated and the attached Dirac string spontaneously breaks the rotation symmetry. Then, the $4\pi$ uncertainty of $\Omega_{\mathcal{P}}$ is just the contribution of the Dirac string to the Aharonov-Bohm phase, and its natural anisotropy leads to the exchange anisotropy. 

So, can a fractional charge see a $2\pi$ quantized Dirac string? The Aharonov-Bohm phase is well-defined for any charge and flux in the absolute continuum limit. However, the fundamental quantization of charge and flux is defined only with a compact regularization. A Dirac string is truly deprived of any physical content in a compact lattice gauge theory -- it can be erased by a gauge transformation that looks singular in the continuum limit, but cannot be sharply distinguished from a non-singular transformation on a lattice. Let us explicitly construct such a transformation. Consider a straight quantized string given by the continuum-limit gauge field ${\bf A}(r,\theta,z) = \hat{\boldsymbol{\theta}}/r$ expressed in cylindrical coordinates. If we place this gauge field on a lattice, then we can carry out a gauge transformation to collect all of its $\oint d{\bf y} {\bf A} = 2\pi$ into a single lattice link on any loop that encloses the string. The resulting gauge field lives inside factors $e^{i A_{ij}}$ in a compact gauge theory, and hence can be trivially removed due to the equivalence of $A_{ij} \in \lbrace 0, 2\pi \rbrace$. At the same time, the wavefunction of a nearby particle that couples to the gauge field will acquire a global $\Delta\varphi = 2\pi$ phase winding by this gauge transformation. The gauge-invariant charge current $J_\mu \sim \partial_\mu\varphi + A_\mu$ is not changed by this Dirac string removal. In this sense, the string specifically associated with the gauge field is not physically observable with gauge-invariant operators.

On the other hand, we are working with artificial gauge fields in this paper. When we apply a singular gauge transformation to extract a gauge field from ordinary matter, that gauge field represents physical currents in a singular gauge. A quantized Dirac string represents a vortex of charge currents, and the analysis of topological ground state degeneracy in Section \ref{secTopDeg} relies on its physical reality.

A simple thought-experiment can further explore the reality of quantized Dirac strings. Consider an infinite superconducting medium of unit-charge particles with a spherical hole that contains a monopole quantum. The superconductor cannot expel the monopole's flux, so it will try to screen it. In typical realistic situations, a superconductor screens magnetic flux by admitting localized Abrikosov vortices via the phase of its order parameter. So, one might naively imagine that a vortex would form near the hole, collect all of the monopole's flux and take it to infinity through a narrow localized tube -- inside which the depletion of the order parameter is physically observable. The first problem with this picture is that an Abrikosov vortex cannot terminate at a point surrounded by the superconductor, such as any point on the surface of the hole. The monopole-screening vortex would necessarily have to stretch between the exterior boundaries of the superconductor, and only pass through the hole. One arm of the vortex would collect the actual monopole's flux, while the other continuation arm would be an avatar of the ``unobservable'' Dirac string (i.e. it would have phase winding without a gauge flux in the core). This is a frustrated situation, the vortex arms must spontaneously choose arbitrary directions. Clearly, one should carefully consider dynamics in order to find out how this frustration is resolved.

The proper approach is to first solve the Schrodinger equation for the superconductor's charged particles in the presence of a monopole. Assuming that we may neglect interactions between the particles, the solution is given by monopole harmonics. The ground state is always degenerate by rotational symmetry: it is an eigenstate of the total angular momentum $L$, but the minimum orbital quantum number is $l=1/2$ due to the angular momentum of the electromagnetic field. This is reviewed in Appendix \ref{appZwanziger}. The superconductor must condense in one of these degenerate single-particle states and effectively break the rotation symmetry by choosing the quantization axis for $L$. Therefore, the physical superconductor's state is biased with respect to rotations according to the quantum dynamics of angular momentum. The ultimate resolution of monopole screening is not very different from the first qualitative picture we built: the gauge-invariant currents of monopole harmonics (\ref{MonHarmonicCurrents}) indeed look much like an Abrikosov vortex that passes through the hole, and build a magnetic moment consistent with the angular momentum $\langle {\bf L}\rangle$ direction.

The lesson learned from this thought experiment is that Dirac strings are only as real as the physical states of particles that espouse them. A physical excitation that carries a quantized line or loop singularity is a vortex, but we can describe it using a singular gauge field. Such excitations can have non-trivial braiding statistics with fractional charges.

Two point-like dyons can also have non-trivial braiding statistics. A part of this braiding statistics is the Aharonov-Bohm phase. However, the Dirac string attached to the monopole appears to make the Aharonov-Bohm phase gauge-dependent. Now we know how to resolve this problem. Solve the Schrodinger equation of a dyon to find the gauge-invariant charge currents in the dyon's ground states. These are given by (\ref{MonHarmonicCurrents}) for an elementary quantized dyon, and strictly related to the physical spin $L=1/2$ of the dyon. Then, carry out the usual singular gauge transformation ${\bf j} = |Y|^2 {\bf a}$ to extract the topological defect into a gauge field. The outcome is:
\begin{equation}
{\bf a}_{+} = -\frac{1}{2r}\frac{\sin\theta}{1-\cos\theta}\,\hat{\boldsymbol{\phi}} \quad,\quad
{\bf a}_{-} = \frac{1}{2r}\frac{\sin\theta}{1+\cos\theta}\,\hat{\boldsymbol{\phi}}
\end{equation}
for the dyon's spin states $L_z = \pm 1/2$. These gauge fields describe the original dyon's monopole, but the attached Dirac string does not have an arbitrary orientation any more -- its orientation is determined by the dyon's spin.



%

\end{document}